\documentclass[twoside,12pt]{elsarticle}
\usepackage{lineno,hyperref}
\usepackage{amsmath}
\usepackage{braket}

\let\imaginary=\Im
\newcommand{\fpKpi}{f_+^{K\pi}(0)}
\newcommand{\fpDK}{f_+^{DK}(0)}
\newcommand{\fpDpi}{f_+^{D\pi}(0)}

\journal{Progress in Particle and Nuclear Physics}

\begin{document}

\begin{frontmatter}

\title{ \vspace{1cm} Experimental Status of the CKM Matrix}
\author{Frank C. Porter}
\address{Lauritsen Laboratory of Physics\\
California Institute of Technology\\
Pasadena, California, USA}

\begin{abstract} 
The CKM matrix, $V$, relates the quark mass and flavor bases. In the standard model, $V$ is unitary $3\times3$, and specified by four arbitrary parameters, including a phase allowing for $CP$ violation.
We review the experimental determination of $V$, including the four parameters in the standard model context. This is an active field; the precision of experimental measurements and theoretical inputs continues to improve. The consistency of the determination with the standard model unitarity is investigated. While there remain some issues the overall agreement with standard model unitarity is good.
\end{abstract}

\end{frontmatter}

\tableofcontents

\section{Introduction}

According to the standard model, quarks come in three families,
with ``up''-type and ``down''-type flavors represented in the mass basis by:
\begin{equation}
U = \begin{pmatrix} u\\ c\\ t \end{pmatrix},\quad 
D = \begin{pmatrix} d\\ s\\ b \end{pmatrix}
\end{equation}
The standard model weak charged current for quarks
may be written
\begin{equation}
 {J}^{\mu+}_{qW} = \textstyle{\frac{1}{\sqrt{2}}}\bar U_L\gamma^\mu V D_L.
\end{equation} 
The hermitian conjugate gives $J^{\mu-}_{qW}$. The $L$ subscript indicates the left-handed projection: $Q_L = \frac{1}{2}(1-\gamma^5)Q$. The quantity of interest to us here is the matrix $V$, called the Cabibbo-Kobayashi-Maskawa (CKM) matrix~\cite{Cabibbo1963,Kobayashi1973}, describing how the mass states are mixed in the weak interaction.

The standard model does not predict $V$; it must be evaluated experimentally. Otherwise,
some physics beyond the standard model is necessary to predict $V$. However, it is not an arbitrary $3\times 3$ complex matrix. In the standard model with 3 generations, it relates two bases, and must be unitrary. Nine real parameters are required to describe an arbitrary $3\times3$ unitary matrix.
Not all of these parameters are of physical significance, five of them can be absorbed as arbitrary phases in
the definition of the quark fields. This leaves four real physical parameters required to completely specify the
CKM matrix in the standard model.  As only three parameters are sufficient to specify a real orthogonal matrix (i.e., a rotation matrix), the matrix is in general complex. This yields a standard model mechanism for $CP$ violation.

The unitarity constraint permits testing the self-consistency of the standard model through
precise measurements of the elements of $V$. If any deviation from unitarity is detected, that is
evidence for new physics, such as the presence of additional generations. Alternatively, if two measurements, via different processes, of the same element in the standard model yield different results, that is also
evidence for new physics. An important class of such searches for new physics
is the possibility of new physics at a high energy scale contributing to some processes via new virtual particle exchange, perhaps in semileptonic decays or in ``penguin diagrams''. For example, this could show up in $CP$ asymmetry measurements. Thus, the measurement of $V$ has two thrusts: First, the four parameters
are fundamental in the standard model, and must be determined by measurement. 
Second, measurements of the CKM matrix elements provide a means to search for
physics beyond  the standard model, or to constrain such theories.

In this article we review the present experimental status of our knowledge of $V$.
For reliability, we include in our averages only published, or accepted for publication, results, with a nominal cutoff of December 31, 2015, although we often mention preliminary work. This remains an active area, both experimentally and theoretically.
Theoretical issues are included only to the extent that they are relevant to the 
measurements of $V$. There are many excellent theoretical reviews; we reference several at appropriate places.

The Review of Particle Properties~\cite{Olive2014} is used for constants such as $G_F$ and particle properties such as masses, lifetimes, etc., except as otherwise noted. In addition, there are several other averaging groups providing extremely useful services, including 
FLAG (Flavor Lattice Averaging Group)~\cite{Aoki2014}, FlaviaNet Kaon Working Group~\cite{Antonelli2010}, and HFAG (Heavy Flavor Averaging Group)~\cite{Amhis2012, Amhis2014, Gershon2015}. In some cases, we have used unpublished averages from these groups, as long as the primary results are published.
A comprehensive review of the physics of the $B$-factories (BaBar and Belle) is available in~\cite{Bevan2014}.
 
In the next section, we'll discuss the parameterization of the CKM matrix. It will be
convenient to separate the discussion of the measurements into the magnitudes of the
elements (Section 3) and the phases via $CP$ violation (Section 4). Then in Section 5 we will briefly discuss three extensive efforts at providing global fits to available information concerning the CKM matrix and possible extensions. We conclude with some fits of our own with discussion 
in Section 6.

\section{Parameters}

There are several useful parameterizations of the CKM matrix. We review here 
the ones that we shall be concerned with.

The conventional labeling for the general $3\times3$ flavor mixing matrix is
\begin{equation}
V = 
\begin{pmatrix}
V_{ud} & V_{us} & V_{ub}\\
V_{cd} & V_{cs} & V_{cb}\\
V_{td} & V_{ts} & V_{tb}\\
\end{pmatrix}
\end{equation}

\begin{center}
 \begin{figure}[h]
\begin{center}
  \includegraphics{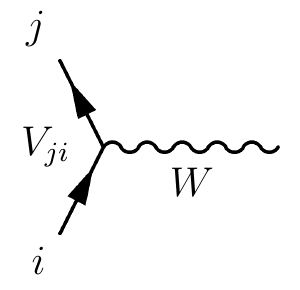}
\end{center}
  \caption{Appearance of CKM matrix element $V_{ji}$ in the Feynman rule for a charged current vertex, where $i$ and $j$
are quark flavor labels.
  \label{fig:vertex}}
 \end{figure}
\end{center}

With this labelling, the charged current vertex gets a factor $V_{ji}$ as in Fig.~\ref{fig:vertex}.
In terms of the flavor labels, $V_{ji}=V_{ij}^*$,  for example,
$V_{du} = V_{ud}^*$.

Standard model unitarity implies nine independent equations relating the elements of $V$:
\begin{itemize}
\item The sum of the absolute squares of the elements in each row (or column) is one:\begin{subequations}
\begin{equation}
|V_{ud}|^2 + |V_{us}|^2 + |V_{ub}|^2 = 1
\end{equation}
\begin{equation}
|V_{cd}|^2 + |V_{cs}|^2 + |V_{cb}|^2 = 1
\end{equation}
\begin{equation}
|V_{td}|^2 + |V_{ts}|^2 + |V_{tb}|^2 = 1
\end{equation}
\label{eq:unitarityEquations}
\end{subequations}
\item The dot product of a column with the complex conjugate of a different column is zero.
This yields  the remaining six equations (considering real and imaginary terms separately):
\begin{subequations}
\begin{equation}
V_{ud}V_{us}^* + V_{cd}V_{cs}^* + V_{td}V_{ts}^* = 0
\end{equation}
\begin{equation}
V_{ud}V_{ub}^* + V_{cd}V_{cb}^* + V_{td}V_{tb}^* = 0
\label{eq:unitarityTriangle}
\end{equation}
\begin{equation}
V_{us}V_{ub}^* + V_{cs}V_{cb}^* + V_{ts}V_{tb}^* = 0
\label{eq:sunitarityTriangle}
\end{equation}
\label{eq:unitarityEquationsOffDiagonal}
\end{subequations}

\noindent Alternatively, we could have taken the dot products by rows, but this does not yield new independent equations (since $V^\dagger V = I \iff VV^\dagger = I$). 
\end{itemize}

These conditions reduce the number of parameters to nine real parameters required to 
define an arbitrary unitary $3\times3$ matrix.  However, there are five arbitrary phases defining the relative quark fields.
When these are chosen, there remain four physical real parameters needed to specify
the CKM matrix.

Since a rotation (orthogonal) matrix in three dimensions is specified by three angles (e.g.,
the Euler angles), it is intuitive to think of the CKM matrix as described by three angles
$\theta_{ij}; i=1,2; i<j\le 3$ corresponding to a rotation, plus an additional phase angle, $\delta$,
giving complex elements to the matrix. Conventionally, the $\theta_{ij}$ are chosen to be in the first quadrant.
The complex phase provides a mechanism for $CP$ violation
in the standard model. Perhaps the most common convention for such a parameterization of
the matrix is~\cite{Chau1984,Olive2014}:
\begin{equation}
\label{eq:Vangles}
V = 
\begin{pmatrix}
c_{12}c_{13} & s_{12}c_{13} & s_{13}e^{-i\delta} \\
-s_{12}c_{23}-c_{12}s_{23}s_{13}e^{i\delta} & c_{12}c_{23}-s_{12}s_{23}s_{13}e^{i\delta} & s_{23}c_{13}\\
s_{12}s_{23}-c_{12}c_{23}s_{13}e^{i\delta} & -s_{23}c_{12}-s_{12}c_{23}s_{13}e^{i\delta} & c_{23}c_{13}\\
\end{pmatrix},
\end{equation}
where $s_{ij}\equiv\sin\theta_{ij}$ and $c_{ij}\equiv\cos\theta_{ij}$.

Empirically, $V$ is approximately diagonal, and this suggests a useful parameterization as an expansion in powers of one of the parameters~\cite{Wolfenstein1983}. Following~\cite{Buras1994, Schmidtler1992}, we define a new set of four real parameters ($\lambda$, $A$, $\rho$, $\eta$, which we shall refer to as Wolfenstein parameters) according to:
\begin{equation}
\begin{split}
\lambda &\equiv s_{12}\\
A\lambda^2 &\equiv s_{23}\\
A\lambda^3(\rho-i\eta) &\equiv s_{13}e^{-i\delta}.\\
\end{split}
\label{eq:Wpars}
\end{equation}
The parameter $\lambda\approx0.22$ then functions as an expansion parameter for 
describing $V$, and we have (e.g., \cite{Hoecker2001}):
\begin{equation}
\label{eq:Vexpansion}
V = 
\begin{pmatrix}
1-\frac{\lambda^2}{2} -\frac{\lambda^4}{8}& \lambda & A\lambda^3(\rho-i\eta) \\
-\lambda\left[1+A^2\lambda^4(\rho+i\eta-\frac{1}{2})\right] & 1-\frac{1}{2}\lambda^2-\frac{1}{8}\lambda^4(1+4A^2) & A\lambda^2\\
A\lambda^3\left[1-(\rho+i\eta)(1-\frac{1}{2}\lambda^2)\right] & -A\lambda^2\left[1+\lambda^2(\rho+i\eta-\frac{1}{2})\right] & 1-\frac{1}{2}A^2\lambda^2\\
\end{pmatrix} + {\cal O}(\lambda^6),
\end{equation}

Measurements of rates are typically sensitive to magnitudes of particular CKM matrix elements,
$|V_{ij}|$. Thus, we shall discuss these magnitudes extensively. In the standard model, they are related according to the unitarity constraints above, but experimentally they provide nine independent quantities, hence allowing for tests of the standard model. If the experimental results do not fit the constraints, that is evidence for new physics.

On the other hand, measurements of $CP$-violating processes are sensitive to the phase $\delta$,
or alternatively the parameter $\eta$. The connection with experiment is often done in the context of the
``unitarity triangles''. Eqs.~\ref{eq:unitarityEquationsOffDiagonal} describe triangles in the complex plane. In the limit of no $CP$-violation, the relative phases of all the elements would be 0 or 180$^\circ$, and the triangles would have zero area. In fact, all of the triangles have the same area, related to a phase-convention invariant quantity called the Jarlskog invariant~\cite{Jarlskog1985}:
\begin{equation}
J = (-1)^{a+b}\imaginary (V_{ij}V_{kl}V_{kj}^*V_{il}^*)
\label{eq:Jinvariant}
\end{equation}
where one row (index $a$) and one column (index $b$) of $V$ is crossed out to obtain the $2\times2$ matrix
\begin{equation}
V_{(ab)} =  
\begin{pmatrix}
V_{ij} & V_{il}\\
V_{kj} & V_{kl}
\end{pmatrix},
\end{equation}
defining indices $i,j,k,l$. The magnitude of $J$, or equivalently the area of the triangles, is a measure of how much $CP$ violation there is in the standard model. It has more recently been pointed out that the Jarlskog invariant can also be expressed as~\cite{Kim2012, Kim2015}:
\begin{equation}
J = \imaginary (V^*_{31}V^*_{22}V^*_{13}),
\end{equation}
assuming $V$ has been expressed in a form with determinant one by multiplying by an overall phase as needed. In the Wolfenstein parameterization, at the order of Eq.~\ref{eq:Vexpansion},
\begin{equation}
J = A^2\lambda^6\eta.
\label{eq:Jwolf}
\end{equation}

\begin{center}
 \begin{figure}[h]
\begin{center}
  \includegraphics{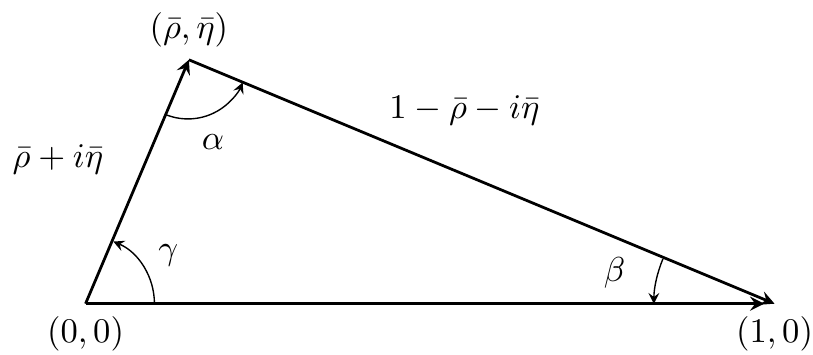}
\end{center}
  \label{fig:unitarityTriangle}
  \caption{The vectors forming the standard unitarity triangle.}
 \end{figure}
\end{center}

One triangle in particular, Eq.~\ref{eq:unitarityTriangle}, has the feature that all of the sides have length of order $A\lambda^3$.
This is usually what we mean when we say the  (standard) ``unitarity triangle''. Dividing through by  $V_{cd}V_{cb}^*$ gives:
\begin{equation}
 1= -\frac{V_{ud}V_{ub}^*}{V_{cd}V_{cb}^*} - \frac{V_{td}V_{tb}^*}{V_{cd}V_{cb}^*}
\label{eq:normalizedUnitarityTriangle}
\end{equation}
It is conventional to define the complex apex point as
\begin{equation}
\label{eq:Vapex}
\bar\rho+i\bar\eta \equiv  -\frac{V_{ud}V_{ub}^*}{V_{cd}V_{cb}^*} = \frac{\sqrt{1-\lambda^2}(\rho+i\eta)}{\sqrt{1-A^2\lambda^4}+\sqrt{1-\lambda^2}A^2\lambda^4(\rho+i\eta)}\approx(\rho+i\eta)(1-\lambda^2/2).
\end{equation}
We may graph this triangle, as in Fig.~\ref{fig:unitarityTriangle}. Three angles are thus defined:
\begin{align}
\alpha &\equiv \arg\left(-\frac{1-\bar\rho-i\bar\eta}{\bar\rho+i\bar\eta}\right)
  =\arg\left(-\frac{V_{td}V_{tb}^*}{V_{ud}V_{ub}^*}\right)\label{eq:alpha}\\
\beta &\equiv\arg\left(-\frac{1}{1-\bar\rho-i\bar\eta}\right)=\arg\left(-\frac{V_{cd}V_{cb}^*}{V_{td}V_{tb}^*}\right)\label{eq:beta}\\
&\approx-\arg V_{td}\\
\gamma &\equiv\arg\left(\bar\rho+i\bar\eta\right)=\arg\left(-\frac{V_{ud}V_{ub}^*}{V_{cd}V_{cb}^*}\right).
\label{eq:gamma}\end{align}
The angles $\beta$, $\alpha$, and $\gamma$ are also commonly called $\phi_1$, $\phi_2$, and $\phi_3$, respectively.

A second unitarity triangle has more recently become experimentally accessible. This is the relation in Eq.~\ref{eq:sunitarityTriangle}, which, dividing by $V_{cs}V_{cb}^*$, is:
\begin{equation}
1 + \frac{V_{ts}V_{tb}^*}{V_{cs}V_{cb}^*}+ \frac{V_{us}V_{ub}^*}{V_{cs}V_{cb}^*} =0.
\end{equation}
Considering the Wolfenstein parameterization in powers of $\lambda$, the first two terms are of order one, while the third term is $O(\lambda^2)$. Hence, this triangle has at least one very small angle,
called $\beta_s$:
\begin{align}
\beta_s&\equiv \arg\left(-\frac{V_{ts}V_{tb}^*}{V_{cs}V_{cb}^*}\right)\sim\lambda^2\bar\eta
\label{eq:betas}\\
&\approx\arg(-V_{ts}).
\end{align}

\section{Magnitudes of CKM elements}

As already noted, rates for weak transitions provide for relatively direct measurements of the magnitudes of the CKM matrix elements. In the following, we review what is known about
each of the nine magnitudes in turn, without assuming unitarity of the matrix. This is a convenient organization also because the results are largely uncorrelated.
We consider the upper left $2\times2$ submatrix elements first, as the best known quantities,
followed by $|V_{ub}|$ and $|V_{cb}|$, then finish with the top-quark elements.

\subsection{$|V_{ud}|$}

The magnitude of $V_{ud}$ is measured in weak $u\leftrightarrow d$ transitions. Thus, the most
promising candidates to study are nuclei, neutrons, and pions. The principal pion decay, $\pi^+\to\mu^+\nu_\mu(\gamma)$, has a rate proportional to $|V_{ud}|^2$, given by:
\begin{equation}
\Gamma\left(\pi^+\to\mu^+\nu_\mu(\gamma)\right) = \frac{G_F^2}{8\pi}f_\pi^2m_\mu^2m_\pi\left(1-\frac{m_\mu^2}{m_\pi^2}\right)^2|V_{ud}|^2\left(1+\frac{\alpha}{\pi}C_\pi\right),
\label{eq:Gpimunu}
\end{equation}
where $G_F$ is the Fermi constant, $f_\pi$ is the pion decay constant, $m_\pi$ is the charged pion mass, $m_\mu$ is the muon mass, and $C_\pi$ allows for radiative corrections, including both virtual and real photons. 
Radiative corrections are not negligible considering available experimental precision, and there is a large literature on this subject. A convenient discussion, though by no means the last word, is provided in~\cite{Marciano1993}. Our main concern here is the dependence on $f_\pi$, which is not perturbatively calculable. The present situation on the lattice calculation of $f_{\pi}$ is reviewed in~\cite{Rosner2015}, where a precision of 1.3\% is quoted. Thus, while the pion lifetime, $\tau_\pi = (2.6033\pm0.0005)10^{-8}$ s~\cite{Olive2014}, is precisely known, it measures the combination $f_\pi^2|V_{ud}|^2$, and
the extraction of $|V_{ud}|$ is hampered by the inability to precisely calculate $f_\pi$. Instead, the
pion decay is typically used to measure $f_\pi$ given other more precise determinations of $|V_{ud}|$. 

A method that is theoretically clean is to look at the rare pion beta decay process $\pi^+\to\pi^0 e^+\nu_e$~\cite{Marciano1987}. The world average branching fraction is dominated by the
most recent measurement from the PIBETA experiment at the Paul Scherrer Institute, yielding $|V_{ud}| = 0.9728(30)$\cite{pocanic2004}. Using the theoretical value for the normalizing branching fraction for $\pi^+\to e^+\nu_e(\gamma)$ instead of the less precise world average, Towner and Hardy~\cite{Towner2010} recommend the value $|V_{ud}|= 0.9742\pm0.0026$. In spite of the
robustness of this approach, the small branching fraction has so far kept it from being as precise as the measurement via nuclear transitions. It provides, however, a valuable consistency check, being independent of nuclear structure.

The lifetime of the neutron is measured to be $\tau_n = 880.3\pm1.1$ s~\cite{Olive2014}. 
There has been substantial movement in the neutron lifetime in recent years, a synopsis of the situation appears in~\cite{Olive2014}. Due to this fluidity, it seems prudent to wait for further developments before reaching firm conclusions based on this quantity.

Presently, the most precise value for $|V_{ud}|$ comes from superallowed $0^+\to 0^+$ nuclear beta transitions. This is largely because these are vector transitions and the conserved vector current hypothesis provides reduced hadronic uncertainties. Over the last several decades there has been considerable experimental effort to improve the rate measurements, and concomitant
improvements in the theoretical evaluation of the few percent corrections for symmetry breaking and radiative diagrams. The corrected $ft$ values (comparative half-lifes) are independent of nucleus, and inversely proportional
to $|V_{ud}|^2$. The most recent update~\cite{Hardy2015} uses an average of the 14 most precise $ft$ values measured on different nuclei, and quotes   
\begin{align}
|V_{ud}|&= 0.97417(21)\\ 
|V_{ud}|^2 &=0.94900(42),
\end{align}
where we quote also the square as being more directly related to the measurements.
This represents a slight shift and slight improvement in precision from the five-year earlier evaluation~\cite{Towner2010} of $|V_{ud}|= 0.97425\pm0.00022$. The dominant uncertainty is theoretical, from the nucleus-independent portion of the radiative corrections~\cite{Marciano2006}.

The measured value of $|V_{ud}|^2$ is many standard deviations different from one, the value
for a world with a single generation. Hence, this measurement alone implies at least two generations are required.

\subsection{$|V_{us}|$}
\label{sec:Vus}

The measurement of the magnitude of $V_{us}$ requires $s\leftrightarrow d$ transitions. The most promising sources are kaon and hyperon decays, as well as $\tau$ decays with strangeness in the final state. 
So far, kaon decays provide the most precise measurements. They may be classified as the purely leptonic decays, $K\to\mu\nu$ (including radiative decays) and semileptonic decays,
$K^0_L\to\pi\ell\nu$, $K^0_S\to\pi \ell\nu$, and $K^\pm\to\pi^0\ell^\pm\nu$. The leptonic $K\to e\nu$
decay rate is suppressed relative to $K\to \mu\nu$ by $\left(m_e/m_\mu\right)^2$, as seen by considering the kaon decay equation analogous to Eq.~\ref{eq:Gpimunu}. 

Extraction of $|V_{us}|$ using the leptonic decay suffers from the same uncertainty, now for the kaon decay constant $f_K$, as for the pion in our discussion of $V_{ud}$. However,
the ratio $f_K/f_\pi$ may be computed rather precisely using lattice QCD (LQCD), and the precise value
of $|V_{ud}|$ may then be used to evaluate $|V_{us}|$:
\begin{equation}
|V_{us}|^2 = |V_{ud}|^2\frac{\Gamma\left(K\to\mu\nu(\gamma)\right)}{\Gamma\left(\pi\to\mu\nu(\gamma)\right)} \left(\frac{f_\pi}{f_K}\right)^2 \frac{m_\pi}{m_K} \left[\frac{1-\left(\frac{m_\mu}{m_\pi}\right)^2}{1-\left(\frac{m_\mu}{m_K}\right)^2}\right]
\frac{1+\frac{\alpha}{\pi}C_\pi}{1+\frac{\alpha}{\pi}C_K}.
\label{eq:VusSquared}
\end{equation}
The ratio of radiative correction factors is taken from~\cite{Antonelli2010}, using the
chiral perturbation theory results in~\cite{Cirigliano2007}. This corresponds to the same central value, but half the uncertainty of the ratio in~\cite{Marciano2004}. Thus, we use
\begin{equation}
\frac{1+\frac{\alpha}{\pi}C_\pi}{1+\frac{\alpha}{\pi}C_K} = 1.0070(18),
\end{equation}
the dominant corrections being the same for the pion and kaon. Two recent four flavor ($N_f=2+1+1$) lattice caculations
of $f_K/f_\pi$ quote a precision of around $0.2$\% (see also~\cite{Bazavov2013}):
\begin{equation}
f_{K^+}/f_{\pi^+} = 
\begin{cases} 1.1916(21)\quad&\hbox{HPQCD~\cite{Dowdall2013}}\\
                     1.1956^{+(27)}_{-(24)}\quad&\hbox{Fermilab Lattice and MILC~\cite{Bazavov2014a}}.
\end{cases}
\end{equation} 
Symmetrizing the second interval and averaging, we use\begin{equation}
f_{K^+}/f_{\pi^+} = 1.1935(21),
\end{equation} 
where we have not reduced the uncertainty below the smaller error because
of potential systematic correlations.

The measured input is summarized in Table~\ref{tab:Vusinput}. Using Eq.~\ref{eq:VusSquared} we find:
\begin{equation}
\begin{split}
\left|\frac{V_{us}}{V_{ud}}\right|^2 &=0.05347(14)_{\Gamma}(10)_C(19)_f,\\
|V_{us}|^2 &= 0.05074(13)_{\Gamma}(2)_{V_{ud}}(9)_C(18)_f,\hskip2cm\smash{\raise9pt\hbox{$(K\to\mu\nu)$}}\\
|V_{us}| &= 0.22526(29)_{\Gamma}(5)_{V_{ud}}(20)_C(40)_f,
\end{split}
\end{equation}
or, quadratically combining the uncertainties,  $|V_{us}|^2 = 0.05074(24)$ and $|V_{us}| = 0.2253(5)$.
This result is close to other determinations, such as in~\cite{Olive2014}, but the 
quoted uncertainty is smaller due to the use of~\cite{Dowdall2013, Bazavov2014a} for the
ratio of decay constants. Nevertheless, this ratio remains the dominant source of uncertainty in $|V_{us}|$. This method introduces a correlation with $|V_{ud}|$. However, the contribution to the
uncertainty from $|V_{ud}|$ is small, and the linear correlation coefficient is only 
$\rho(|V_{us}|^2,|V_{ud}|^2)=\frac{{\rm cov}(|V_{us}|^2,|V_{ud}|^2)}{\sigma_{|V_{ud}|^2}{\sigma_{|V_{us}|^2}}}
=
|\frac{V_{us}}{V_{ud}}|^2\frac{\sigma_{|V_{ud}|^2}}{\sigma_{|V_{us}|^2}}\approx0.09$.
\begin{table}
\caption{Experimental rates in the evaluation of $|V_{us}|$ from leptonic decays. ``Fit'' refers to the constrained fit performed by the Particle Data Group.\label{tab:Vusinput}}
\begin{center}
\begin{tabular}[h]{lrc}
\hline
\multicolumn{1}{c}{Quantity} & \multicolumn{1}{c}{Value} & Reference\\
\hline
$\tau(\pi^+)$ & $2.6033(5)\times10^{-8}$ s & \cite{Olive2014}\\
$B\left(\pi^+\to\mu^+\nu(\gamma)\right)$ & $0.9998770(4)$ & \cite{Olive2014} (fit)\\
$\Gamma\left(\pi^+\to\mu^+\nu(\gamma)\right)$ & $3.8408(7)\times10^{7}$ s$^{-1}$ & \\
$\Gamma\left(K^+\to\mu^+\nu(\gamma)\right)$ & $5.133(13)\times 10^7$ s$^{-1}$ & \cite{Olive2014} (fit)\\
\hline
\end{tabular}
\end{center}
\end{table}

The value of $|V_{us}|$ is also measured in $K\to\pi\ell\nu$, $\ell=e,\mu$ decays, avoiding the dependence on the decay constant, but introducing the form factor parameter $\fpKpi$. 
The total decay
rate for this process may be expressed in terms of the product $|V_{us}|\fpKpi$ according to~\cite{Leutwyler1984, Antonelli2010,Cirigliano2008}:
\begin{equation}
\Gamma_{K_{\ell3}} = \frac{G_F^2m_K^5}{192\pi^3}C_K^2 |V_{us}|^2\fpKpi^2 I_{K\ell}S_{\rm EW}(1+\delta_{\rm EM}+\delta_{SU(2)}),
\label{eq:Kl3rate}
\end{equation}
where $C_K^2$ is 1 or 1/2 for $K^0$ or $K^+$, respectively, and $I_{K\ell}$ is a form factor dependent phase space integral (see~\cite{Antonelli2010} for discussion). The remaining $S_{\rm EW}(1+\delta_{\rm EM}+\delta_{SU(2)})\sim O(1)$ factor includes corrections for short distance electroweak, long distance electromagnetic, and isospin-breaking effects. The correction depends on whether it
is a neutral or charged kaon, and whether the lepton is $e$ or $\mu$. The form factor $\fpKpi$ is by
convention that of the neutral kaon decay.

The FlaviaNet Working Group on Kaon Decays reviewed the $K_{\ell3}$ measurement in 2010~(\cite{Antonelli2010} and references therein). We use the most recent update from~\cite{Moulson2014}, which includes more recent results from KLOE~\cite{Ambrosino2011}, KTeV~\cite{Abouzaid2011}, and NA48/2~\cite{Wanke2012}. It should be remarked that the NA48/2 results are actually unpublished, in violation of our selection criteria. However, the main difference with the 2010 average
is actually in the correction for strong isospin breaking rather than in the additional data, and the influence of the new NA48/2 analysis on the result is very small. The result
after radiative corrections is:
\begin{equation}
\fpKpi|V_{us}| = 0.2165(4).
\end{equation}
This is an average over neutral and charged kaon decays. The largest uncertainty is from
measurement, in the lifetime for $K^0_L$, and in the branching fractions for 
$K^0_S$ and $K^\pm$.

We use the recent lattice result~\cite{Bazavov2014} $\fpKpi=0.9704(32)$ to obtain
\begin{equation}
\begin{split}
|V_{us}|^2 &=0.04978(38) ,\\
|V_{us}| &= 0.2231(8).\hskip2cm\smash{\raise9pt\hbox{$(K_{\ell3})$}}
\end{split}
\end{equation}
The uncertainty is dominated by the uncertainty in $\fpKpi$.
This result is about 2.5 standard deviations
below the $K_{\ell2}$ result.

 We may average the $K_{\ell2}$ and $K_{\ell3}$ results to obtain:
\begin{equation}
\begin{split}
|V_{us}|^2 &=0.05046(20) ,\\
|V_{us}| &= 0.2247(5),
\end{split}
\end{equation}
In this average, the correlation with $|V_{ud}|$ reduces to $\rho(|V_{us}|^2,|V_{ud}|^2)\approx0.08$.
Including the Particle Data Group scaling factor procedure~\cite{Olive2014} the uncertainties
enlarge with a scale factor of $S=2.2$ to (44) and (10), respectively.

The $|V_{us}|$ matrix element is also measured in hyperon decays and in tau decays to strangeness (e.g., see Blucher and Marciano in~\cite{Olive2014} and references therein). The precision and theoretical understanding of these measurements is not competitive with the $K_{\ell2}$ and $K_{\ell3}$ measurements at this time.
In particular, there has been a long-standing discrepancy between $|V_{us}|$ determined from inclusive $\tau\to X_s\nu_\tau$ decays compared with the kaon results. An evaluation in~\cite{Lusiani2014}, based on sum rule and flavor breaking theoretical work described in~\cite{Gamiz2007},  yields $|V_{us}|(\tau\to X_s\nu_\tau) =
0.2176(21)$, which is 3.3 standard deviations smaller than the $K_{\ell2}$ and $K_{\ell3}$ average. This value would also imply a 3.6$\sigma$ deviation from three generation unitarity when combined with the values for $|V_{ud}|^2$ and $|V_{ub}|^2$ in this review.
 Improvements in the theoretical framework have recently been suggested~\cite{Maltman2015}, and a new evaulation of the $\tau$ data yields
$|V_{us}|(\tau\to X_s\nu_\tau) = 0.2228(23)_{\rm exp}(5)_{\rm thy}$.
This value (which uses some unpublished preliminary data from BaBar on $\tau\to K^-\pi^0\nu$~\cite{Adametz2011} that also helps to improve the agreement)
 is consistent with the kaon determination.
The experimental uncertainty dominates, and there is room for the $\tau$ determination
of $|V_{us}|$ to improve, e.g., with Belle-II.

With $|V_{ud}|$ and $|V_{us}|$ we may ask whether a third generation is required.
We find (without using the scaled error for $|V_{us}|$):
\begin{equation}
1-|V_{ud}|^2-|V_{us}|^2 = (0.00054\pm0.00047).
\end{equation}
This is consistent with zero, hence there is no evidence that a third generation exists based on these
values. However, a constraint is obtained on how large the mixing with
the third generation could be or on new physics scenarios such as additional generations.

\subsection{$|V_{cs}|$}
\label{sec:Vcs}

The $c\to s$ transition is the ``Cabibbo-favored'' decay channel for charm. The value of $|V_{cs}|$ is best measured in $D$ and $D_s$ decays analogous to the kaon decays discussed for $|V_{us}|$.
The analog of the $K_{\mu2}$ decay is $D_s\to\ell\nu$ with $\ell=\mu$ or $\tau$ and the analog of the $K_{\ell3}$ decay
is $D\to K\ell\nu$. There are other processes dependent on $|V_{cs}|$, such as charmed baryon  
decays to strangeness and $W\to c\bar s$, but the available precision as measurements of $|V_{cs}|$ is not competitive. 

The Heavy Flavor Averaging Group (HFAG)~\cite{Amhis2014} has averaged the published
results from BaBar~\cite{delAmoSanchez2010b}, Belle~\cite{Zupanc2013}, and CLEO-c~\cite{Alexander2009, Onyisi2009,Naik2009}
for  $D_s\to\ell\nu$, corrected for $\tau$ branching fractions in~\cite{Olive2014} where
relevant, obtaining:
\begin{equation}
\begin{split}
B(D_s\to\mu\nu) &= 5.57(24)\times10^{-3}\\
B(D_s\to\tau\nu) &= 5.55(24)\times10^{-2}
\end{split}
\end{equation}
Using the appropriately relabled Eq.~\ref{eq:Gpimunu}, neglecting the radiative corrections, but
including experimental correlations, these results are combined by HFAG~\cite{Amhis2014}, obtaining:
\begin{equation}
f_{D_s}|V_{cs}| = 250.6\pm4.5 \hbox{ MeV}.
\label{eq:fDsVcs}
\end{equation}

The value of the $f_{D_s}$ decay constant is evaluated in lattice QCD. We use the 
HPQCD collaboration result from~\cite{Davies2010} and the Fermilab Lattice (FNAL)/MILC collaboration
result from~\cite{Bazavov2014a}. We do not include the also recent but less accurate HPQCD result
in~\cite{Na2012}. The average is performed assuming both no correlation and completely
correlated systematics. The central value we quote is the no correlation value, and the uncertainty
is increased linearly by the difference between the correlated and uncorrelated averages,
with the result
\begin{equation}
f_{D_s} = 248.6\pm 1.6\hbox{ MeV}.
\label{eq:fDs}
\end{equation}
Combining Eqs.~\ref{eq:fDsVcs} and \ref{eq:fDs} gives
\begin{equation}
|V_{cs}| = 1.008\pm0.019.
\end{equation}
The lattice calculation has improved to the point where the dominant uncertainty
comes from the experimental measurement.

We may also determine the value of $|V_{cs}|$ using the semileptonic
$D\to K\ell\nu$ process and the analog of Eq.~\ref{eq:Kl3rate}.  Several methods have been
employed and compared to evaluate the form factor dependent integral $I_{D\ell}$, 
including dispersion relations~\cite{Burdman1997}, pole parameterizations~\cite{Becirevic2000}, 
``$z$-expansion''~\cite{Hill2006} (and references therein), and ISGW2~\cite{Scora1995}. Recent analyses typically settle on the $z$-expansion for quoting results.  The basic idea of the $z$-expansion is to map $t=q^2$ to a variable ($z$) such that a Taylor series with good
convergence properties can be used. The chosen mapping is
\begin{equation}
z(t,t_0) \equiv \frac{\sqrt{t_+-t}-\sqrt{t_+-t_0}}{\sqrt{t_+-t}+\sqrt{t_+-t_0}},
\end{equation}
where $t_+=(m_D+m_K)^2$ is the threshold for $DK$ production and $t_0\in(-\infty,t_+)$ is the value of $t$ corresponding to $z=0$, and may be chosen for desirable properties.
The form factor is then expanded in a Taylor series in $z$.
Three terms in the expansion is
usually found to be sufficient.

The most precise published results on the semileptonic decay 
$D\to K\ell^+\nu$ ($\ell = e, \mu$) come from BaBar,
Belle, BESIII, and CLEO-c, see Table~\ref{tab:VcsSLinput}. 
On a 2.9  fb$^{-1}$ dataset taken at the $\psi(3770)$, BESIII reports a very precise result~\cite{Ablikim2015a}, $\fpDK|V_{cd}| = 0.7172(25)(35)$ from $D^0\to K^-e^+\nu$ decays.

\begin{table}
\caption{Experimental measurements of $|V_{cs}|\fpDK$. The first error is statistical, the second systematic. The notation ``10 GeV'' refers to data taken in the $B\bar B$ threshold region, mostly on the $\Upsilon(4S)$.\label{tab:VcsSLinput}}
\begin{center}
\begin{tabular}[h]{llll}
\hline\hline
\multicolumn{1}{c}{Quantity} & Data &\multicolumn{1}{c}{$|V_{cd}|\fpDK$} & Reference\\
\hline
$D^0\to K^-e^+\nu$ & 10 GeV, 75 fb$^{-1}$ & $0.720(7)(7)$ & BaBar~\cite{Aubert2007, Amhis2014}\\
$D^0\to K^-(e^+,\mu^+)\nu$ & 10 GeV, 282 fb$^{-1}$ & $0.692(7)(22)$ & Belle \cite{Widhalm2006, Amhis2014} \\
$D^+\to K_Le^+\nu$ & $\psi(3770)$, 2.92 fb$^{-1}$ & $0.728(6)(11)$ & BESIII~\cite{Ablikim2015}\\
$D^0\to K^-e^+\nu$ & $\psi(3770)$, 2.92 fb$^{-1}$ & $0.7172(25)(35)$ & BESIII~\cite{Ablikim2015a}\\
$D^0\to K^-e^+\nu$ & $\psi(3770)$, 281 pb$^{-1}$ & $0.747(9)(9)$ & CLEO-c untagged~\cite{Dobbs2008}\\
$D^+\to \bar K^0e^+\nu$ & $\psi(3770)$, 281 pb$^{-1}$ & $0.733(14)(11)$ & CLEO-c untagged~\cite{Dobbs2008}\\
$D\to K e^+\nu$ & $\psi(3770)$, 818 pb$^{-1}$ & $0.719(6)(5)$ & CLEO-c tagged~\cite{Besson2009}\\
\quad$D^0\to K^-e^+\nu$ & $\psi(3770)$, 818 pb$^{-1}$ & $0.726(8)(4)$ & CLEO-c tagged~\cite{Besson2009}\\
\quad$D^+\to\bar K^0e^+\nu$ & $\psi(3770)$, 818 pb$^{-1}$ & $0.707(10)(9)$ & CLEO-c tagged~\cite{Besson2009}\\
\hline
$D\to K\ell^+\nu$ &  & $0.7208(33)$ & Our average\\
\hline\hline
\end{tabular}
\end{center}
\end{table}

In Table~\ref{tab:VcsSLinput} the 818 pb$^{-1}$ dataset from CLEO-c includes the 281 pb$^{-1}$ dataset, and there is
some correlation between the tagged and untagged analyses in both statistical and systematic
errors. The correlations, including as well those between the charged and neutral $D$ channels, have been derived by CLEO-c on the 281 pb$^{-1}$ dataset~\cite{Ge2009}.
We use the correlation information (including the assumption that all of the 818 pb$^{-1}$
dataset has the same correlation in the systematic errors with the untagged analysis) to obtain
an average of the data in Table~\ref{tab:VcsSLinput}: $|V_{cd}|\fpDK = 0.7208(33)$. The 
$p$ value for the $\chi^2$ statistic is 0.004. The correlation between the two BESIII results is small, and is neglected.
This result is close to the HFAG average~\cite{Amhis2012, Amhis2014} (0.728(5)),
except that we now use the published BESIII result for $D^0\to K^-e^+\nu$ and we also include the BESIII $D^+\to K_Le^+\nu$ measurement.

It is noted in~\cite{Antonelli2010} that the HFAG averaging
does not remove the final-state Coulomb correction in the $D^0$ channel prior to averaging.
This and other corrections are potentially important now that the measurements have become
precise. However, this determination of $|V_{cs}|$ is limited by the uncertainty
in $\fpDK$, hence neglect of these corrections is presently safe enough.

The smallness of the $p$ value is of some concern. It is possible that we are misestimating
the correlations in our treatment of the CLEO data. However, it may also be an indication
that the neglect of the different electromagnetic corrections between $D^+$ and $D^0$ is
no longer justified. The incorporation of such corrections should be performed as part of the analysis. However, they naively could be as large as $O(\%)$~\cite{Atwood1990}. To investigate the possible effect, ``correcting'' the $D^0$ numbers by a factor of 0.99 improves the
$p$ value to 2\%, while changing the average to 0.7194(33); a factor of 0.98 improves it to 5\%, while changing the average to 0.7181(33). As already noted the uncertainty for $|V_{cs}|$ remains dominated by the uncertainty in $\fpDK$, so this remains a secondary issue for our
discussion. Nevertheless, we suggest that future evaluations of $\fpDK|V_{cs}|$ include
such corrections in the analysis over the Dalitz plot. 

To extract $|V_{cs}|$, the form factor at zero recoil, $\fpDK$, is required, and
lattice calculations are available. The FLAG evaluation~\cite{Aoki2014} is
\begin{equation}
\fpDK = 0.747\pm 0.019.
\end{equation}
Due to the FLAG quality and publication requirements, this is just the HPQCD evaluation
from Ref.~\cite{Na2010}. This yields $|V_{cs}|=0.965(25)$.

The leptonic and semileptonic results are consistent ($p(\chi^2)= 0.24$), averaging them yields:
\begin{equation}
\begin{split}
|V_{cs}|^2 &=0.983(30) ,\\
|V_{cs}| &= 0.992(15).
\end{split}
\end{equation}

The use of the  detailed $q^2$ dependence in $D\to K\ell\nu$ decays to improve
the measurement of $|V_{cs}|$ is discussed in a preprint~\cite{Koponen2013} from
the HPQCD Collaboration, in which a value of $|V_{cs}|=0.973(5)_{\rm exp}(14)_{\rm lattice}$
is obtained. 
It should be noted that the lattice QCD calculations are also improving for semileptonic decays
to vector mesons, for example $D_s\to\phi\ell\nu$. A recent evaluation~\cite{Donald2014} extracts
$|V_{cs}|=1.017(63)$ using the branching fraction measured by BaBar~\cite{Aubert2008}.
The error is dominated by the theoretical uncertainties, but not by much.

\subsection{$|V_{cd}|$}

Completing the upper left $2\times2$ submatrix is the ``Cabibbo-suppressed'' $c\to d$ transition. Early measurements of $|V_{cd}|$ were performed in neutrino production of charm, and this method remains in principle competitive. Lattice calculation of form factors has advanced such that the semileptonic
$D\to\pi\ell\nu$ decay is also useful, analogous to $D\to K\ell\nu$ for $|V_{cs}|$. Also similarly with
other elements, the leptonic decay $D^+\to\ell^+\nu$ provides a measurement of $|V_{cd}|$ if
lattice calculations are used for the $f_D$ decay constant.

The neutrino measurements consist in measuring di-muon production, where one muon
is the result of a charged current interaction (providing a $d\to c$ transition, hence dependence
on $|V_{cd}|$), and the second muon tags the decay of a charmed hadron. The measurement is
reported as the product $B_\mu|V_{cd}|^2$, where $B_\mu$ is the semileptonic branching fraction of charmed hadrons, as appropriate to the experimental conditions.  

There has not been much recent development in this area in the neutrino experiments. The 2004
Review of Particle Properties (RPP)~\cite{Gilman2004} quotes an average for the CDHS~\cite{Abramowicz1982}, CCFR~\cite{Rabinowitz1993,Bazarko1995}, and CHARM II~\cite{Vilain1999} measurements of $B_\mu|V_{cd}|^2=0.00463(34)$. An evaluation~\cite{Lellis2004} of $B_\mu$ for the nominal kinematic regime (visible energy $>\sim 30$ GeV) is combined in the 2014 RPP~\cite{Olive2014} with
a measurement from CHORUS (nuclear emulsion)~\cite{KayisTopaksu2005} obtaining $B_\mu=0.087(5)$, and thence $|V_{cd}|=0.230(11)$. There is
a further result from CHORUS based on events produced in the lead-scintillating fiber calorimeter~\cite{KayisTopaksu2008}, which has not been included in this average. 

However, the review in~\cite{Lellis2004} notes the inconsistency in combining leading order (LO)
and next-to-leading order (NLO) determinations, and quotes separate results. The LO results
from CDHS, CCFR, and CHARM II are averaged obtaining  $|V_{cd}|_{\rm LO}=0.232(10)$, while
the NLO result from CCFR yields $|V_{cd}|_{\rm NLO}=0.246(16)$. We do not include the neutrino in our average for $|V_{cd}|$, opting instead for the in any event presently more precise results from $D$ meson decays.

As with earlier elements, $|V_{cd}|$ may be measured in leptonic and semileptonic decays,
in particular, $D\to\ell\nu$ and $D\to\pi\ell\nu$. For the leptonic channel, the $f_{D^+}$ decay
constant is required, and this is computed in lattice QCD. These calculations are becoming rather precise (see~\cite{Rosner2015} for a recent summary), the recent result from the Fermilab Lattice/MILC colaboration~\cite{Bazavov2014a} is
\begin{equation}
f_{D^+} = 212.6\pm0.4^{+1.0}_{-1.2}\hbox{ MeV}.
\end{equation}
The first uncertainty is statistical, and the second is systematic, dominated by the uncertainty in the
continuum extrapolation. We note that we are entering the precision regime where the
distinction between $f_{D^+}$ and $f_{D^0}$ is becoming important.

The most precise measurements of the $D\to\mu\nu$ branching fraction are from
CLEO~\cite{Eisenstein2008}, $(3.82\pm0.32\pm0.09)\times10^{-4}$ and BESIII~\cite{Ablikim2014}, $(3.71\pm0.19\pm0.06)\times10^{-4}$, which may be averaged to obtain~\cite{Amhis2014}
\begin{equation}
B(D\to\mu\nu) = (3.74\pm0.17)\times10^{-4}.
\end{equation} 
With the above value for $f_{D^+}$, this yields, with the appropriately modified Eq.~\ref{eq:Gpimunu} and neglecting the radiative correction term,  $|V_{cd}|^2 = 0.0467(22)$, or  $|V_{cd}| = 0.216(5)$. The error is completely dominated by the uncertainty on the branching fraction.

The most precise published results on the semileptonic decay 
$D\to \pi\ell^+\nu$ ($\ell = e, \mu$) have come from BaBar,
Belle, BESIII, and CLEO-c, see Table~\ref{tab:VcdSLinput}. 
On a 2.9  fb$^{-1}$ dataset taken at the $\psi(3770)$, BESIII reports a very precise result~\cite{Ablikim2015a}, $\fpDpi|V_{cd}| = 0.1435(18)(9)$ from $D^0\to\pi^-e^+\nu$ decays. 
Just as for the $D\to K\ell\nu$ case, the 818 pb$^{-1}$ dataset from CLEO-c includes the 281 pb$^{-1}$ dataset, and there is
some correlation between the tagged and untagged analyses in both statistical and systematic
errors. 
As before, we use the correlation information~\cite{Ge2009} to obtain
an average of the data in Table~\ref{tab:VcdSLinput}: $|V_{cd}|\fpDpi = 0.1432(16)$. The 
$p$ value for the $\chi^2$ statistic is 0.53.
This result is very close to the HFAG average~\cite{Amhis2014} (0.1425(19)), except that we now have the published BESIII result. 

\begin{table}
\caption{Experimental measurements of $|V_{cd}|\fpDpi$. The first error in statistical, the second systematic. The third error for the BaBar result is from the uncertainty in the normalizing channel. The notation ``10 GeV'' refers to data taken in the $B\bar B$ threshold region, mostly on the $\Upsilon(4S)$.\label{tab:VcdSLinput}}
\begin{center}
\begin{tabular}[h]{llll}
\hline\hline
\multicolumn{1}{c}{Quantity} & Data &\multicolumn{1}{c}{$|V_{cd}|\fpDpi$} & Reference\\
\hline
$D^0\to\pi^-e^+\nu$ & 10 GeV, 384 fb$^{-1}$ & $0.1374(38)(22)(9)$ & BaBar~\cite{Lees2015}\\
$D^0\to\pi^-(e^+,\mu^+)\nu$ & 10 GeV, 282 fb$^{-1}$ & $0.140(4)(7)$ & Belle \cite{Widhalm2006, Amhis2014} \\
$D^0\to \pi^-e^+\nu$ & $\psi(3770)$, 2.92 fb$^{-1}$ & $0.1435(18)(9)$ & BESIII~\cite{Ablikim2015a}\\
$D^0\to\pi^-e^+\nu$ & $\psi(3770)$, 281 pb$^{-1}$ & $0.140(7)(3)$ & CLEO-c untagged~\cite{Dobbs2008}\\
$D^+\to\pi^0e^+\nu$ & $\psi(3770)$, 281 pb$^{-1}$ & $0.138(11)(4)$ & CLEO-c untagged~\cite{Dobbs2008}\\
$D\to\pi e^+\nu$ & $\psi(3770)$, 818 pb$^{-1}$ & $0.150(4)(1)$ & CLEO-c tagged~\cite{Besson2009}\\
\quad$D^0\to\pi^-e^+\nu$ & $\psi(3770)$, 818 pb$^{-1}$ & $0.152(5)(1)$ & CLEO-c tagged~\cite{Besson2009}\\
\quad$D^+\to\pi^0e^+\nu$ & $\psi(3770)$, 818 pb$^{-1}$ & $0.146(7)(2)$ & CLEO-c tagged~\cite{Besson2009}\\
\hline
$D\to\pi\ell^+\nu$ &  & $0.1432(16)$ & Our average\\
\hline\hline
\end{tabular}
\end{center}
\end{table}

Using the lattice QCD result for $\fpDpi$, we may extract a measurement of $|V_{cd}|$.
The FLAG collaboration~\cite{Aoki2014}  identifies one calculation as meeting their
criteria, that of~\cite{Na2011}, with $\fpDpi = 0.666(29)$.
Using this value, we obtain the semileptonic result: $|V_{cd}|^2 = 0.0462(42)$, or  $|V_{cd}| = 0.215(10)$,
where the dominant uncertainty is from the lattice calculation. The nearly 10\% uncertainty in $|V_{cd}|^2$ justifies the neglect of higher order corrections.

The leptonic and semileptonic results are consistent, combining them we obtain:
\begin{equation}
\begin{split}
|V_{cd}|^2 &=0.0466(19) ,\\
|V_{cd}| &= 0.216(5).
\end{split}
\end{equation}

It may be noted that there are small correlations between $|V_{cs}|$ and
$|V_{cd}|$ due to the presence of correlated systematic uncertainties in some of the measurements. We neglect these correlations.

\subsection{$|V_{cb}|$}
\label{sec:Vcb}

Weak decays of hadrons containing a $b$ quark decay most often to charmed hadrons, so
we consider $|V_{cb}|$ next. There are two general approaches to measuring $|V_{cb}|$ with competing strengths, namely inclusive and exclusive semileptonic $B$ decays to charm. The idea is very similar to the elements already discussed; however, there is an important difference in
practice. Both the $c$ and $b$ quarks are ``heavy'', on the scale of $\Lambda_{\rm QCD}$, and
Heavy Quark Effective Field Theory (HQET; for a brief review, see Bauer and Neubert in~\cite{Olive2014}) provides a useful theoretical framework for confronting the soft physics.

The most precise measurements in exclusive semileptonic decays are in $B\to D^*\ell\nu$ and $B\to D\ell \nu$, with $\ell = e$ or $\mu$.
The differential rate for these decay modes may be expressed in terms of a kinematic variable, $w$, and a decay-specific form factor (Kowalewski and Mannel in~\cite{Olive2014}, \cite{Caprini1998}, \cite{Neubert1994}): 
\begin{equation}
\frac{d\Gamma(\bar B\to D^{(*)}\ell\nu)}{dw} = \frac{G_F}{48\pi^3}|V_{cb}|^2m_{D^{(*)}}^3
\sqrt{w^2-1}\eta_{\rm EW}^2
  \begin{cases}
    (m_B+m_D)^2(w^2-1)G(w)^2& \text{$D$},\\
    (m_B-m_{D^*})^2(w+1)^2\left[1+\frac{4w}{w+1}\frac{1-2wr+r^2}{(1-r)^2}\right]F(w)^2& \text{$D^*$.}
  \end{cases}
\label{eq:BtoDlnu}
\end{equation}
The variable $w=v\cdot v^\prime$ is the scalar product of the four-velocities of the $B$ and $D^{(*)}$ mesons. In the $B$ rest frame, $w =E_{D^{(*)}}/m_{D^{(*)}}$ is just the gamma-factor of the $D^{(*)}$ meson. The quantity $r$ is the mass ratio $r=m_{D^*}/m_B$. Higher order electroweak
corrections are contained in $\eta_{\rm EW}=1.00662(16)$~\cite{Bailey2014, Aoki2014, Sirlin1982}. Coulomb corrections in neutral $B$ decays have begun to be important on the current scale of precision,
and $\eta_{\rm EW}$ is now replaced with the symbol $\bar\eta_{\rm EW}$ to incorporate these corrections.  The lepton masses
are assumed to be negligible, i.e., we are assuming $\ell = e$ or $\mu$. We note that the $w^2-1$
factor for the $\bar B\to D\ell\nu$ case is a reflection of the helicity suppression at zero recoil, since the
leptonic system must then have angular momentum zero; for $\bar B\to D^*\ell\nu$ there is no such suppression.

The normalization is such that, in the limit of infinite $c$ and $b$ quark masses, the form factors 
$G$ and $F$ equal one at zero recoil, that is at $w=1$. Measuring the differential decay rates in Eq.~\ref{eq:BtoDlnu} provides measurements of $|V_{cb}|G(w)$ and $|V_{cb}|F(w)$. In order to precisely determine $|V_{cb}|$, it is necessary to account for finite mass effects and to in practice perform an extrapolation to $w=1$. That is, the measurements are quoted in terms of $|V_{cb}|G(1)$ and $|V_{cb}|F(1)$, and then lattice calculations are used to provide $G(1)$ and $F(1)$. This approach is attractive also because the $O(\Lambda_{\rm QCD}/m_Q)$ corrections for $\bar B\to D^*\ell\nu$ vanish by Luke's theorem~\cite{Luke1990}.

For the $w$-dependence of the form factors it is conventional to use the formalism in~\cite{Boyd1995, Boyd1996, Boyd1997, Caprini1998}, where unitarity and analyticity are used to derive constraints on the form factors.
This is reminiscient of the $z$-expansion employed for $|V_{cs}|$ (Section~\ref{sec:Vcs}).
Expansions about the zero recoil point are used. In the ``CLN'' approach~\cite{Caprini1998} for $G(w)$, an expansion depending on two
parameters, $G(1)$ and ``slope'' $\rho^2$, provides a good approximation. For $F(w)$, two
additional parameters, form factor ratios denoted as $R_1(1)$ and $R_2(1)$, are needed. These parameters are 
determined in fits to the data. It may be remarked that the precision in $|V_{cb}|$ has improved to the point that the approximate expansion in~\cite{Caprini1998} may no longer be adequate~\cite{Gambino2016}, and use of~\cite{Boyd1995, Boyd1996, Boyd1997} preferred. A recent result from Belle~\cite{Glattauer2016} obtains $\eta_{\rm EW}|V_{cb}| = 40.12(1.34)\times10^{-3}$ using~\cite{Caprini1998} and $\eta_{\rm EW}|V_{cb}| = 41.10(1.14)\times10^{-3}$ using~\cite{Boyd1995, Boyd1996, Boyd1997} (truncating after $z^3$ in the expansion). The shift between these two values is approaching one standard deviation.  For consistency, we have used the results obtained with~\cite{Caprini1998} in our average.

The most precise measurements of $|V_{cb}|$ in $B\to D\ell\bar\nu$ use data from BaBar~\cite{Aubert2010} and a very recent analysis from Belle~\cite{Glattauer2016}. A recent unquenched lattice evaluation of the form factors at non-zero recoil applied to the BaBar data yields~\cite{Bailey2015a} $|V_{cb}|=0.0396(17)(2)$, where the first uncertainty is experimental plus lattice and the second uncertainty is for the omitted QED corrections, in particular the Coulomb correction for the $B^0$ channel. With slight differences, the Belle result is $\eta_{\rm EW}|V_{cb}|=0.0401(13)$, corresponding, with our above value for $\eta_{\rm EW}$, to $|V_{cb}|=0.0399(13)$. The average of
the BaBar and Belle results is  $|V_{cb}|=0.0398(11)$. This average is computed assuming 100\% 
correlation from the lattice input, although this is negligible here. The error estimate is dominated by the experimental uncertainties.

There is also a recent lattice calculation of the form factor at zero recoil for $B\to D^*\ell\bar\nu$~\cite{Bailey2014}, $F(1) = 0.906(4)(12)$, where the first uncertainty is statistical and the second systematic. This reference has also made an estimate for the Coulomb corrections for the
neutral $B$ mode. We update their result for $|V_{cb}|$ with the most recent HFAG average~\cite{Amhis2014} of the experimental measurements from ALEPH~\cite{Buskulic1997}, BaBar~\cite{Aubert2008c, Aubert2008d, Aubert2009d}, Belle~\cite{Dungel2010}, CLEO~\cite{Adam2003}, DELPHI~\cite{Abreu2001, Abdallah2004}, and OPAL~\cite{Abbiendi2000},  
$|V_{cb}|\bar\eta_{\rm EW}F(1)=0.03581(45)$, obtaining
$|V_{cb}|= 0.03894(49)(53)(19)$, where the first uncertainty is experimental, the second from the lattice, and the third for the Coulomb correction.

We may combine the $D\ell\bar\nu$ and $D^*\ell\bar\nu$ results to obtain the measured
value of $|V_{cb}|$ from exclusive channels:
\begin{equation}
\begin{split}
|V_{cb}|^2(\hbox{excl})&=0.00154(4)(3)\\
|V_{cb}|(\hbox{excl})&=0.0392(5)(3),\\
\end{split}
\label{eq:Vcbex}
\end{equation}

\noindent where the first error bar is experimental and the second theoretical. We anticipate that precise results on $|V_{cb}|$ from $B_s\to D_s$ semileptonic decays
will become available from the LHC and Belle II.

Inclusive measurements of $|V_{cb}|$ have comparable precision. In this case, 
the measurement is performed in the context of the operator product expansion (OPE).
The non-perturbative contributions are described in terms of a small number of parameters
that are estimated in fits to the data. The data inputs are moments of kinematical
quantities, such as the lepton energy and the mass of the hadronic system.
A review of both the theoretical background and experimental methods may be found in~\cite{Bevan2014}. 

A recent calculation,~\cite{Alberti2015} and references therein, includes the complete
$O(\alpha_s\Lambda^2_{\rm QCD}/m_b^2)$ effects in the theoretical semileptonic rate. Using published data~\cite{Gambino2014} on the moments of the inclusive $b\to c$ semileptonic spectra from BaBar~\cite{Aubert2010a, Aubert2004}, Belle~\cite{Schwanda2007, Urquijo2007}, CDF~\cite{Acosta2005}, CLEO~\cite{Csorna2004}, and Delphi~\cite{Abdallah2005}, the value for $|V_{cb}|$
obtained is~\cite{Alberti2015}:
\begin{equation}
\begin{split}
|V_{cb}|^2(\hbox{incl})&=0.00178(3)(6)\\
|V_{cb}|(\hbox{incl})&=0.0422(3)(7),\\
\end{split}
\end{equation}
where the first uncertainty approximates the experimental uncertainty and the second the theoretical uncertainty.

The inclusive value for $|V_{cb}|$ is larger than the exclusive value by three standard
deviations (two-sided $p$ value of 0.003 if Gaussian sampling is assumed). This
has been a long-standing discrepancy with no clear resolution, although there may be
unincluded systematic effects in both experiment and theory (e.g.,~\cite{Ricciardi2014}). The approximate $z$-expansion in the exclusive result, as discussed above, might be partly responsible for the difference, however the results
in~\cite{Gambino2016} and~\cite{Glattauer2016} yield contradictory directions to the correction.
Attempts to explain the discrepancy between exclusive and inclusive determinations
of $|V_{cb}|$ and of $|V_{ub}|$ (next section) in terms of new physics run into
other experimental constraints  and are hence implausible~\cite{Crivellin2015}. We'll return to this issue in Section~\ref{sec:discussion}.
 
Finally, we briefly comment on the channels $B\to D^{(*)}\tau\nu$, which is also sensitive to $|V_{cb}|$. This is experimentally
much more challenging than the corresponding decays with an electron or muon, because of the difficulty in distinguishing the tau
from background. It is also sensitive to possible new physics, for example in the Higgs sector. Indeed, experimental results, from
BaBar~\cite{Lees2012c}, Belle~\cite{Huschle2015, Abdesselam2016}, and LHCb~\cite{Aaij2015i}
show some deviation from the standard model. This is an area of ongoing investigation, with a need for more experimental input;
conclusive input may have to await Belle-II and LHCb upgrade running.

\subsection{$|V_{ub}|$}

The furthest off-diagonal element of the first row describes the weak coupling of the $b$ quark to the $u$ quark. The measurement of $|V_{ub}|$ is analogous with the measurement of
$|V_{cb}|$, but more difficult due to the smaller branching fractions and larger relative backgrounds. Again, both exclusive and inclusive approaches are taken.

So far, the most precise exclusive measurement of $|V_{ub}|$ is obtained with 
$B\to\pi\ell\nu$ semileptonic decays. However, the LHCb collaboration has recently
published~\cite{Aaij2015d} a competitive result for the ratio $|V_{ub}/V_{cb}|$ using semileptonic
$\Lambda_b$ decays to baryons, $\Lambda_b\to p\mu^-\bar\nu_\mu$ and $\Lambda_b\to \Lambda_c^+\mu^-\bar\nu_\mu$:
\begin{equation}
\left|\frac{V_{ub}}{V_{cb}}\right| = 0.083(6).
\end{equation}
The error estimate has equal contributions from experimental uncertainties and LQCD uncertainties~\cite{Detmold2015}. Using our exclusive average for $|V_{cb}|$, Eq.~\ref{eq:Vcbex},
we obtain $|V_{ub}| = 0.00325(16)(16)$, separating the theoretical and experimental uncertainties. This result is correlated with the result for $V_{cb}$, with correlation coefficient $\rho\sim0.2$.

Two recent lattice QCD calculations of the form factors for $B\to\pi\ell\nu$ semileptonic decays
have been published, one using the MILC asqtad (2+1)-flavor ensembles~\cite{Bailey2015} and the other using domain-wall light quarks~\cite{Flynn2015}. Both evaluations are applied to
measurements from BaBar~\cite{delAmoSanchez2011,Lees2012} and Belle~\cite{Ha2011,Sibidanov2013} to obtain $|V_{ub}|$. Their results are close to each other, $|V_{ub}|=0.00372(16)$~\cite{Bailey2015} and $|V_{ub}|=0.00361(32)$~\cite{Flynn2015},
we adopt the more precise determination. We infer an experimental uncertainty of (10) and an theoretical uncertainty of (12) from the discussion in~\cite{Bailey2015}.

The RBC/UKQCD collaboration~\cite{Flynn2015} also computes the $B_s\to K\ell\nu$
form factors, which can be used to measure $|V_{ub}|$ from that channel once experimental
results are available.

Other measurements of $|V_{ub}|$ in exclusive reactions have been made
in $B\to\rho\ell\nu$, $B\to\omega\ell\nu$, and $B\to\tau\nu$.
Both BaBar~\cite{delAmoSanchez2011, Lees2013} and Belle~\cite{Sibidanov2013} report 
measurements in $B\to\rho\ell\nu$ and $B\to\omega\ell\nu$. However, more theoretical work is needed to bring these evaluations of $|V_{ub}|$ to the same level as for the $\pi\ell\nu$ channel. It has been suggested~\cite{Kang2014} that the model dependence in $B\to\rho\ell\nu$ may be avoided by a model-independent analysis of the $\pi\pi\ell\nu$ final state, using dispersion theory for the $\pi\pi$ form factors. We do not include these measurements in our average. We also note a recent analysis~\cite{Hsiao2016} using the branching fractions for the baryonic decay modes $B^-\to p\bar p \pi^-$ and $\bar B^0\to p\bar p D^0$ to extract the ratio $|V_{ub}/V_{cb}|=0.088^{+0.022}_{-0.016}\pm0.010$, where the first error is theoretical and the second is experimental.

The decay $B\to\tau\nu$ provides for a measurement of the product $|V_{ub}|f_B$, where $f_B$ is the $B$ meson decay constant, according to the appropriately relabeled Eq.~\ref{eq:Gpimunu}.
The branching fraction for this decay has been measured by BaBar~\cite{Aubert2010b,Lees2013a} and Belle~\cite{Hara2013, Kronenbitter2015} and earlier references cited in these. Both experiments provide experiment averages of their hadronic and semileptonic $B$-tag results. The ``tag'' $B$ refers to the other $B$ in the event, not decaying to $\tau\nu$. Signatures of the tag $B$ are used to reduce background. This may either be through a full or partial reconstruction of the tag $B$ in a hadronic decay mode, or by detecting a semileptonic decay with an identified $e$ or $\mu$.
We average the experiments to obtain $B(B^+\to\tau^+\nu_\tau)=1.06(20)\times10^{-4}$, with
a $\chi^2$ probability for consistency between the experiments of 10\%. 
This translates to $|V_{ub}|^2f_B^2=60(11)\times10^{-8}$ GeV$^2$, using Eq.~\ref{eq:Gpimunu} (relabeled), but neglecting the radiative correction term.

To extract a value for $|V_{ub}|$, we need to divide out the $B^+$ decay constant. For $f_B$ we use lattice QCD, where an $N_f=2+1+1$ calculation~\cite{Aoki2014, Dowdall2013a}  gives $f_B=0.184(4)$ GeV. Thus, the $B\to\tau\nu$ measurement implies
$|V_{ub}|=0.0042(4)$, consistent, with a large uncertainty, with the $B\to\pi\ell\nu$ semileptonic value.

 \begin{table}[h]
\caption{Summary of measurements of $|V_{ub}|$ in exclusive channels. The first uncertainty is experimental, the second thoeretical.\label{tab:VubExcl}}
\begin{center}
\begin{tabular}[h]{lll}
\hline\hline
\multicolumn{1}{c}{Channel} & $|V_{ub}|$ & References\\
\hline
$B\to\pi\ell\nu$ &  $0.00372(10)(12)$ & \cite{delAmoSanchez2011,Lees2012, Ha2011,Sibidanov2013, Bailey2015}\\
$\Lambda_b\to p\mu\nu$  & $0.00325(16)(16)$ & \cite{Aaij2015d, Detmold2015}\\
$B\to\tau\nu$ & $0.00422(40)(9)$ & \cite{Lees2013a, Kronenbitter2015, Dowdall2013a} \\
\hline\hline
\end{tabular}
\end{center}
\end{table}

The exclusive measurements of $|V_{ub}|$ that go into our average are summarized in Table~\ref{tab:VubExcl}. We note that the lattice evaluation performed in~\cite{Bailey2015} uses a blinded method to avoid subjective bias.
The resulting average is: 
\begin{equation}
\begin{split}
|V_{ub}|^2(\hbox{excl})&=1.29(7)(6)\times10^{-5}\\
|V_{ub}|(\hbox{excl})&=0.00359(9)(8),\\
\end{split}
\label{eq:Vubex}
\end{equation}
where the first uncertainty is experimental and the second theoretical, both here and in Eq.~\ref{eq:Vubexnotau} below.
The $p$ value for consistency of the three measurements is 8\%.
If we exclude the purely leptonic decay from our exclusive average, as has so far been conventional (at least partly because $B\to\tau\nu$ may be more sensitive to new physics contributions), we obtain
\begin{equation}
\begin{split}
|V_{ub}|^2(\hbox{excl})&=1.26(6)(7)\times10^{-5}\hbox{\ \ no $B\to\tau\nu$}\\
|V_{ub}|(\hbox{excl})&=0.00354(9)(10)\hbox{\hskip.78cm no $B\to\tau\nu$}\\
\end{split}
\label{eq:Vubexnotau}
\end{equation}
with a consistency $p$ value of 9\%. Our estimate for the correlation coefficient with 
$|V_{cb}|^2(\hbox{excl})$ is $\rho\left[|V_{ub}|^2(\hbox{excl}),|V_{cb}|^2(\hbox{excl})\right] \approx 0.14$.

As with $|V_{cb}|$, $|V_{ub}|$ may be measured with an inclusive semileptonic approach as well. The measurement is complicated by the large background from $B\to X_c\ell\nu$ decays.
In principle, this can be dealt with by selecting leptons above the endpoint for $b\to c$ transitions. However, this leads to theoretical difficulties with the operator product expansion,
and the determination of $|V_{ub}|$ in inclusive semileptonic decays is a story in
tradeoffs between the theoretical difficulties near the endpoint and experimental backgrounds
far from the endpoint. 
Several different theoretical calculations of the inclusive partial decay rate in QCD have been proposed,
most labeled by authors' initials: ADFR~\cite{Aglietti2004, Aglietti2007, Aglietti2009}, BLL~\cite{Bauer2001}, BLNP~\cite{Bosch2004, Bosch2004a, Lange2005}, DGE (``dressed-gluon exponentiation'')~\cite{Andersen2006}, and GGOU~\cite{Gambino2007}.
Reviews of the experimental and theoretical issues may be found in~\cite{Bevan2014}, Kowalewski and Mannel in~\cite{Olive2014}, and~\cite{Amhis2014}. The HFAG compilation~\cite{Amhis2012, Amhis2014} considers data from BaBar~\cite{Aubert2005, Aubert2006, Lees2012}, Belle~\cite{Kakuno2004,Limosani2005, Bizjak2005, Urquijo2010}, and CLEO~\cite{Bornheim2002}. BaBar~\cite{Lees2012} and Belle~\cite{Urquijo2010}
have published results on their entire datasets with a large portion of the phase space included; these results provide for the smallest theoretical uncertainties. With one exception, all of the measurements have been compiled by HFAG under the four schemes, ADFR, BLNP, DGE, and GGOU. We present the 2014 updated averages for $|V_{ub}|$ under these schemes in the middle column of Table~\ref{tab:VubIncl}.

\begin{table}
\caption{Summary of average $|V_{ub}|$ from inclusive semileptonic $B$ decays,
according to different theoretical calculations~\protect{\cite{Amhis2014}}.
The first uncertainty is from the experimental measurement, the second is the theoretical uncertainty.\label{tab:VubIncl}}
\begin{center}
\begin{tabular}[h]{lll}
\hline\hline
\multicolumn{1}{c}{Theoretical scheme} & $|V_{ub}|(10^{-3})$~\cite{Amhis2014} & $|V_{ub}|(10^{-3})$ (see text)\\
\hline
ADFR &  $4.05\pm0.13^{+0.18}_{-0.11}$ & $4.42\pm0.19\pm0.19$\\
BLNP  & $4.45\pm0.15^{+0.20}_{-0.21}$ & $4.40\pm0.18\pm0.21$\\
DGE & $4.52\pm0.16^{+0.15}_{-0.16}$ & $4.53\pm0.18\pm0.13$\\
GGOU & $4.51\pm0.16^{+0.12}_{-0.15}$ & $4.50\pm0.18\pm0.11$\\
\hline\hline
\end{tabular}
\end{center}
\end{table}

It is concerning that the ADFR average is lower than that obtained with the other three,
especially noting that at least the experimental uncertainties are highly correlated. 
If we restrict to the BaBar ($M_X, q^2$ fit in~\cite{Lees2012}) and Belle results~\cite{Urquijo2010} using the greatest phase space selections, this difference disappears. 
The average of these two measurements, for the four theoretical calculations, are shown in the left column of Table~\ref{tab:VubIncl}, where a consistent picture is apparent. We thus view these averages as potentially more reliable. To quote 
an overall inclusive measurement, people typically take a simple average of the results and their uncertainties under the four calculations. However, in the spirit of this review, we prefer to use the most precise result, in the absence of any other discriminating factor. Hence, we find:
\begin{equation}
\begin{split}
|V_{ub}|^2(\hbox{incl})&=2.02(16)(9)\times10^{-5},\\
|V_{ub}|(\hbox{incl})&=0.00450(18)(11),\\
\end{split}
\label{eq:Vubinc}
\end{equation}
where the first uncertainty is experimental and the second theoretical.
 
Comparing~\ref{eq:Vubinc} and~\ref{eq:Vubex} we see, simliarly with $|V_{cb}|$, that the inclusively measured
value is substantially higher than the exclusive value. The $p$ value for consistency, assuming
Gaussian statistics, is only $4\times10^{-4}$. Note that it matters whether Gaussian sampling
is assumed for the estimate of $|V_{ub}|^2$ or for $|V_{ub}|$; we assume it for the square, as that is the quantity that is most directly related to the experimental measurement. However, the contribution to the uncertainty from the theoretical uncertainties may not have a simple statistical interpretation; we should interpret the $p$ values with some caution. Nevertheless, the differences between
inclusive and exclusive measurements of $|V_{cb}|$ and $|V_{ub}|$ are not likely to be due solely to statistical fluctuations.

We have already remarked in Section~\ref{sec:Vcb} that the discrepancy is difficult to
explain in terms of new physics. Most likely, there is a bias in either the inclusive or
exclusive determination, plausibly from similar sources for both $|V_{cb}|$ and $|V_{ub}|$.
The question is which is wrong. We may attempt to answer this under the assumption that the CKM matrix is unitary. We will discuss this possibility in Section~\ref{sec:discussion}.

\subsection{$|V_{tb}|$}
\label{sec:Vtb}

Finally, we get to the elements on the third row of the matrix, describing couplings of the top quark. We begin with the flavor ``diagonal'' element, $V_{tb}$. In the three family standard model, the
smallness of the elements in the upper two rows in the third column already imply that this element is of order one. Independent measurement is consistent with this, but still with large uncertainty.

\begin{center}
 \begin{figure}[h]
  \includegraphics{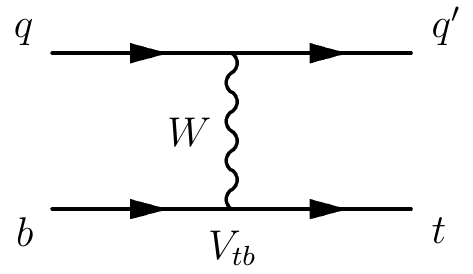}
  \raise4pt\hbox{\includegraphics{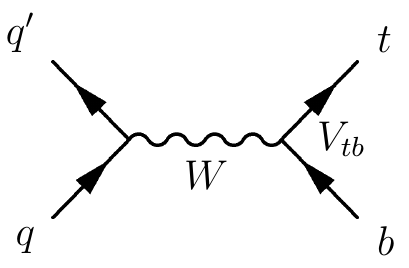}}
  \includegraphics{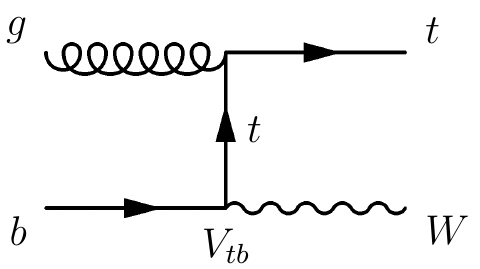}
  \raise4pt\hbox{\includegraphics{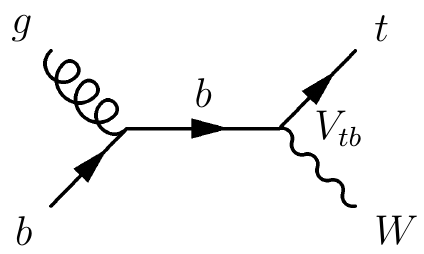}}\break
\vbox{\hskip2.05cm(a)\hskip4.15cm (b)\hskip3.94cm(c)\hskip4.25cm(d)}
  \caption{Single top production via (a) the $t$-channel; (b) the $s$-channel; (c) and (d) $Wt$ associated production.\label{fig:texchangetop}}
 \end{figure}
\end{center}

So far the best method to independently measure $|V_{tb}|$ is in single top production at the hadron colliders. The production cross section is proportional to $|V_{tb}|^2$, the dominant
graphs are shown in Fig.~\ref{fig:texchangetop}. This assumes that $|V_{tb}|\gg |V_{td}|, |V_{ts}|$, which gains support from the results in Section~\ref{sec:Vts},
so that the
$b$ quark dominates at the top production vertex. We also remark that possible contributions from
heavy fourth generation quarks may reasonably be neglected because such quarks would already
have been seen and/or would be kinematically highly suppressed in associated production with the top quark. 
The value for $|V_{tb}|$ is extracted by comparing with the theoretical cross section (for $|V_{tb}|=1$):
\begin{equation}
 |V_{tb}| =\sqrt{\sigma_{\rm measured}/\sigma_{\rm theory}}.
\end{equation}
The theoretical cross section is computed at next-to-next-to-leading-order (NNLO) in QCD~\cite{Kidonakis2010, Kidonakis2010a, Kidonakis2011,Martin2009}

Recently, an analysis combining the Tevatron results in~\cite{Aaltonen2014, Aaltonen2014a, Aaltonen2014b, Aaltonen2014d, Abazov2013} has been performed~\cite{Aaltonen2015} to obtain a cross section for single top production in the $s$ and $t$ channels at 1.96 GeV:
\begin{equation}
\sigma_{s+t} = 3.30^{+0.52}_{-0.40}\hbox{ pb}.
\end{equation}
The value of $|V_{tb}|$ extracted from the measurement is
\begin{equation}
|V_{tb}| = 1.02^{+0.06}_{-0.05}.
\end{equation}

At the LHC, the $t$-channel and $Wt$ associated production mechanisms are the best, since the $s$ channel
suffers in the absence of a valence antiquark in the initial state. A measurement, with large uncertainty, of the
$s$-channel cross section at 8 TeV has been reported however~\cite{Aad2015}. The LHC results for the $t$-channel and $Wt$ associated production are shown in Table~\ref{tab:sigmatLHC}.
We exclude the unpublished results from our average. Averaging the ATLAS and CMS 7 GeV
cross sections is complicated by the presence of correlations, as is the combination of
the results for $|V_{tb}|$ at different energies and for different channels. Since the cross sections for different processes and different energies cannot be usefully averaged, we work with $|V_{tb}|^2$. This introduces an additional correlation through the theoretical cross section.  The ATLAS and CMS collaborations have begun to consider correlations in 
averaging the results of the two experiments~\cite{ATLASCMS2013}. The correlation coefficient at 8 TeV
for the $t$ channel is estimated to be $\rho=0.38$. We will assume this holds also at 7 TeV.
We'll further assume this correlation for the $Wt$ channel between the two experiments; the uncertainties are here much larger, so an error in this assumption will not have much effect on our
final average. We note that the correlated uncertainty from the top quark mass is perhaps being treated inconsistently in our average, but this has a small effect on the result.

CMS has computed $|V_{tb}|$ for the combined 7 and 8 TeV results for the $t$-channel,
obtaining~\cite{Khachatryan2014a} $|V_{tb}| = 0.998\pm0.038\pm0.016$, where the first uncertainty is experimental and the second theoretical. With this information, we deduce
a correlation in the experimental uncertainties in $|V_{tb}|^2$ of about 40\% between the
7 and 8 TeV $t$-channel results. We will assume the same correlation for the $Wt$ production; the uncertainties are here much larger, so an error made will not have much effect on our
final average. The correlation between $Wt$ and $t$-channel  is taken to be zero; this is probably an underestimate, but again the effect of this error on the final average is small.

The theoretical uncertanties are treated, perhaps conservatively, as having 100\% correlation.
The uncertainties are often quoted as asymmetric, but the asymmetry is never large; we
just take the average of the upper and lower values. Our average of the published LHC data in Table~\ref{tab:sigmatLHC} is $|V_{tb}| = 1.00\pm0.04$. We caution that the $\chi^2$ probability is greater than 99\%, suggesting that the covariance matrix may be misestimated. Combining this result with the Tevatron average (neglecting experimental correlations, but assuming 100\% correlation in the
theoretical cross section, yields:

\begin{equation}
\begin{split}
|V_{tb}|^2 &= 1.01(7)\\
|V_{tb}| &= 1.007(36).\\
\end{split}
\label{eq:Vtb}
\end{equation}

\begin{table}
\caption{Experimental measurements of single top quark (both $t$ and $\bar t$) production cross sections at the LHC. 
The first error is statistical, the second is systematic (including theory and luminosity uncertainties). \label{tab:sigmatLHC}}
\begin{center}
\begin{tabular}[h]{lllc}
\hline\hline
Experiment & Channel, $\sqrt{s}$ & Cross section (pb) & Reference\\
\hline
ATLAS & $t$-channel, 7 GeV & $68\pm2\pm8$ & \cite{Aad2014}\\
% superseded by Aad2014 & $t$-channel, 7 GeV & $83\pm4^{+20}_{-19}$ &\cite{Aad2012}\\
 & $Wt$ production, 7 GeV & $16.8\pm2.8\pm4.9$ &\cite{Aad2012a}\\
 & $t$-channel, 8 GeV & $82.6\pm1.2\pm12.0$ &\cite{ATLAS2014} (unpublished)\\
 & $Wt$ production, 8 GeV & $27.2\pm2.8\pm5.4$ & \cite{ATLAS2013} (unpublished)\\
CMS & $t$-channel, 7 GeV & $67.2\pm3.7\pm4.8$ &\cite{Chatrchyan2012}\\
& $Wt$ production, 7 GeV & $16\pm3\pm4$ & \cite{Chatrchyan2013}\\
& $t$-channel, 8 GeV & $83.6\pm2.3\pm7.4$ &\cite{Khachatryan2014a}\\
& $Wt$ production, 8 GeV & $23.4\pm1.9\pm5.2$ & \cite{Chatrchyan2014}\\
\hline\hline
\end{tabular}
\end{center}
\end{table}

Under the assumption that $|V_{td}|^2+|V_{ts}|^2+|V_{tb}|^2=1$ it is also possible to
measure $|V_{tb}|$ in $t\bar t$ production channels, by measuring the ratio
$B(t\to Wb)/B(t\to Wq)$, where $q=d,s$, or $b$. This has been done at both
the Tevatron \cite{Aaltonen2013, Aaltonen2014c, Abazov2011} and the LHC~\cite{Khachatryan2014}.
Because of the assumption, however, we do not include these measurements in our average.

\subsection{$|V_{ts}|$ and $|V_{td}|$}
\label{sec:Vts}

The remaining third-row elements $V_{td}$ and $V_{ts}$ are very small, and it is experimentally difficult to precisely measure the $t\to d$ and $t\to s$ cross sections
in single top production at hadron colliders. Instead, these elements are currently best measured
in virtual processes involving loop diagrams or box diagrams.

\begin{figure}[h]
\begin{center}
\includegraphics{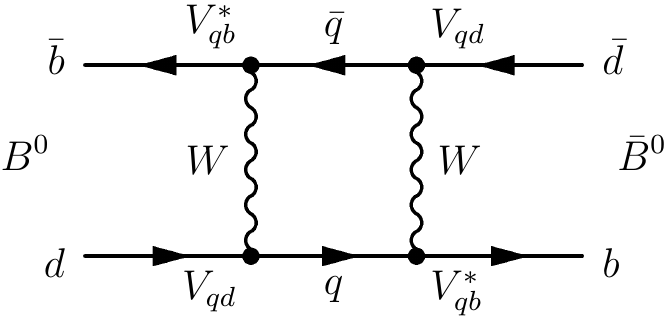}\hskip1cm
\includegraphics{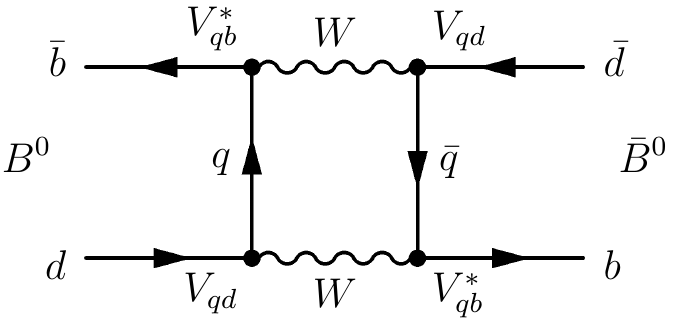}
\caption{Feynman graphs of box diagrams for $B^0-\bar B^0$ mixing. The label $q$ is $u,c$, or $t$.\label{fig:boxes}}
\end{center}
\end{figure}

The mixing of $B^0$ or $B^0_s$ mesons provides for measuring $V_{td}$ or $V_{ts}$.
Figure~\ref{fig:boxes} shows the box diagrams giving rise to $B^0-\bar B^0$ mixing; for
$B^0_s-\bar B^0_s$ mixing the $d$ quark is replaced with an $s$ quark.
The dominant contribution in these graphs is from the $t$-quark exchange, hence the sensitivity to the desired matrix elements. 

Experimentally, the idea is
to measure the probability that a $B^0$ state at time $0$ will decay as a $\bar B^0$ at some later time $t$. This probability is given by
\begin{equation}
 P\left[B^0(0)\to \bar B^0(t)\right] = \frac{1}{2\tau}e^{-t/\tau}\left(1-\cos\Delta m t\right),
\end{equation}
where $\tau$ is the $B^0$ lifetime (inverse of the average width $\Gamma$ of the
two mass eigenstates) and $\Delta m = m_H - m_L$ is the mass of the heavy $B^0$ mass eigenstate minus the mass of the light eigenstate. Time $t$ is measured in the $B$ rest frame.
Thus, we measure $\Delta m$, which is 
related to $|V_{tq}|$ according to (for a review, see, e.g.,~\cite{Buras1998}, also Section~\ref{sec:MICPB} below):
\begin{equation}
\Delta m = \frac{G_F^2m_W^2m_{B}}{6\pi^2}\eta_Bf_{B}^2\hat B_B S_0(x_t)|V_{tq}V_{tb}|^2,
\label{eq:Bmixing}
\end{equation}
where $f_B$ is the $B$ decay constant, $\hat B_B$ is the (renormalization group invariant) $B$ bag parameter, $x_t\equiv (m_t/m_W)^2$, and~\cite{Inami1981}
\begin{equation}
\label{eq:InamiLimS0x}
S_0(x) = \frac{4x-11x^2+x^3}{4(1-x)^2}- \frac{3x^3\log x}{2(1-x)^3}.
\end{equation}
QCD corrections to $S_0$ are contained in $\eta_B=0.551(7)$~\cite{Buchalla1996, Charles2005}.
With the appropriate substitutions, Eq.~\ref{eq:Bmixing} holds for $B_s$ mixing as well.
It is conventional to define dimensionless parameters for the mixing frequencies, $x_d\equiv\Delta m/\Gamma$ and $x_s\equiv\Delta m(B_s)/\Gamma(B_s)$. 

There is some ambiguity concerning the meaning of $m_t$, the top quark mass, where various quantities exist. However, in the present context the running mass $m_t(m_t)$ in the $\overline{\hbox{MS}}$ scheme is needed~\cite{Buchalla1996}. A recent evaluation is $m_t(m_t)=162.3(2.4)$ GeV, based on fits to the Tevatron and LHC cross sections~\cite{Alekhin2014}. However, we use the more precise evaluation using direct measurements with matching from pole mass to $\overline{\hbox{MS}}$
mass as in~\cite{Bevan2014}, obtaining $m_t(m_t)=163.3(0.9)$ GeV.

The measurement of $\Delta m$ or $\Delta m_s$ may be performed by explicitly measuring the
time-dependence, or by measuring a time-integrated mixing probability.
However, the oscillation for the $B_s$ system is rapid, so integrating greatly dilutes the
available precision. Even for the $B$ system the time-dependent measurements are more 
precise, and we restrict to these results. 
 
There are many published results on time-dependent $B^0$ mixing from both $e^+e^-$ (ALEPH, BaBar, Belle, DELPHI, L3, OPAL) and hadron collider (CDF, D0, LHCb) experiments (see~\cite{Amhis2014} for a compilation and references).  
We use the HFAG average, which takes into account correlations, with the result~\cite{Amhis2014}:
\begin{equation}
 \Delta m = 0.510(3)(2)\hbox{ ps}^{-1},
\end{equation}
where a $B^0$ lifetime of 1.520(4) ps has been used. Similarly, for $\Delta m(B_s)$, assuming an average lifetime of 1.509(4) ps, the combined (CDF~\cite{Abulencia2006} and LHCb~\cite{Aaij2012e, Aaij2013e, Aaij2015c, Aaij2013f}) result is~\cite{Amhis2014}:
\begin{equation}
 \Delta m(B_s) = 17.757(20)(7)\hbox{ ps}^{-1}.
\end{equation}

The $B^0$ decay constant and bag constant are computed using lattice QCD~\cite{Aoki2014}:
\begin{equation}
 f_B\sqrt{\hat B_B} = 216(15) \hbox{ MeV}.
\end{equation}
Likewise, 
the $B_s$ decay constant and bag constant are evaluated in lattice QCD~\cite{Aoki2014, Gamiz2009}:
\begin{equation}
 f_{B_s}\sqrt{\hat B_{B_s}} = 266(18) \hbox{ MeV}.
\end{equation}
There is substantial correlation in the quoted uncertainty between these two quantities.
The uncertainties are still being significantly reduced; a recent preprint~\cite{Bazavov2016}
quotes $f_B\sqrt{\hat B_B} = 229.4(9.3) \hbox{ MeV}$ and $f_{B_s}\sqrt{\hat B_{B_s}} = 276.0(8.5) \hbox{ MeV}$.

With Eq.~\ref{eq:Bmixing} we have:
\begin{align}
|V_{td}V_{tb}|^2 &=7.2(1.0)\times10^{-5} & |V_{td}V_{tb}| &=  0.0085(6),\\
|V_{ts}V_{tb}|^2 &=1.62(22)\times10^{-3}    & |V_{ts}V_{tb}| &=  0.0403(27).
\end{align}
Values for $|V_{td}|$ and $|V_{ts}|$ are often quoted assuming $|V_{tb}|=1$, as is essentially the case if the matrix is unitary; that is
the above values are used. However, we wish to see what we know without the unitary assumption. Using the measured result for $|V_{tb}|$ (Eq.~\ref{eq:Vtb}), we find:
 \begin{align}
|V_{td}|^2 &=7.1(1.1)\times10^{-5} & |V_{td}| &=  0.0084(7),\\
|V_{ts}|^2 &=1.61(25)\times10^{-3}    & |V_{ts}| &=  0.0401(31).
\end{align}
The uncertainty is presently dominated by the uncertainties from the lattice calculations.
We caution that the $|V_{td}|$ and $|V_{ts}|$ estimates are highly correlated in their
dominant uncertainties, from the lattice, and from $|V_{tb}|$.
We assume 100\% correlation, i.e., $\rho(|V_{td}V_{tb}|^2,|V_{ts}V_{tb}|^2)=1$.

The lattice calculation of the ratio $\xi\equiv(f_{B_s}/f_B)\sqrt{B_{B_s}/B_B}$ is more precise than the terms for either meson because some of the uncertainties cancel. The lattice result for this ratio is~\cite{Aoki2014} $\xi=1.268(63)$, from which we
deduce
\begin{equation}
\left|\frac{V_{td}}{V_{ts}}\right|^2 = \xi^2\frac{\Delta m}{m}\frac{m(B_s)}{\Delta m(B_s)} =
0.047(5), \qquad \left|\frac{V_{td}}{V_{ts}}\right| = 0.217(11).
\label{Eq:VtdVts}
\end{equation}
Finally, we note a recent interesting calculation~\cite{Du2016} in which the penguin decays $B\to K\ell^+\ell^-$ and $B\to\pi\ell^+\ell^-$ are used to obtain
$|V_{td}/V_{ts}| = 0.201(20)$, based on measurements from LHCb~\cite{Aaij2014g, Aaij2015g}. This may be regarded as a test of the standard model in comparison
with the value in Eq.~\ref{Eq:VtdVts}.

\section{Phases}

If the CKM matrix is $3\times3$ unitary, then there is one complex phase, $\delta$, needed in addition
to the $\theta_{12}$, $\theta_{13}$, and $\theta_{23}$ parameters in Eq.~\ref{eq:Vangles}. If $V$ is not unitary, e.g., if it is a submatrix of a higher dimension unitary matrix, then there may
be additional phases. It is thus a test of the standard model whether one complex phase
is sufficient to describe all $CP$ violating phenomena in the quark sector. Our discussion of the magnitudes of the CKM elements has avoided assumptions about the $3\times 3$ unitarity of the matrix.
In principle, it would be appropriate to take the corresponding approach and report the measurements of the relative phases of each of the matrix elements. However, the available information remains incomplete, and it is also not so convenient to express measured quantities in this form. Instead, it is convenient and conventional to couch much of the discussion in terms of the $3\times3$ unitary parameterization in Eqs.~\ref{eq:Vexpansion} and~\ref{eq:Vapex}. Deviations from $3\times3$ unitarity may then show up as internal inconsistencies. Theoretical reviews of the material in this section may be found in, for example,~\cite{Buchalla1996,Fleischer2004,Nir2005, Nierste2009}.

\subsection{$CP$ violation in kaon mixing}
\label{sec:epsilon}

$CP$ violation may be observed in both mixing (``indirect'') and in decay amplitudes (``direct'') in kaon decays.
The larger and theoretically cleanest to relate to the CKM matrix appears in mixing, and we concentrate on this here. If there were no $CP$ violation, the neutral kaon mass eignestates would
be $CP$ eigenstates. With $CP$ violation, the mass eigenstates are mixtures of the $CP$ eigenstates (denoted $K_1$ and $K_2$):
\begin{align}
K_S &= \frac{1}{\sqrt{1+|\tilde\epsilon|^2}}(K_1+\tilde\epsilon K_2)
\label{eq:KmixS}\\
K_L &= \frac{1}{\sqrt{1+|\tilde\epsilon|^2}}(K_2+\tilde\epsilon K_1).
\label{eq:KmixL}
\end{align}
Thus $\tilde\epsilon\ne0$ is a measure of $CP$ violation in neutral kaon mixing. 

The measurement of $|\tilde\epsilon|$ is best done in the two pion decays of the neutral kaons. Defining
\begin{align}
\eta_{+-} &=\frac{A(K_L\to\pi^+\pi^-)}{A(K_S\to\pi^+\pi^-)}\\
\eta_{00} &=\frac{A(K_L\to\pi^0\pi^0)}{A(K_S\to\pi^0\pi^0)},
\end{align}
we see that both $\eta_{+-}$ and $\eta_{00}$ are approximately given by $\tilde\epsilon$. However, there may be direct $CP$ violation in the amplitudes (indeed there is, coming from penguin amplitudes). The interfering amplitudes in direct $CP$ violation come from the $\Delta I=1/2$ and $\Delta I=3/2$ processes. Thus, to isolate the $CP$ violation in mixing, we look at only $\Delta I = 1/2$ processes, and define
a parameter $\epsilon$ (e.g.,~\cite{Buras1998}):
\begin{equation}
\epsilon = \frac{A\left[K_L\to(\pi\pi)_{I=0}\right]}{A\left[K_S\to(\pi\pi)_{I=0}\right]}, 
\end{equation}
which is equal to $\tilde\epsilon$ in the absence of direct $CP$ violation. This quantity may be measured by taking the linear combination (reference~\cite{Winstein1993}, where the conventions and approximations may also be noted):
\begin{equation}
\epsilon \approx \frac{1}{3}(2\eta_{+-}+\eta_{00}).
\end{equation}
The current experimental value for $|\epsilon|$ is obtained in a fit to kaon data as described by Wolfenstein, Lin, and Trippe in~\cite{Olive2014}, with the result:
\begin{equation}
|\epsilon| = 2.228(11)\times10^{-3}.
\label{eq:epsilonmeas}
\end{equation}

Theoretically, $\epsilon$ is predicted in the standard model by considering the effective Hamiltonian for $\Delta S=2$ transitions~\cite{Buchalla1996}. In terms of the CKM matrix elements~\cite{Bevan2014},
\begin{equation}
\label{eq:epsilonK}
|\epsilon| = \frac{G_F^2m_W^2m_{K^0}f_K^2\hat B_K}{12\sqrt{2}\pi^2\Delta m_K}\left\{\eta_{cc}S_0(x_c)\Im\left[\left(V_{cs}V_{cd}^*\right)^2\right] + 
\eta_{tt}S_0(x_t)\Im\left[\left(V_{ts}V_{td}^*\right)^2\right] +
2\eta_{ct}S_0(x_c,x_t)\Im\left(V_{cs}V_{cd}^*V_{ts}V_{td}^*\right)\right\},
\end{equation}
where $\hat B_K$, or the kaon ``bag'' parameter, measures the strength of the
four quark $\Delta S = 2$ hadronic matrix element~\cite{Gaiser1981}, in renormalization group invariant form. It is calculated using lattice QCD, with present
value (average for $N_f=2+1$)~\cite{Aoki2014}: $\hat B_K = 0.7661(99)$, consistent with a more recent preliminary FLAG average of  $\hat B_K = 0.7627(97)$~\cite{Vladikas2015}. It may be remarked that the present lattice average slightly exceeds the upper bound of $\hat B_K \le 0.75$ deduced in the $1/N_c$ expansion~\cite{Gerard2011, Buras2014, Blanke2016}; this is something to watch.
For the kaon form factor we also use the lattice QCD calculation (for $N_f=2+1$, in the isospin-symmetric limit)~\cite{Aoki2014}: $f_K = 0.1563(9)$ GeV. The $K_L - K_S$ mass difference is~\cite{Olive2014} $\Delta m_K = 3.484(6)\times10^{-15}$ GeV. The Inami-Lim function
$S_0(x)$ is defined in Eq.~\ref{eq:InamiLimS0x} (note that $S_0(x_c)\approx x_c\equiv (m_c/m_W)^2$), and~\cite{Inami1981, Buras1998}
\begin{equation}
S_0(x_c,x_t) =x_c\left[\log\frac{x_t}{x_c}-\frac{3x_t}{4(1-x_t)}-\frac{3x_t^2}{4(1-x_t)^2}\log x_t\right].
\end{equation}
With $m_t(m_t) =163.3(0.9)$ GeV, as in Section~\ref{sec:Vts}, and $m_c(m_c)=1.275(25)$ GeV, where we use the Particle Data Group average including their inflation of the uncertainty~\cite{Olive2014}, we obtain
\begin{align}
S_0(x_t) &=2.317(20)\\
S_0(x_c,x_t) &=2.22(8)\times10^{-3}\\
S_0(x_c) &=2.52(10)\times10^{-4}.
\end{align}

The quantities $\eta_{cc}$, $\eta_{ct}$, and $\eta_{tt}$ (also known as $\eta_1$, $\eta_3$, and $\eta_2$, respectively) are short distance QCD corrections:
\begin{align}
\eta_{cc} &= 1.87(76) & \hbox{NNLO~\cite{Brod2012}}\\
\eta_{ct} &=0.496(47) &\hbox{NNLO~\cite{Brod2010}}\\
\eta_{tt} &=0.5765(65) &\hbox{NLO~\cite{Buras2008}}.
\end{align}
To account for long distance effects, Eq.~\ref{eq:epsilonK} is nowadays multiplied by
the factor $k_\epsilon = 0.94\pm0.02$~\cite{Buras2010}.

Using the parameterization of the CKM matrix in Eq.~\ref{eq:Vexpansion} we obtain, with
appropriate translation from $(\rho,\eta)$ to $(\bar\rho,\bar\eta)$ at the same order in $\lambda$~\cite{Buras2008, Cirigliano2012}:
\begin{equation}
\label{eq:epsilonKrhoeta}
|\epsilon| = \frac{G_F^2m_W^2m_{K^0}f_K^2\hat B_K}{6\sqrt{2}\pi^2\Delta m_K} k_\epsilon A^2\lambda^6\bar\eta\left[A^2\lambda^4(1-\bar\rho)\eta_{tt}S_0(x_t) +\eta_{ct}S_0(x_c,x_t)-\eta_{cc}S_0(x_c)\right].
\end{equation}
We evaluate the constant
\begin{equation}
C_\epsilon\equiv \frac{G_F^2m_W^2m_{K^0}f_K^2}{6\sqrt{2}\pi^2\Delta m_K} = 3.663(43)\times 10^4.
\label{eq:Cepsilon}
\end{equation}

The theoretical uncertainties from charm contributions, especially through $\eta_{cc}$, presently restrict the available precision. Some mitigation for this
has recently been proposed in~\cite{Ligeti2016}, where it is shown that a rephrasing of the neutral kaon fields can eliminate the dependence on $\eta_{cc}$. The gain in precision is somewhat offset by an increased uncertainty from the long distance contribution, hence it becomes important to
improve the uncertainty from this source.

The $|\epsilon|$ measurement, Eq.~\ref{eq:epsilonmeas}, and its relation to the CKM elements via Eq.~\ref{eq:epsilonK} (including the $k_\epsilon$ factor) will be incorporated  into our
discussion of the unitarity of $V$ in Section~\ref{sec:discussion}.

\subsection{Mixing-induced $CP$ violation in $B$ decays}
\label{sec:MICPB}

The phenomenom of $B^0$--$\bar B^0$ mixing may be used to study $CP$ violation in the $B$ system. In this case, the $CP$ violating phase is extracted from the interference of two amplitudes 
leading to a given final state, $f$. One of the amplitudes may be a direct $B^0\to f$ tree-level process, and the other amplitude involves the mixing $B^0\to\bar B^0\to f$ process. This is referred to as ``mixing-induced $CP$ violation''. An early discussion of the formalism may be found in~\cite{Bigi1981}.

For neutral $B$ mesons (we could be talking about either $B_d$ or $B_s$), the flavor basis
is given by $B^0$ and $\bar B^0$. We adopt the convention 
\begin{equation}
CP\ket{B^0} = -\ket{\bar B^0},\quad CP\ket{\bar B^0} = -\ket{B^0}
\label{eq:CPconvention}
\end{equation}
for a meson at rest. The mass eigenstates (with $H$ for ``heavy'' and $L$ for ``light'') are combinations of these, which we may
express as
\begin{align}
\ket{B_L} &= p\ket{B^0} + q\ket{\bar B^0}\\ 
\ket{B_H} &= p\ket{B^0} - q\ket{\bar B^0},
\end{align}
normalized with $|p|^2 + |q|^2 = 1$. If $|q/p|\ne1$, $CP$ is violated, and this is referred to as ``$CP$ violation in mixing''.
For the neutral kaon system, we see $CP$ violation in mixing in the
real part of $\epsilon$. If $|q/p|=1$, we may still have $CP$ violation via interference in mixing and decay amplitudes as noted above, that is, mixing-induced $CP$ violation. This source is important in the neutral $B$ system.

Assuming CPT conservation (so far supported by experiment, including for the $B^0$ system~\cite{Aubert2006c, Aubert2004b, Higuchi2012}), the effective Hamiltonian for this two-state system may be written in the
flavor basis as
\begin{equation}
H_{\rm eff} = 
\begin{pmatrix}
 M - \frac{i}{2}\Gamma & M_{12} - \frac{i}{2}\Gamma_{12}\\
M_{12}^* - \frac{i}{2}\Gamma_{12}^* & M - \frac{i}{2}\Gamma
\end{pmatrix}.
\end{equation}
This is not a Hermitian matrix, hence $B_L$ and $B_H$ are not orthogonal states in general.

To solve the eigenvalue problem, we write down the secular equation and find that the difference
in eigenvalues is
\begin{equation}
 \left(\Delta m+\frac{i}{2}\Delta\Gamma\right)^2 = 4\left(M_{12}-\frac{i}{2}\Gamma_{12}\right)\left(M_{12}^*-\frac{i}{2}\Gamma_{12}^*\right),
\end{equation}
where $\Delta m\equiv m_H-m_L > 0$ and $\Delta\Gamma\equiv \Gamma_L-\Gamma_H$.
Our conventions are such that we define $\Delta m$ to be positive, and $\Delta\Gamma$ is expected to be positive for the $B$ system in the standard model, though this has been confirmed only for the $B_s$ system~\cite{Aaij2012d} (see Section~\ref{sec:betas}). The literature does not have a consistent convention for the definition of $\Delta\Gamma$.
Letting $m\equiv (m_H+m_L)/2=M$ and $\Gamma\equiv (\Gamma_H+\Gamma_L)/2$, the meson masses and widths are:
\begin{align}
m_{H,L} &= m\pm\Re\sqrt{\left(M_{12}-\frac{i}{2}\Gamma_{12}\right)\left(M_{12}^*-\frac{i}{2}\Gamma_{12}^*\right)}\label{eq:BmHL}\\
\Gamma_{L,H} &= \Gamma\pm2\Im\sqrt{\left(M_{12}-\frac{i}{2}\Gamma_{12}\right)\left(M_{12}^*-\frac{i}{2}\Gamma_{12}^*\right)}
\label{eq:BGHL}
\end{align}
With basis transformation $T=\begin{pmatrix} p&p\\q&-q\end{pmatrix}$ from the mass basis to the flavor basis, we obtain
\begin{equation}
 \left(\frac{q}{p}\right)^2 = \frac{M_{12}^*-\frac{i}{2}\Gamma^*_{12}}{M_{12}-\frac{i}{2}\Gamma_{12}}
\label{eq:qoverp}
\end{equation}

The off-diagonal elements of $H_{\rm eff}$ may be computed in the standard model. The weak interaction provides for $\Delta B=2$ transitions as shown already in the box diagrams in Fig.~\ref{fig:boxes}.
The dispersive part, $M_{12}$, is computed from the
$|\Delta B| = 2$ Hamiltonian. 
However, there is no absorptive contribution from this operator, and determining $\Gamma_{12}$ involves
a second order $\Delta B = 1$ calculation~\cite{Hagelin1981}. The prediction is that $|\Gamma_{12}|\ll |M_{12}|$ ($|\Gamma_{12}/M_{12}|\sim O(0.005)$ for both $B_d$ and $B_s$, e.g.,~\cite{Lenz2007}). 
Referring to Eq.~\ref{eq:qoverp} we thus expect that $\left|q/p\right|\approx1$. In this limit, the mass eigenstates are also $CP$ eigenstates, and there is no $CP$ violation in mixing in the $B^0$ system, in contrast with $CP$ violation in mixing being an important contribution to $CP$ violation in the neutral kaon system (Eqs.~\ref{eq:KmixS}, \ref{eq:KmixL}). Experimental measurements of $\left|q/p\right|\approx1$ in the $B_d$ system bear this out. Reference~\cite{Bevan2014} (see also~\cite{Lees2016}) gives a $B$-factory average of $|q/p|-1 =0.3(2.8)\times10^{-3}$, from measurements of $CP$ asymmetries in both semileptonic and hadronic $B$ decays.
LHCb~\cite{Aaij2015h} obtains a consistent precise result from the semileptonic asymmetry, $|q/p|-1 = -0.1(9)(1.5)\times10^{-3}$. 
Using semileptonic decays with $D_s$ instead of $D$, LHCb~\cite{Aaij2014h} obtains a corresponding result for the $B_s$ system: $|q/p|_s-1 = -0.6(5.0)(3.6)\times10^{-3}$. It should be remarked that a measurement by D0~\cite{Abazov2014}, including both $B$ and $B_s$ contributions, obtained a result for the dilepton asymmetry differing from the standard model by 3.6 standard deviations.

As the overall phase is not physical, mixing can be described in terms of the
magnitudes of $M_{12}$ and $\Gamma_{12}$ and their relative phase. Thus, we define the ``mixing phase'' as 
\begin{equation}
\varphi_q\equiv\arg\left(-\frac{M_{12}}{\Gamma_{12}}\right),
\end{equation}
where $q=d$ or $s$ depending on whether the $B_d$ or $B_s$ system is being discussed. The reader is cautioned that notation here varies; we use $\varphi$ to distinguish it from other uses for $\phi$. From Eq.~\ref{eq:qoverp}, if $\varphi\ne 0,\pi$, then $|q/p|\ne 1$ and we have $CP$ violation in mixing. This phase is thus accessible in $CP$ asymmetry in mixing, for example in the semileptonic asymmetry mentioned above. This asymmetry is expected to be very small
in the standard model, suppressed by $|\Gamma_{12}/M_{12}|$. 

The standard model prediction for $M_{12}$ and $\Gamma_{12}$ is dominated by the top quark in the box diagrams (Fig.~\ref{fig:boxes}),
with the result (e.g.,~\cite{Hagelin1981, Buras1984, Nierste2009} and Schneider in~\cite{Olive2014}):
\begin{align}
M_{12} &= \frac{G_F^2m_W^2m_{B}}{12\pi^2}\eta_Bf_{B}^2\hat B_B S_0(x_t)(V_{tb}V_{tq}^*)^2,\label{eq:BmixingPhase}\\
\Gamma_{12} &= -\frac{G_F^2m_{B}}{8\pi}\eta_B^\prime f_{B}^2\hat B_B m_b^2 \left[(V_{tb}V_{tq}^*)^2 + 
  V_{tb}V_{tq}^*V_{cb}V_{cq}^*O\left(\frac{m_c^2}{m_b^2}\right) +
  (V_{cb}V_{cq}^*)^2O\left(\frac{m_c^4}{m_b^4}\right)\right],
\label{eq:Gamma12}
\end{align}
where the quantities are as defined for Eq.~\ref{eq:Bmixing}, except QCD correction $\eta_B^\prime\approx 1$ is given as $\eta_4^{(B)}$ in~\cite{Buras1984}. 

Examining the CKM terms in equations~\ref{eq:BmixingPhase} and~\ref{eq:Gamma12} we see first that the ratio $|\Gamma_{12}/M_{12}|$ is expected to be about the same for both the $B_d$ and $B_s$ systems, as already suggested above. Second, since $\Gamma_{12}$ and $M_{12}$  are, to lowest order, proportional to $\lambda^4$ for the $B_s$ system and to $\lambda^6$ for the $B_d$ system, we expect both the
mixing frequency (as in Section~\ref{sec:Vts}) and the width difference to be of order $1/\lambda^2$ larger in the $B_s$ system. We further confirm that the phase $\varphi_q$ is expected to be small in both neutral $B$ sytems as the leading phases of $\Gamma_{12}$ and $M_{12}$ differ by $\pi$. This also provides the standard model expectation that
$\Delta\Gamma >0$ in our convention.

We note that the mixing frequency $\Delta m$ in Eq.~\ref{eq:Bmixing} is obtained, in the limit $\Gamma_{12}\to0$, from Eqs.~\ref{eq:BmHL} and~\ref{eq:BmixingPhase}.
In this approximation we also have, using Eq.~\ref{eq:qoverp}, 
\begin{equation}
\frac{q}{p} \approx -\frac{M^*_{12}}{|M_{12}|}.
\label{eq:qoverpapprox}
\end{equation}
The minus square root has been taken here, according to the convention we have adopted for the $B_H$ and $B_L$ states
and requiring $\Delta m > 0$~\cite{Nierste2009}. The reader is again cautioned that the conventions are not universal.
Thus, measuring $q/p$ measures the phase of $(V_{tb}V_{tq}^*)^2$. 

As mentioned earlier, the $CP$ violation can be measured in the interference  between the amplitudes for $B^0\to f$ and $B^0\to \bar B^0\to f$, or the charge conjugate reactions, where $f$ is a $CP$ eigenstate. Two important examples are $f=J/\psi K_S$ for $B^0_d$ decays, and $f=J/\psi\phi$ for $B_s^0$ decays. Experimentally, we prepare our $B$ mesons in flavor eigenstates. Thus, if we initially prepare a pure $B^0$ state, it may eventually decay to $f$
either directly or via mixing as a $\bar B^0$. Assuming $\Gamma_{12}=0$ the time evolution operator in the flavor basis is
\begin{equation}
U(t) = e^{-itH_{\rm eff}} = e^{-imt-\frac{\Gamma}{2}t}\begin{pmatrix}
\cos \frac{\Delta m t}{2} & i\frac{p}{q}\sin\frac{\Delta m t}{2}\\
  i\frac{q}{p}\sin\frac{\Delta m t}{2} & \cos\frac{\Delta m t}{2}
\end{pmatrix}.
\end{equation}
A state that is pure $B^0$ at time $t=0$ thus evolves in time as:
\begin{equation}
\ket{B^0(t)} = e^{-imt}e^{-\Gamma t/2}\left(\cos\frac{\Delta m t}{2}\ket{B^0} + i\frac{q}{p}\sin\frac{\Delta m t}{2}\ket{\bar B^0}\right).
\end{equation}

Let $A_f$ be the $\Delta B=1$ amplitude for a $B^0$ to decay to $f$, and $\bar A_f$ be the amplitude for
a $\bar B^0$ to decay to $f$. We assume for simplicity here that a single tree-level process dominates. It is convenient and conventional to define
\begin{equation}
\lambda_f\equiv \frac{q}{p}\frac{\bar A_f}{A_f}.
\end{equation}
Then the time-dependent decay rate for a $B^0$ at $t=0$ to decay into
final state $f$ is:
\begin{equation}
\Gamma(B^0(t)\to f) \propto |A_f|^2 e^{-\Gamma t} \frac{1+|\lambda_f|^2}{2}\left(1+\frac{1-|\lambda_f|^2}{1+|\lambda_f|^2}\cos\Delta mt-\frac{2\Im\lambda_f}{1+|\lambda_f|^2}\sin\Delta mt\right).
\label{eq:Btof}
\end{equation}
The corresponding rate for a $\bar B^0$ at $t=0$ to decay to $f$ is:
\begin{equation}
\Gamma(\bar B^0(t)\to f) \propto |\bar A_f|^2 e^{-\Gamma t} \frac{1+|\lambda_f|^2}{2|\lambda_f|^2}\left(1-\frac{1-|\lambda_f|^2}{1+|\lambda_f|^2}\cos\Delta mt+\frac{2\Im\lambda_f}{1+|\lambda_f|^2}\sin\Delta mt\right).
\label{eq:Bbartof}
\end{equation}
If $f$ is a $CP$ eigenstate and the rates in Eqs.~\ref{eq:Btof} and~\ref{eq:Bbartof} are different, then $CP$ is not conserved. 
In the case where $\Delta\Gamma$ is not negligible (i.e., $\Gamma_{12}\ne 0$, Eq.~\ref{eq:BGHL}), the 
above two equations must be modified by replacing the ``1'' (that is, the first term in the brackets) with
\begin{equation}
\cosh\frac{\Delta\Gamma t}{2} - \frac{2\Re\lambda_f}{1+|\lambda_f|^2}\sinh\frac{\Delta\Gamma t}{2}.
\label{eq:DGmod}
\end{equation}

Here we come to an important experimental consideration. We must have some way of determining whether we started at $t=0$ with a $B^0$ or a $\bar B^0$. Two methods have so far been used to address this: ({\it i\/}) The most common technique has been the ``opposite-side tag'', in which associated production of $b\bar b$ occurs. In this case we use the flavor of one of the mesons (as determined, eg, by the sign of the lepton in semi-leptonic decays) to tag the flavor of the other meson at the time the tagging meson decays, using the coherence of the wave function; ({\it ii\/}) Long used in charm physics, and adapted to $B$ physics is the ``same-side tag''~\cite{Gronau1993, Abe1998, Affolder2000, Aaij2015a, Sato2012}, in which the meson flavor is identified using the charge of associated particles in the production process (for example, the decay of an excited $B$ state).

In method {(\it ii\/)}, the origin of time ($t=0$) is determined by the vertex with the extra particles. The time associated with the $B$ traveling from this production vertex to its decay vertex is always positive (up to possible resolution effects). In method {(\it i\/)}, however, $t=0$ is determined by the vertex of the tagging $B$ decay. This decay could happen before or after the decay of the ``signal'' $B$. Thus, the signal decay time $t$ could equally probably be positive or negative. In quantum mechanics, the wave function may be propagated forward or backward in time, so both signs of $t$ have equal utility. 
It has the consequence, however, that if we imagine doing an experiment where we don't measure the time, then the coefficient of the $\sin\Delta m t$ term in Eq.~\ref{eq:Btof} is inaccessible. We must measure the time, or at least its sign, to measure this coefficient. In any event, the most precise measurements of the coefficients of $\cos\Delta mt$ and $\sin \Delta mt$ are obtained using time information. 

Experimentally, it is convenient (i.e., systematic effects tend to cancel) to form a normalized $CP$ asymmetry between the $B^0$ and $\bar B^0$ decay rates in Eqs.~\ref{eq:Btof} and~\ref{eq:Bbartof}:
\begin{equation}
{\cal A}_f(t)\equiv\frac{\Gamma(\bar B^0(t)\to f)-\Gamma(B^0(t)\to f)}{\Gamma(\bar B^0(t)\to f)+\Gamma(B^0(t)\to f)} = S_f\sin\Delta m t - C_f\cos\Delta m t,
\label{eq:CPasymmetry}
\end{equation}
where
\begin{equation}
S_f\equiv\frac{2\Im \lambda}{1+|\lambda|^2},\quad C_f\equiv\frac{1- |\lambda|^2}{1+|\lambda|^2}.
\end{equation}
Another notation employed in the literature is $-A_f$ for $C_f$ (not to be confused with the amplitude $A_f$).

\subsubsection{The angle $\beta$}

\begin{figure}[h]
\begin{center}
\hskip1pt\raise.3cm\hbox{\includegraphics[width=5cm]{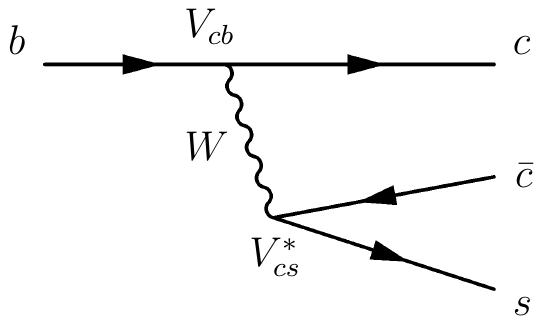}}\hskip.9cm
\includegraphics[width=4.8cm]{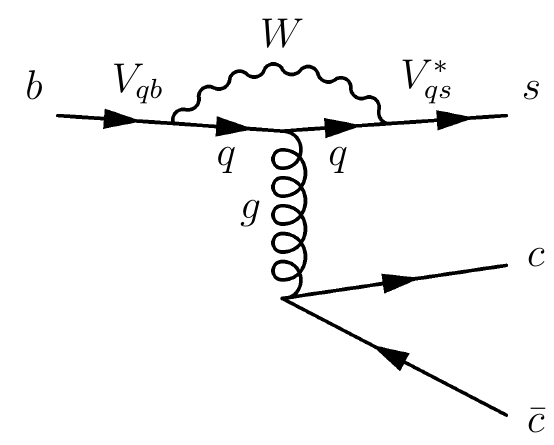}\break
\vbox{\hskip-.3cm(a)\hskip5.5cm (b)}
\caption{\label{fig:bccs}
The $b\to c\bar c s$ process: (a) at tree level, (b) penguin diagram.}
\end{center}
\end{figure}

We may classify the $CP$ violation measurements in hadronic $B$ decays according to the
quark-level sub-process involved. The $b\to c\bar c s$ process, Fig.~\ref{fig:bccs}, with the final state, $f$, a $CP$ eigenstate admits a clean theoretical
treatment, as well as being practical experimentally. A $B^0$ can decay to a $CP$ eigenstate of charmonium and a $K^0_S$ or a $K^0_L$ $CP$ eigenstate providing an overall final eigenstate of $CP$. The decay can happen in tree-level  $B^0\to f$ decays and via the mixing  $B^0\to\bar B^0\to f$, providing two interfering amplitudes. This can be done for both initial $B^0$ and $\bar B^0$ states and the results compared.

For an explicit, and important, example, consider the measurement of $CP$ violation in $B\to J/\psi K^0_S$. The final state is here a $CP$ eigenstate, with $\eta_{J/\psi K_S} = -1$, neglecting the small $CP$ violation in mixing in the $K^0$ system. 

For amplitude $\bar A_f$, we include the CKM factors for the process $\bar B^0\to J/\psi \bar K$, $\bar K\to\pi\pi$, and for
$A_f$ the corresponding factors for $B^0\to J/\psi K$, $K\to\pi\pi$.
We pick up a minus sign from~\ref{eq:CPconvention}, as well as $\eta_{J/\psi K_S}$, obtaining finally:
\begin{equation}
\frac{\bar A_{J/\psi K^0_S}}{A_{J/\psi K^0_S}} = -\eta_{J/\psi K_S}\frac{V_{cb}V_{cs}^*}{V_{cb}^*V_{cs}}\frac{V_{us}V_{ud}^*}{V_{us}^*V_{ud}}.
\end{equation}
We also have, from Eqs.~\ref{eq:BmixingPhase} and~\ref{eq:qoverpapprox}
\begin{equation}
\frac{q}{p} = -\frac{V_{tb}^*V_{td}}{V_{tb}V^*_{td}}
\end{equation}
 Incorporating all the factors gives:
\begin{equation}
\lambda_{J/\psi K_S} = \eta_{J/\psi K_S}
\frac{V_{td}V_{tb}^*}{V^*_{td}V_{tb}}
\frac{V_{us}V^*_{ud}}{V^*_{us}V_{ud}} 
\frac{V_{cb}V^*_{cs}}{V^*_{cb}V_{cs}}
\label{eq:lambdaA}
\end{equation}
Recall from Eq.~\ref{eq:beta} that 
\begin{equation}
\beta=\arg\left(-\frac{V_{cd}V^*_{cb}}{V_{td}V_{tb}^*}\right)
\label{eq:betaA}
\end{equation}
With $\eta_{J/\psi K_S}=-1$ and our adopted conventions in Eq.~\ref{eq:Vexpansion}, we find, up to terms of order $A^2\lambda^4$, that Eqs.~\ref{eq:lambdaA} and~\ref{eq:betaA} give
\begin{equation}
\lambda = -e^{-2i\beta}.
\label{eq:lambdabeta}
\end{equation}
We have used unitarity in making this correspondence; for more general interpretations we measure the combination in Eq.~\ref{eq:lambdaA}. The treatment as measuring $\beta$ may be regarded as a test of unitarity when compared with other constraints.
Thus, the expected time-dependent $CP$ asymmetry for this process is
\begin{equation}
{\cal A}_f(t) = \sin2\beta\sin\Delta m t,
\end{equation}
i.e., $S_{J/\psi K_S} =\sin2\beta$, $C_{J/\psi K_S} = 0$. This changes sign if the $K_S$ is replaced by $K_L$ with opposite $CP$. 

The $J/\psi$ is especially attractive experimentally because of its frequent decay to $e^+e^-$ or $\mu^+\mu^-$. This provides the most precise measurement. However, other charmonium states ($\psi(2S)$, $\eta_c$, $\chi_{c1}$) are also used, somewhat improving the statistical precision. The even $CP$ channel $J/\psi K_L$ is also used, again improving the precision of the result somewhat. Hadronic $J/\psi$ decays have been used as well~\cite{Aubert2004a}. Belle~\cite{Sato2012} has also made a measurement in $\Upsilon(5S)$ data with tagging based on the sign of the pion in channels with a charged pion, a charged $B$, and a neutral $B\to J/\psi K_S$. 

Averages are maintained by HFAG. The 2014 average~\cite{Amhis2014} for the charmonium channels is dominated by results
from the $B$ factories BaBar~\cite{Aubert2009, Aubert2009a, Aubert2004a} and Belle~\cite{Adachi2012, Sato2012}. Early, not very precise measurements from ALEPH~\cite{Barate2000} and OPAL~\cite{Ackerstaff1998} at LEP and CDF~\cite{Affolder2000} are included, as is an early result from LHCb~\cite{Aaij2013} using 1 fb$^{-1}$ of data. 
This average has been updated~\cite{Gershon2015} with one more recent measurement; LHCb has now analyzed their full Run I dataset (3 fb$^{-1}$ at 7 and 8 TeV center of mass energies)~\cite{Aaij2015a} with a precision now approaching that of the $B$ factories. The resulting average is
\begin{equation}
 \sin2\beta = 0.691\pm0.017\quad\hbox{(charmonium modes)}.
\end{equation}
The dominant uncertainty is statistical, and further precision improvements can be expected from LHCb and Belle-II.

An important question is whether other diagrams exist for the $B^0\to J/\psi K^0_S$ and related decays, in particular standard model penguin processes, such as Fig.~\ref{fig:bccs}b (though for the present purposes the term is broadened beyond particular diagrams). Such processes could pollute the $\sin 2\beta$ measurement with additional phases. While such contributions are expected to be small (an early estimate~\cite{Gronau1989} implies a penguin amplitude less than 1\% of the tree amplitude for $B\to J/\psi K_S$), the measurement precision is steadily improving, and this is a subject of ongoing investigation. A traditional approach to checking possible penguin pollution has been to look at $SU(3)$-related decays in which the penguin diagrams are not so strongly Cabibbo suppressed, such as
$B_s\to J/\psi K_S$, $B^0\to J/\psi\pi^0$, and $B^+\to J/\psi (K^+,\pi^+)$, as discussed, e.g., in~\cite{Fleischer1999, Ciuchini2005, Faller2009, Gronau2009, Ligeti2015}.  The results of reference~\cite{Ligeti2015} favor larger values of $\beta$ (and the inclusive value for $|V_{ub}|$), although remaining consistent at the 1-2$\sigma$ level. Another approach is to avoid $SU(3)$ symmetry arguments and perform standard model calculations, for example~\cite{Beneke2005,Frings2015}. The recent calculation in~\cite{Frings2015} conservatively evaluates a maximum penguin effect of $\pm0.34^\circ$ on $\beta$ in the
$J/\psi K_S$ mode, with somewhat larger potential effects in other modes. 
An alternative strategy for measuring penguin effects has been proposed~\cite{Dadisman2016} using a generalization of the method used by BaBar to
demonstrate $T$ violation~\cite{Lees2012b}.

\begin{figure}[h]
\begin{center}
\includegraphics[width=9.9cm]{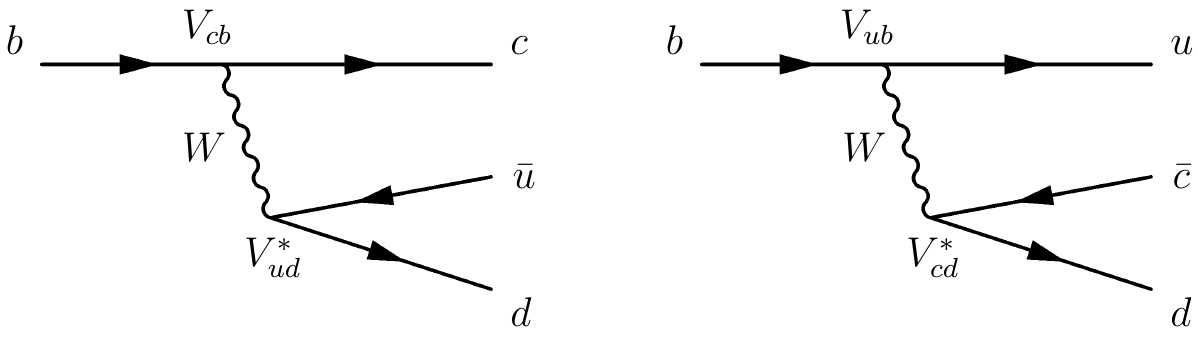}\break
\vbox{\hskip.2cm(a)\hskip4.8cm (b)}
\caption{\label{fig:bcud}
The (a) $b\to c\bar u d$ and (b) $b\to u\bar c d$ processes.}
\end{center}
\end{figure}

It is possible to measure $\sin2\beta$ in other processes that do not go through charmonium states, although it is hard to find other channels that are both theoretically clear and experimentally competitive. The $b\to c\bar u d$ tree-level process (Fig.~\ref{fig:bcud}a) $\bar B^0\to D^{(*)0}h^0$, where $h^0$ is a light neutral hadron ($\pi^0$, $\eta$, or $\omega$) and the $D$ decay is observed in a $CP$ eigenstate has such a clear interpretation, with no competing penguin diagram. A different phase comes in through the $b\to u\bar c d$ transition, Fig.~\ref{fig:bcud}b, but this amplitude is CKM suppressed by a factor of about 50.  A measurement has recently been published from a joint analysis of BaBar and Belle data~\cite{Abdesselam2015}, with the
result $\sin2\beta =0.66\pm0.12$, a statistically significant observation that is consistent with the charmonium average.

We have assumed that $C_f=0$ in our interpretation of
the above measurements as $\sin2\beta$, so it is important to check this assumption.
The analyses used to obtain $\sin2\beta$ can be used to measure the presence of a possible $\cos\Delta m t$ term in the
$CP$ asymmetry,~Eq.~\ref{eq:CPasymmetry}.  The HFAG average for
measurements of $C_f$ is~\cite{Gershon2015}
\begin{equation}
C_f = -0.004\pm 0.015\quad\hbox{(charmonium modes)},
\end{equation}
supporting the assumption.

There are many other channels in which the $CP$ violation is sensitive to $\sin2\beta$. These other channels typically have difficulties with contributions from possible additional phases that complicate the interpretation. The $b\to s$ ``penguin'' transition has approximately the same weak phase as the
$b\to c(\bar c s)$ transition we have been discussing (e.g.,~\cite{Chua2006}). However, this process is more interesting from the perspective of searches for new physics as new virtual particles may contribute to the loop. The issue for these is how different the effective value of $\sin2\beta$ can be in these channels from $\sin2\beta$ as measured in $J/\psi K_S$, according to the standard model. If the observed difference is greater, then that is a sign of new physics.

The measurement of $\sin2\beta$ has a four-fold ambiguity as a measurement of the angle $\beta\in(0,2\pi)$, since for any $\theta\in(0,\pi/2)$, the angles $\theta,\pi/2-\theta,\pi+\theta,3\pi/2-\theta$ all have the same sine of twice the angle. This ambiguity is not entirely resolved by direct measurements, although the unitarity constraint with other measurements picks out the solution at
$\beta = 21.85(67)$ degrees. 

However, there is some experimental information favoring $\cos2\beta>0$, which eliminates two of the four values of $\beta$.
One approach uses the time-dependent angular measurement of $B^0\to J/\psi K\pi$ decays in the $K^*(892)$ region~\cite{Dunietz1991}.
A second approach  is a time-dependent Dalitiz plot analysis of $B^0\to \bar D^0 h^0$, $h = \pi,\eta,\omega$, $\bar D^0\to K^0_S\pi^+\pi^-$~\cite{Bondar2005}, in which the interference between $D^0$ and $\bar D^0$ channels yields sensitivity to $2\beta$.
The published data from these two approaches is shown and averaged in Table~\ref{tab:cos2beta}.

A third approach to determing $\cos2\beta$ is to measure the time-dependent distribution of $B^0\to D^{*+}D^{*-}K^0_S$ events in the Dalitz plot variables~\cite{Browder2000}. Such a measurement has been carried out by
BaBar~\cite{Aubert2006a}, disfavoring the $\cos2\beta<0$ hypothesis with a $p$ value of 0.06, with some assumptions, and by Belle~\cite{Dalseno2007}, measuring $\cos2\beta$ consistent with zero (hence either sign).

\begin{table}
\caption{Measurements of $\cos2\beta$.
The first error is statistical, the second is systematic, and if present, the third is Dalitz plot model uncertainty. The averages are computed by first symmetrizing the intervals. It is important to note that the 68\% confidence intervals for these measurements can not be reliably used to infer other confidence intervals. The final average will be more accurate in this respect, but caution is still advised in the tails of the distribution.\label{tab:cos2beta}}
\begin{center}
\begin{tabular}[h]{llc}
\hline\hline
Experiment & $\cos2\beta$ & Reference\\
\hline
$B\to J/\psi K\pi$ & & \\
\quad BaBar & $2.72^{+0.5}_{-0.79}\pm0.27$ & \cite{Aubert2005a}\\
\quad Belle & $0.56\pm0.79\pm0.11$ &\cite{Itoh2005}\\
\quad Average & $1.7\pm0.5$ &\\
\hline
$B^0\to D^{(*)}h^0, D\to K_S\pi^+\pi^-$ & & \\
\quad BaBar & $0.42\pm0.49\pm0.09\pm0.13$ & \cite{Aubert2007a}\\
\quad Belle & $1.87^{+0.40}_{-0.53}\ ^{+0.22}_{-0.32}$ &\cite{Krokovny2006}\\
\quad Average & $1.1\pm0.4$ &\\
\hline
Average, both channels & $1.3\pm0.3$ &\\
\hline\hline
\end{tabular}
\end{center}
\end{table}

The average in Table~\ref{tab:cos2beta} for $\cos2\beta$ may be compared with the possible values $\pm0.723$ from the measured value of $\sin2\beta$.
While none of the measurements alone is compelling that $\cos2\beta>0$, the average  provides strong evidence, even allowing that there probably remain
substantial non-Gaussian tails in its sampling distribution. Thus, we conclude that
\begin{equation}
\beta = 21.9(7) \hbox{ or } 201.9(7) \hbox{ degrees}.
\end{equation}
We note that all of these measurements of $\cos2\beta$ are based on early partial datasets; the opportunity exists to significantly improve the situation by using the presently available data.

\subsubsection{The angle $\alpha$}

The angle $\alpha$ measures the phase of $V_{ud}V^*_{ub}$ relative to $V_{td}V_{tb}^*$ (Eqn.~\ref{eq:alpha}), thus we may determine $\alpha$ by looking at time-dependent $CP$ violation in the $b\to u\bar u d$
process, the analog of $b\to c \bar c d$ for $\beta$. As with $b\to c\bar c d$, one must worry about the contribution of penguin amplitudes carrying a different phase.

\begin{figure}[h]
\begin{center}
\includegraphics[width=9.9cm]{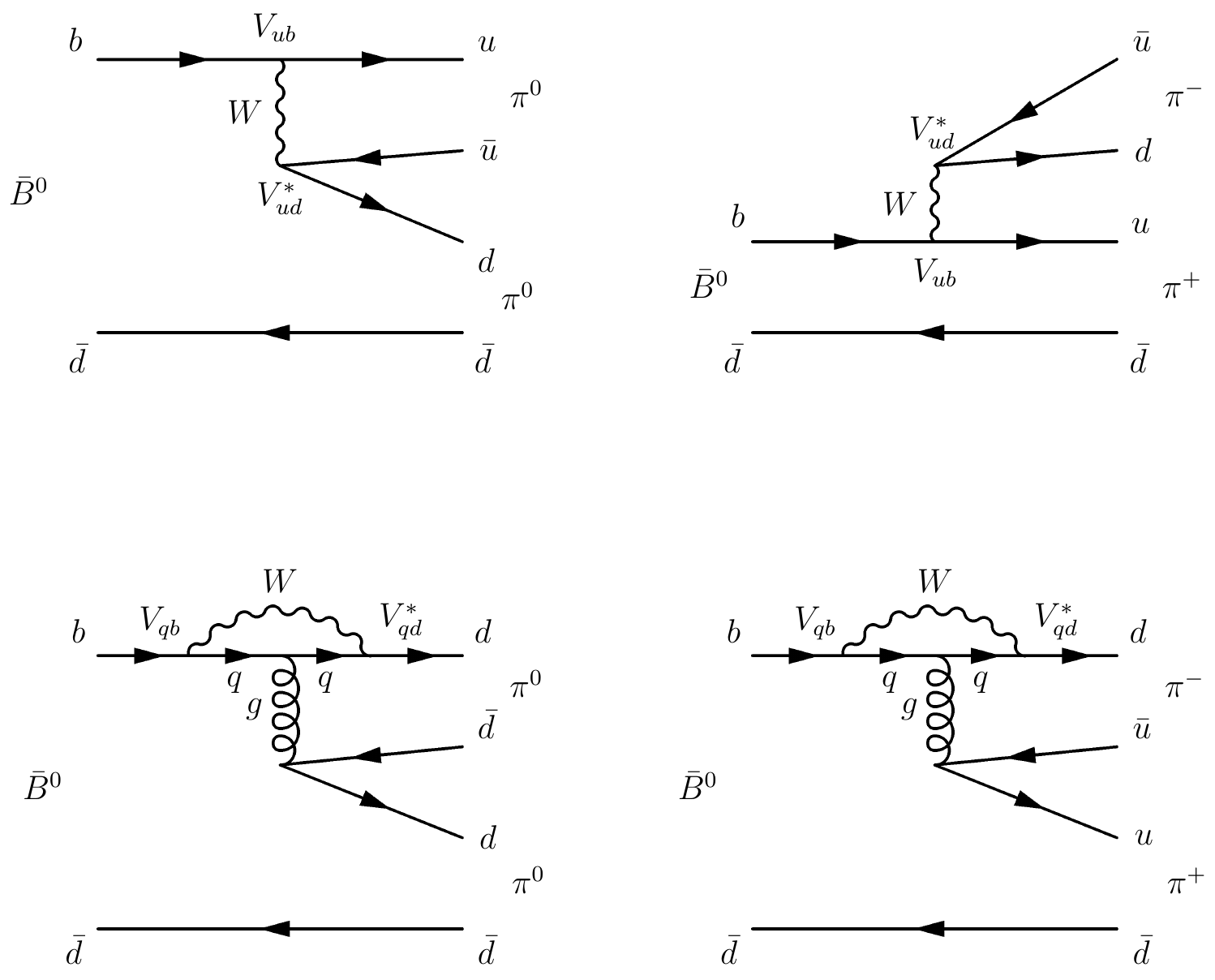}\break
\vskip-5.4cm
\hbox{\hskip6.5cm(a)\hskip4.8cm (b)}
\vskip4.4cm
\hbox{\hskip6.5cm(c)\hskip4.8cm (d)}
\caption{\label{fig:Bpipi}
Graphs depicting $\bar B^0\to\pi\pi$: (a) tree-level $\bar B^0\to\pi^0\pi^0$ (color-suppressed, $I(\pi\pi)=0,2$); (b) tree-level $\bar B^0\to\pi^+\pi^-$ ($I(\pi\pi)=0,2$); (c) penguin $\bar B^0\to\pi^0\pi^0$ ($I(\pi\pi)=0$); (d) penguin $\bar B^0\to\pi^+\pi^-$ ($I(\pi\pi)=0$). }
\end{center}
\end{figure}

The simplest channel to consider here is $B\to\pi\pi$ with tree-level and penquin diagrams illustrated in Fig.~\ref{fig:Bpipi}. It is in principle possible to separate the contributions from the tree-level amplitude from the penguin amplitude using measurements of $B^0\to \pi^+\pi^-$, $B^0\to\pi^0\pi^0$, and $B^+\to\pi^+\pi^0$~\cite{Gronau1990}. The idea is as follows: First, we note that the $\pi\pi$ system must be either in an $I=0$ or $I=2$ state, from Bose statistics. The tree-level process may have contributions from both isospins. However, the gluonic penguin amplitude can only have $I=0$ contributions since the gluon is $I=0$. For example, $B^+\to\pi^+\pi^0$ has no gluonic penguin contribution. At present we may safely neglect electroweak penguin amplitudes. Thus, isospin may be used to extract the desired tree process as described in~\cite{Gronau1990}. The small rates and an eight-fold ambiguity make this so far experimentally challenging to carry out. BaBar~\cite{Lees2013b}, Belle~\cite{Dalseno2013}, and LHCb~\cite{Aaij2013a} all report measurements of $CP$ asymmetry in
$B\to \pi^+\pi^-$. BaBar and Belle perform the isospin analysis to extract $\alpha$. Both experiments
find that the region $\sim 23^\circ <\alpha < \sim 67^\circ$ is disfavored but there is little further discrimination without making additional standard model assumptions. The uncertainties are dominantly statistical, and improved results may be anticipated in future larger datasets. LHCb has measured the $CP$ violation parameters ($S_{\pi^+\pi^-}$ and $C_{\pi^+\pi^-}$) in  $B^0\to\pi^+\pi^-$~\cite{Aaij2013a} with results consistent with those from BaBar and Belle.

We may replace a pion with a rho meson and consider the $B\to\rho\pi$ channel. The isospin analysis for the $\pi\pi$ channel unfortunately does not suffice in this case, as there are now too many amplitudes to determine~\cite{Gronau1990}. However, it is possible to measure $\alpha$ with a time-dependent Dalitz plot analysis of $B^0,\bar B^0\to \pi^+\pi^-\pi^0$~\cite{Lipkin1991, Snyder1993, Quinn2000}. The idea is that one can once again separate out the tree and penguin amplitudes, now using the distribution of events over the three-pion Dalitz plot. Both BaBar~\cite{Lees2013c} and Belle~\cite{Kusaka2007,Kusaka2008} have carried out such an analysis. Belle finds, using a dataset with $449\times10^6$ $B\bar B$ events at the $\Upsilon(4S)$, several disconnected regions in $\alpha$ in the $68\%$ confidence set, including a standard model preferred region of $68^\circ<\alpha<95^\circ$. BaBar, with $471\times10^6$ $B\bar B$ events, performs a study of statistical robustness of the determination of $\alpha$ and finds that the available statistics is insufficient for a reliable extraction of $\alpha$.

The most precise results for $\alpha$ are presently obtained in the time-dependent analysis of $B\to\rho\rho$ events. The basic idea for distinguishing tree and penguin amplitudes is the same as for $B\to\pi\pi$ and $B\to\rho\pi$~\cite{Gronau1990, Lipkin1991}. However, this turns out to be a fortunate channel. The $B^0\to\rho^0\rho^0$ tree diagram is color-suppressed compared with the other charge modes, and with the penguin diagram. Thus, the
$\rho^0\rho^0$ decay is expected to be more strongly influenced by the penguin diagram. It is observed that~\cite{Olive2014} (see also~\cite{Aaij2015e}):
\begin{equation}
{\cal B}(B^0\to\rho^0\rho^0) = 0.73(29)10^{-6}\ll {\cal B}(B^0\to\pi^+\pi^-) =2.42(31)10^{-5}\approx {\cal B}(B^+\to\pi^+\pi^0) = 2.40(19)10^{-5}.
\end{equation}
Hence, the penguin diagram evidently contributes at a small level compared with the tree diagram.
A further happy circumstance concerns the $CP$ structure. The $\rho\rho$ final state can be either $CP$-even or $CP$-odd. However, the longitudinal polarization fraction $f_L$ for $\rho^+\rho^-$ and $\rho^+\rho^0$ is observed to be large, implying that they are nearly pure $CP$-even states.
BaBar~\cite{Aubert2005b, Aubert2008a, Aubert2009b} and Belle~\cite{Zhang2003, Somov2006, Somov2007, Vanhoefer2014} (also a recently published Belle analysis of
$B\to\rho^+\rho^-$~\cite{Vanhoefer2016} obtaining $\alpha=93.7(10.6)^\circ$ up to a $\pi+\alpha$ ambiguity) have carried out isospin analyses to measure $\alpha$ in $B\to\rho\rho$. 

Other channels also provide information on $\alpha$, generally with additional complexity and assumptions.
Most notable perhaps is $B\to a_1(1260) \pi$ and $SU(3)$-related channels in an analysis using approximate
$SU(3)$ symmetry~\cite{Gronau2006, Aubert2007b, Dalseno2012}.  

BaBar and Belle have combined their results (excluding Belle's recent $\rho^+\rho^-$ result~\cite{Vanhoefer2016}) on $\alpha$ in the $\pi\pi$, $\pi^+\pi^-\pi^0$ and $\rho\rho$ channels~\cite{Bevan2014}. The 68\% confidence interval obtained in this combination, including the $\pi+\alpha$ amiguity, is
\begin{equation}
\alpha = 88(5)^\circ,\quad 268(5)^\circ.
\end{equation}
The uncertainty due to isospin breaking is of order one degree.
The region around $\alpha=0$ is not entirely ruled out, with $p$ values of a few per cent.

\subsubsection{The angle $\gamma$}

The third angle of the unitarity triangle is $\gamma$:
\begin{equation}
\gamma \equiv\arg\left(\bar\rho+i\bar\eta\right)=\arg\left(-\frac{V_{ud}V_{ub}^*}{V_{cd}V_{cb}^*}\right).
\label{eq:gamma2}
\end{equation}
The situation is a bit different than for $\beta$ and $\alpha$ with no dependence on CKM elements $V_{tq}$.
Measurements of $\gamma$ are not done with the aid of $B^0 -\bar B^0$ mixing, involving $t$ quarks in the box diagram loops of Fig.~\ref{fig:boxes}. Instead, $\gamma$ is accessible in interference between various pairs of tree-level diagrams. This does not imply an easy measurement, unfortunately, and $\gamma$ remains the most poorly measured of the three unitarity triangle angles, in spite of many possible channels that have been proposed.

\begin{figure}[h]
\begin{center}
\hskip1pt\raise.3cm\hbox{\includegraphics[width=15cm]{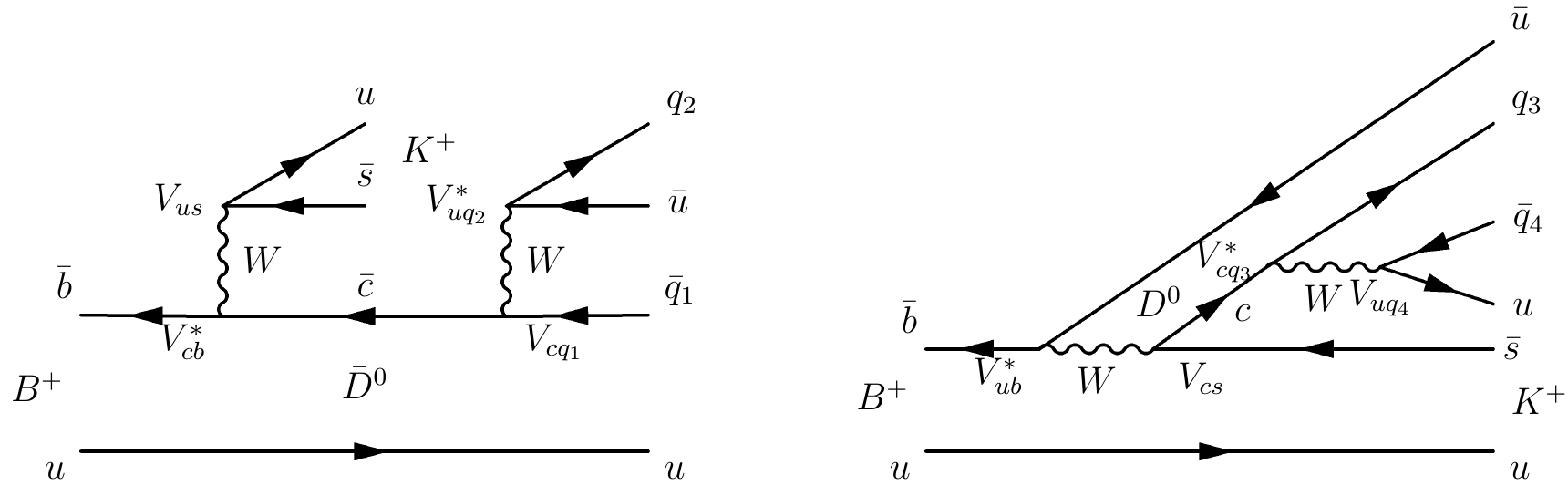}}\break
\vbox{\hskip-.3cm(a)\hskip5.5cm (b)}
\caption{\label{fig:BplusDK}
(a) $B^+\to \bar D^0 K^+$, $\bar D^0\to f(u\bar u\bar q_1q_2)$, (b) $B^+\to D^0 K^+$, $D^0\to f(u\bar u\bar q_4q_3)$.}
\end{center}
\end{figure}

There are three general approaches to the measurement of $\gamma$, each with its own trade-offs.
The basic idea is to use the interference between $b\to c\bar u s$ and $b\to \bar c u s$ processes leading to the 
same final state $f$~\cite{Carter1981, Gronau1991a, Gronau1991b}. For example, consider the decays $B^-\to D^0K^-$ and $B^-\to \bar D^0 K^-$ where the $D^0$ and $\bar D^0$ are observed in a common final state. Thus, we have two interfering amplitudes. Let us investigate the phase difference between the two amplitudes. We may safely neglect $D^0-\bar D^0$ mixing~\cite{Amorim1999, Grossman2005, Giri2003}. Let 
\begin{align}
R_B &\equiv \frac{\hbox{amplitude for $B^+\to D^0 h$}}{\hbox{amplitude for $B^+\to \bar D^0 h$}}\\
R_D &\equiv \frac{\hbox{amplitude for $D^0\to f$}}{\hbox{amplitude for $\bar D^0 \to f$}}\\
R &= R_BR_D
\end{align}
From Fig.~\ref{fig:BplusDK}  we see that
\begin{align}
R &= R_0\frac{V^*_{ub}V_{cs}}{V^*_{cb}V_{us}}\frac{V^*_{cq_3}V_{uq_4}}{V_{cq_1}V^*_{uq_2}}\label{eq:RBDf}\\
&\approx R_0\left(-\frac{V^*_{ub}V_{ud}}{V^*_{cb}V_{cd}}\right),
\label{eq:RBDfapprox}
\end{align}
where $R_0$ is the ratio other than the dependence on the CKM elements, including any strong phase difference.
The quark labels $q_1,\ldots,q_4$ may be $d$ or $s$ with the constraint that the pairs $(q_1,q_2)$ and $(q_3, q_4)$ have the same flavor content. Similarly with Eq.~\ref{eq:lambdabeta}, the approximation in Eq.~\ref{eq:RBDfapprox} holds up to corrections of order $\lambda^4$. As with our discussion of $\beta$, this result assumes unitarity; more generally we use Eq.~\ref{eq:RBDf}. The $D$ decay portion actually plays no role in the weak phase here.
Referring to Eq.~\ref{eq:gamma2} the phase difference is
\begin{equation}
\arg(R) = \arg(R_0) + \gamma.
\end{equation}
The $CP$-violating weak phase changes sign if we look at $B^-$ decays instead, but the $CP$-conserving strong phase does not.
In principle then we can disentangle the weak and strong phases and measure $\gamma$.

It is conventional to define
\begin{equation}
r_B \equiv \left|\frac{\hbox{amplitude for $B^+\to D^0 K^+$}}{\hbox{amplitude for $B^+\to \bar D^0 K^+$}}\right|,
\end{equation}
that is, $r_B = |R_B|$ for $h=K^+$. Noting that $B^+\to D^0K^+$ is color-suppressed compared with $B^+\to \bar D^0 K^+$, we expect $r_B\approx \frac{1}{3}\left|\frac{V^*_{ub}V_{cs}}{V^*_{cb}V_{us}}\right|\sim 0.1$.
It is also conventional to define the strong phase difference between the two $B$ decays as $\delta_B$,
hence $R_B = r_Be^{i(\delta_B+\gamma)}$. Correspondingly, for the ratio of the two $D$ decay amplitudes we
may write $R_D
= r_De^{i\delta_D}$, where $\delta_D$ is a strong phase difference. Neglecting $D$ mixing, the phase $\delta_D$ will cancel out of our expressions.

In the GLW~\cite{Gronau1991a, Gronau1991b} method, the $D$ decays to a $CP$ eigenstate $f$, such as $K^+K^-$, $\pi^+\pi^-$, or $K^0_S\pi^0$. Consider the $CP$ eigenstates of the neutral $D$ meson (neglecting $CP$ violation in the $D$ system):
\begin{align}
D_+ &= \frac{1}{\sqrt{2}}(D^0+\bar D^0)\\
D_- &=\frac{1}{\sqrt{2}}(D^0-\bar D^0).
\end{align}
We define a $(B^+, B^-)$-summed branching ratio for decays to $D$ eigenstates of $CP$ compared with flavor eigenstates:
\begin{equation}
R_\pm \equiv 2\frac{\Gamma(B^-\to D_\pm K^-) +  \Gamma(B^+\to D_\pm K^+)}{\Gamma(B^-\to D^0 K^-) +  \Gamma(B^+\to \bar D^0 K^+)} = 1+ r_B^2\pm2r_B\cos\delta_B\cos\gamma,
\label{eq:Rpm}
\end{equation}
and a corresponding direct $CP$-violation asymmetry:
\begin{equation}
A_\pm \equiv \frac{\Gamma(B^-\to D_\pm K^-) -  \Gamma(B^+\to D_\pm K^+)}{\Gamma(B^-\to D_\pm K^-) +  \Gamma(B^+\to \bar D_\pm K^+)}=\frac{\pm 2r_B\sin\delta_B\sin\gamma}{1+r_B^2\pm2r_B\cos\delta_B\cos\gamma}.
\label{eq:Apm}
\end{equation}
Thus, by measuring the two ratios and the two asymmetries we have four measurements which can be used to
determine the three parameters $r_B,\delta_B$, and $\gamma$. Unfortunately, there is an ambiguity under $(\delta_B,\gamma) \leftrightarrow (\pi-\delta_B,\pi-\gamma)\leftrightarrow (\pi+\delta_B,\pi+\gamma)\leftrightarrow (\gamma,\delta_B)$.

\begin{figure}[h]
\begin{center}
\hskip1pt\raise.2cm\hbox{\includegraphics[width=15cm]{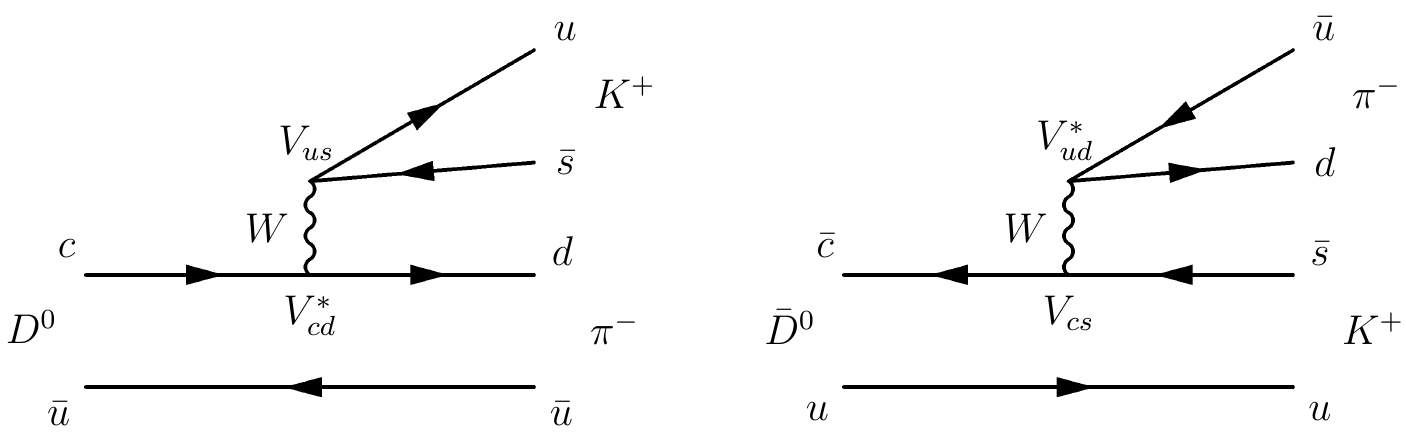}}\break
\vbox{\hskip-.6cm(a)\hskip7.5cm (b)}
\caption{\label{fig:ADS}
(a) Doubly Cabibbo-suppressed decay $D^0\to K^+\pi^-$, (b) Cabibbo-allowed decay $\bar D^0 \to K^+\pi^-$.}
\end{center}
\end{figure}

A difficulty with the GLW method is the smallness of $r_B$, which suppresses the interference terms in Eqs.~\ref{eq:Rpm} and~\ref{eq:Apm}. The ADS method~\cite{Atwood1997, Atwood2001} is designed to mitigate this with interfering amplitudes of comparable magnitude. The idea is, instead of using $D\to f_{CP}$
decays to $CP$ eigenstates, to use decay modes that counter the difference in the $B$ decay amplitudes with different Cabibbo suppression for the $D^0$ and $\bar D^0$ decays. For example (Fig.~\ref{fig:ADS}), $D^0\to K^+\pi^-$ is doubly Cabibbo-suppressed while $\bar D^0\to K^+\pi^-$ is Cabibbo-favored. For $D^0$ and $\bar D^0$ to $K^-\pi^+$ this pattern reverses.
With $r_D\equiv |R_D|$, the interference term is strongest when $r_Br_D\sim 1$. Unfortunately, this also means working with small branching fractions.

Another variation~\cite{Giri2003, Poluektov2004} is sometimes referred to as the GGSZ method. In this approach,
a multibody decay of the $D$ meson is used, such as $D\to K_S\pi^+\pi^-$. This permits the use of only Cabibbo-favored $D$ decays, although the $r_B$ suppression remains. The idea is to study the
interference as a function of the Dalitz plot variables for both $B^+$ and $B^-$ decays. This method presently provides the most precise values for $\gamma$. 

The color-suppression contributing to $r_B$ can also potentially be mitigated by looking at multibody $X_s$ in $B\to X_s D$ decays. The simplest case is perhaps the neutral $B^0\to D^0K^{*0} and \bar D^0K^{*0}$ decay~\cite{Gershon2009, Gershon2009a}. The $K^{*0}$ decay tags the $B$ flavor, and a time-independent analysis may be used. A $DK\pi$ Dalitz plot analysis can be used to extract $\gamma$ from the relative amplitudes. LHCb has recently performed such an analysis~\cite{Aaij2016}, demonstrating the principle.

We remark that the analyses must deal with $D$ hadronic decay 
parameters, such as $r_D$ and $\delta_D$ (including variation over the Dalitz plot). This is handled in various ways, but measurements at the $\psi(3770)$, where the coherence of the $D\bar D$ state may be employed, provide important information, so far from CLEO-c~\cite{Briere2009, Lowery2009, Libby2010}. There is some correlation among $\gamma$ results from
different analyses due to common input; however at present the statistical
uncertainties dominate so this is not a significant issue.

For the average for the ADS/GLW/GGSZ $B\to D^{(*)}K^{(*)}$ analyses, we may consider
the results of the CKMfitter collaboration~\cite{Charles2015}, discussed further in section~{sec:fits}.
The CKMfitter average is $\gamma = \left(73.2^{+6.3}_{-7.0}\right)^\circ$, based on data compiled by HFAG~\cite{Amhis2014} from BaBar, Belle, and LHCb. However, this average contains unpublished results.

We also perform our own average of published results from the three experiments. BaBar has published a combined analysis of the ADS, GLW, and GGSZ methods with the result~\cite{Lees2013d} (modulo an ambiguity of 180 degrees):
\begin{equation}
\gamma = (69^{+17}_{-16})^\circ\qquad\hbox{BaBar (ADS, GLW, GGSZ)}.
\end{equation}
The uncertainty is dominated by statistics.
Belle has unfortunately not yet published their GLW analysis on their full dataset  (which is their most precise measurement), but provides a combination of their published GGSZ~\cite{Poluektov2010} and ADS~\cite{Horii2011} results~\cite{Trabelsi2013}:
\begin{equation}
\gamma = (68\pm22)^\circ\qquad\hbox{Belle (ADS, GGSZ)}.
\end{equation}
LHCb has embarked on an extensive program to measure $\gamma$~with several published ADS, GLW, and GGSZ analyses~\cite{Aaij2012c, Aaij2013b,  Aaij2014a, Aaij2014b, Aaij2014c, Aaij2014d}. This includes the $B_s\to D_s^\mp K^\pm$ channel mentioned below. Their average of these measurements is (``robust combination'', which excludes $B\to D\pi$)~\cite{LHCb2014}:
\begin{equation}
\gamma=(73^{+9}_{-10})^\circ\qquad\hbox{LHCb (ADS, GLW, GGSZ, $B_s\to D_s^\mp K^\pm$)}.
\end{equation}
We make a slight symmetrization adjustment to the intervals and obtain the
average of the BaBar, Belle, and LHCb published results, including the two-fold ambiguity:
\begin{equation}
\gamma = (71\pm8,\ 251\pm 8)^\circ\qquad\hbox{our average}.
\end{equation}
It is of interest that the $\chi^2$ probability ($p$ value) for the three measurements is $97.5$\%. While there are common systematics, the dominant uncertainties are statistical; we may be seeing an effect of non-Gaussian sampling distributions.

The angle $\gamma$ can also be measured in the combination $2\beta+\gamma$ in $b\to c\bar u d, \bar c u d$ and $\bar b\to c\bar u \bar d, \bar c u\bar d$ transitions~\cite{Dunietz1998} in a time-dependent analysis of neutral $B$ decays. The interference is again between a Cabibbo-favored and a Cabibbo-suppressed decay, but in this case there is an additional contribution from $B^0\bar B^0$ mixing, hence the $2\beta+\gamma$ combination. Experimental measurements have been performed by BaBar~\cite{Aubert2005c} using partial reconstruction of $B\to D^{*\mp}\pi^\pm$ decays~\cite{Aubert2006b}, with full reconstruction of $B\to D^{(*)\mp}\pi^\pm$ and $B\to D^{\mp}\rho^\pm$ decays and similarly from Belle~\cite{Bahinipati2011} (partial reconstruction),~\cite{Ronga2006} (full and partial reconstruction). To use these measurements to extract $\gamma$, additional theoretical assumptions are required, such as $SU(3)$ symmetry. The precision obtained is so far not competitive with the best measurements above. 

A similar situation holds in the time-dependent $b\to c\bar u s$ (eg., $B\to D^\pm K_S\pi^\mp$) case~\cite{Aubert2008b}. The LHCb collaboration reports results from $b\to c\bar u s$ in a time-dependent analysis of $B^0_s\to D_s^\mp K^\pm$~\cite{Aaij2014a} decays. This provides a measurement of $\gamma-2\beta_s$, but since $\beta_s$ is small (see Section~\ref{sec:betas} below) on the scale of present precision, effectively a measurement of $\gamma$. The result of this analysis is $\gamma =\left(115^{+28}_{-43}, 295^{+28}_{-43}\right)^\circ$, consistent within large uncertainty with other determinations. Such measurements provide in any event useful consistency cross-checks. With forthcoming data and analysis efforts at LHCb and Belle II it may be expected that the precision on $\gamma$ will soon be substantially improved.
$CP$ violation has also been measured in
$B_s\to K^+K^-$~\cite{Aaij2013a}. This channel is senstive to a combination of $\gamma$ and $\beta_s$, depending on relative tree and penguin contributions.

Summing the measured values of the three angles yields:
\begin{equation}
 \alpha+\beta+\gamma = 181(9)^\circ.
\end{equation}
This is consistent with the standard model triangle prediction of $180^\circ$.

\subsubsection{The angle $\beta_s$ (and $\phi_s$)}
\label{sec:betas}

We may measure the angle $\beta_s$ by considering $B_s\to J/\psi \phi$
(or more generally $B_s\to J/\psi K^+K^-$) in an anagolous way to
$\sin2\beta$ in $B_d\to J/\psi K_S$. The $d$ quark
is replaced by an $s$ quark, and we do not include neutral kaon mixing. Since the $\phi$ is also spin one, the orbital angular momentum can be $\ell=0,1,2$. Allowing for this,
Eq.~\ref{eq:lambdabeta} becomes, in the absence of additional phases from 
penguin diagrams and at the same level of approximation as before,
\begin{equation}
\lambda_{J/\psi\phi} = (-)^\ell\frac{V_{ts}V_{tb}^*}{V_{ts}^*V_{tb}}\frac{V_{cb}V_{cs}^*}{V_{cb}^*V_{cs}}=(-)^\ell e^{2i\beta_s}.
\end{equation}
The $\beta_s$ appears now with a positive sign in the exponent because of the difference in definition compared with $\beta$.

It is conventional to define
\begin{equation}
\phi_s\equiv \arg M_{12}.
\end{equation}
Then, to out present level of approximation, in the standard model $\phi_s=-2\beta_s$. It is presently common to quote the measurement in terms of $\phi_s$, using this relation.

As with the measurement of $\beta$ in $B^0\to J/\psi K_S$, penguin pollution is expected to be small, suppressed by $\sim\left|V_{us}V_{ub}/V_{cs}V_{cb}\right|\approx|\rho+i\eta|\lambda^2\approx0.02$. A discussion of the theoretical and experimental status may be found in~\cite{Amhis2015}.
A recent calculation using an operator product expansion in~\cite{Frings2015} conservatively estimates a maximum penguin effect of $\sim\pm1^\circ$ on $\phi_s$, depending on the polarization state. Since $\phi_s$ is itself small, it will be important to watch this possible pollution as the experimental precision improves.

In the $B_s$ case, it turns out that $\Delta\Gamma_s\equiv\Delta\Gamma(B_s)$ cannot be neglected,
and Eqs.~\ref{eq:Btof},~\ref{eq:Bbartof}, and~\ref{eq:DGmod} give the time-dependent $CP$
asymmetry:
\begin{equation}
{\cal A}_f(t) = \frac{\Gamma(\bar B_s^0(t)\to f)-\Gamma(B_s^0(t)\to f)}{\Gamma(\bar B_s^0(t)\to f)+\Gamma(B_s^0(t)\to f)} = \frac{S_f\sin\Delta m(B_s) t - C_f\cos\Delta m(B_s) t}{\cosh\frac{\Delta\Gamma_s t}{2} - \frac{2\Re\lambda}{1+|\lambda|^2}\sinh\frac{\Delta\Gamma_s t}{2}}.
\end{equation}
Hence, both $\Delta\Gamma_s$ and $\phi_s$ are extracted in time-dependent fits to the decay distributions. This determination is ambiguous under $(\phi_s,\Delta\Gamma_s)\leftrightarrow(\pi-\phi_s,-\Delta\Gamma_s)$. However, the ambiguity has been experimentally resolved~\cite{Aaij2012d} by examining the interference between
the $S$-wave ($CP(J/\psi K^+K^-)=-1$) and $P$-wave ($CP=\pm1$) contributions as a function of $K^+K^-$ mass near the $\phi$ resonance in $B_s\to J/\psi K^+K^-$ decays.
The result is that the solution with $\Delta\Gamma_s>0$ and $\phi_s$ near zero
are preferred, in accordance with the standard model expectation. We thus quote only this solution for our average for $\phi_s$.

Our average of published results for $\phi_s$ is given in Table~\ref{tab:phis}. This includes a new result using the $B_s\to D_s^+D_s^-$ channel~\cite{Aaij2014e}.
There are also results from the decay $B_s\to J/\psi \pi^+\pi^-$~\cite{Aaij2014f}. However, the subleading contributions in this case remain unclear~\cite{Amhis2015} and we do not include the result in our average.
There are additionally preliminary results from ATLAS (both the 7 and 8 GeV datasets) in~\cite{Aad2016} and from
CMS (8 GeV dataset) in~\cite{Khachatryan2015}. Our result for the average is
\begin{align}
 \phi_s &= -0.057(46) \hbox{\quad or\quad} -3.3(2.6)^\circ\\
 \Delta\Gamma_s &= 0.077(8).
\end{align}
It is noticed that the average is dominated by the LHCb measurement in the
$J/\psi K^+K^-$ channel. The value of $\phi_s$ is so far consistent with zero,
as well as with the unitarity expectation (Eq.~\ref{eq:betas}) of $\sim-0.035$ ($-2^\circ$).

\begin{table}
\caption{Experimental measurements of $\phi_s$ and $\Delta\Gamma(B_s)$. Where two errors are given, the first is statistical, the second systematic. The CDF 68\% confidence region is re-quoted here with a central value and symmetric errors. The statistical and systematic uncertainties are quadratically combined before forming the average. Asymmetric intervals are symmetrized before averaging.\label{tab:phis}}
\begin{center}
\begin{tabular}[h]{lllll}
\hline\hline
\multicolumn{1}{c}{Experiment} & Mode &\multicolumn{1}{c}{$\phi_s$ (radians)} & $\Delta\Gamma(B_s)\hbox{ ps}^{-1}$ & Reference\\
\hline
ATLAS & $J/\psi\phi$  4.9 fb$^{-1}$  & $0.12(25)(5)$ & 0.053(21)(10) &  \cite{Aad2014a}\\
CDF & $J/\psi\phi$  9.6 fb$^{-1}$ & $-0.24(36)$&  0.068(26)(9) & \cite{Aaltonen2012}\\
D0 & $J/\psi\phi$  8.0 fb$^{-1}$ & $-0.55^{+0.33}_{-0.36}$ & $0.163^{+0.065}_{-0.064}$ & \cite{Abazov2012}\\
LHCb & $J/\psi K^+K^-$  3.0 fb$^{-1}$ & $-0.058(49)(6)$ & 0.0805(91)(32) & \cite{Aaij2015c}\\
LHCb & $D_s^+D_s^-$  3.0 fb$^{-1}$ & $0.02(17)(2)$ & \hfil--\hfil & \cite{Aaij2014e}\\
\hline
  Our average & & -0.057(46) & 0.077(8)& \\
\hline\hline
\end{tabular}
\end{center}
\end{table}

\section{Global Fits}
\label{sec:fits}

There are at least three substantial efforts, with different approaches, that concentrate on the global 
compatibility of the measurements related to the CKM matrix with the standard model
and possible new physics scenarios. These efforts also often perform their own averages of expermeintal results, and provide best-fit estimates of the standard model CKM parameters ($A$, $\lambda$, $\bar\rho$, $\bar\eta$). The most familiar product of these 
efforts are graphs of allowed regions in the $(\bar\rho,\bar\eta)$ plane, but many other matters are addressed as well. As they all have the same base of measurements, it is not surprising that they usually lead to similar conclusions, but this is not guaranteed and it is useful to have the different approaches to compare.  We briefly discuss and contrast the three
approaches here, as there has been some confusion and controversy about their intent and validity.

The CKMfitter collaboration~\cite{Charles2005,Charles2015} (home page \href{http://ckmfitter.in2p3.fr/}{http://ckmfitter.in2p3.fr/}) is based on frequentist statistics.
The SCAN method~\cite{Eigen2014,Eigen2015} is also based on frequentist statistics, 
but with a different approach for interval estimation. It has its genesis in the methodology of the ``BaBar physics book''~\cite{Harrison1998}.
The UTfit collaboration~\cite{Ciuchini2001, Bona2005, Bona2006}
(home page \href{http://www.utfit.org/UTfit/}{http://www.utfit.org/UTfit/}) is based
on Bayesian statistics. It may be noted that~\cite{Ali2003} also performs a fit using Bayesian statistics.

A point of controversy has been the treatment of ``theoretical errors''. These arise when a theoretical calculation of a quantity (e.g., a bag parameter) is performed, perhaps at some order in an expansion. The uncalculated higher orders are, while presumed small, not known, and hence there is an uncertainty in the true value of the quantity because the calculation
is only approximate. The difference between  the approximate calculation and the true value is a theoretical error. We should remark that distinctions are usually made between
theoretical uncertainties that have a statistical basis, such as sampling statistics in lattice computations, and systematic uncertainties, such as neglect of higher orders.  We'll discuss the different treatments of these errors in the following.
It should perhaps be cautioned here that the different fits don't always use the same theoretical input, as noted in~\cite{Ligeti2016} concerning $\eta_{cc}$.

We first discuss the two frequentist approaches, CKMfitter and SCAN method. We leave aside for
the moment the problematic issue of theoretical uncertainties and start with the distinguishing statistical algorithms. It should be remarked that both use standard frequentist statistical methodology with correct coverage under the assumed model in the limit of no theoretical uncertainty. The model in both cases is that the measurements
are sampled from normal distributions with mean values according to the standard model
(or new physics model) predictions. Because of the normality, the least-squares statistic (using the appropriate covariance matrix) follows a $\chi^2$ distribution.

The different statistical treatment of these two methods is in the interval estimation. The
basic methods are discussed and contrasted in~\cite{Narsky2014}, and generically in standard statistics textbooks, such as~\cite{Shao2003} .
The CKMfitter approach looks for specified changes in $\chi^2$ from the minimum value, e.g., a change of $\Delta\chi^2=1$ for a 68\% confidence interval in one dimension of parameter space. This familiar approach can be regarded as the use of pivotal quantities. It presents the confidence interval in such a way that the size of the interval reflects the underlying precision of the measurement. 

The SCAN method had motivation in the testing of the consistency of measurements with the standard model (or new physics models). This motivation extends to the methodology for the confidence intervals. As the $\chi^2$ test for consistency is used, the $\chi^2$ statistic is again used in the construction of a confidence interval. In this
construction, say for a $1-\alpha = 68\%$ confidence interval, we consider the $\chi^2$ test at the $\alpha$ significance level. The critical value $\chi^2_{\rm crit}$ for such a test is give by solving for it in ${\rm Prob}(\chi^2\ge \chi^2_{\rm crit}|H_0) = \alpha$, where the null hypothesis is the model in which the parameter resides (e.g., the standard model). If a confidence region in a $d$-dimensional subspace of a $p$-dimensional parameter space is desired, the number of degrees of freedom is
$n-p+d$, where $n$ is the number of measurements. 
Knowing $\chi^2_{\rm crit}$ the confidence region is constructed by including all those points in the $d$ dimensional subspace for which the $\chi^2$ is not greater than $\chi^2_{\rm crit}$ (under $H_0$). Intervals constructed in this way provide a sense of
the size of the region that is consistent with the model, at the stated significance level.
It is quite possible that null intervals will arise, indicating inconsistency of the data with the model at the stated significance. Nevertheless, if the model is correct, the quoted intervals have the correct coverage.

The treatment of theoretical uncertainties is implemented differently in the two
frequentist approaches, but this is more a difference in technology than in principle.
Both interpret the theoretical uncertainties as reflecting ranges within which the 
true value is reasonably expected to lie, rather than as providing some sort of ''confidence interval''. In frequentist statistics, there is no sampling distribution for
these theoretical parameters. Any value in the specified range is deemed a possible value (and any
value outside is deemed not possible, although CKMfitter optionally relaxes this).

CKMfitter implements the theoretical uncertainties by including in the likelihood a term for each theoretical uncertainty, in which the term is one for any value in the range and zero for any value
outside the range. Since there is no relative ``penalty'' for different allowed values of the theoretical parameter, the confidence intervals for a quantity of interest (e.g. $\bar\rho$)
will be the union of the confidence intervals for each allowed value of the theoretical parameter. 
The SCAN method implements a theoretical parameter range by trying values throughout the range. The resulting confidence interval is then the union of the confidence intervals
corresponding to each value tried. That is, one ``scans'' over the theoretical parameters
within their allowed ranges. The net result for the two methods is the same, except for the different construction of confidence intervals, as well as other implementation details
for which the reader is referred to the primary references.

It is important to notice that these frequentist approaches (assuming an overall correct model of course) lead to confidence intervals that err toward overcovering (and to statistical tests
that tend to accept $H_0$ with more than the stated probability). This is because
the model space is extended from the true model (the correct values of the theoretically
uncertain parameters) to a set of models that includes the true model. This is generally considered the appropriate direction to err, but it does mean that, for example, our 
statisical tests are not as powerful as they would be with more precise theoretical
parameters.

UTfit is based instead on Bayesian statistics. We consider the case in which a posterior distribution in the $(\bar\rho,\bar\eta)$ plane is desired. The prior distribution for the physically allowed region of this plane is taken to be uniform in UTfit. The likelihoods corresponding to the experimental measurements is in principle the same
as for the frequentist approaches, being based on the same sampling distributions.
Likelihood terms are also included for the theoretically uncertain parameters. These likelihoods
may be uniform, Gaussian, or something else, based on the expert judgement of the
researchers. They represent relative degree-of-belief statements about the possible values of these parameters.

None of these approaches is ``wrong'', although this is unfortunately a controversial statement. The frequentist approaches assert under their stated assumptions that their
results have at least the stated coverage, in the frequency sense. There is no arguing with this. If there is to be an argument, it should be over whether this is useful, since 
further (Bayesian!) steps are required to turn it into statements about truth. The Bayesian approach
states that their posterior provides a probability statement about the truth. If you agree
with the prior distribution and the expert degree-of-belief statements then you come to
the same degree-of-belief conclusions about truth. Again, no argument. It is even useful since it provides your relative betting odds on different versions of truth according to your degree-of-belief. The argument here is whether you agree with the subjective (degree-of-belief) premises. If not, all bets are (literally) off.

Attempts have been made to compare these approaches on common inputs, recognizing that they are not meant to do the same thing~\cite{Bevan2014,Eigen2015}. The results of the fit for $\bar\rho$, $\bar\eta$, $\alpha$, $\beta$, and $\gamma$ are compared.
The central values (point estimates) for all of these are similar for all three approaches,
with differences well within the one standard deviation error bars, which shouldn't be
surprising).  The interval estimates from UTfit are generally smaller than from the frequentist calculations, presumably due to the somewhat more agressive treatment
of the theoretical uncertainties. The SCAN method intervals tend to be larger than 
the CKMfitter intervals. This is probably due to a fluctuation towards consistency with
the standard model, the $p$ value is 0.76, hence the inverted test acceptance region intervals will tend to be large.

\section{Summary and Discussion}
\label{sec:discussion}

Table~\ref{tab:meas} provides a summary of what we consider to be the present status of measurements
of the CKM quantities as reviewed herein. For the magnitudes of the elements, we express the results as the squares, because these are probably closer to being normally-distributed samplings than simply the magnitudes. Thus, when we perform a least squares fit, we'll use these squared values, improving the validity of the assumption of a $\chi^2$ distribution for the least-squares statistic. The quantities in Table~\ref{tab:meas} are selected to avoid correlations as well, although there are two exceptions where correlation coefficients are noted. These correlations are included in our fits, all others are assumed to be zero.

\begin{table}[h]
\caption{Summary of measurements related to the CKM matrix. The 
abbreviations ``ex'' and ``in'' stand for the ``exclusive" and ``inclusive'' averages, respectively.
\label{tab:meas}}
\begin{center}
\begin{tabular}[h]{ll}
\hline\hline
Quantity & Measured value \\
\hline
$|V_{ud}|^2$ &  $0.94900(42)$  \\
$|V_{us}|^2$  & $0.05046(20)$  \\
\quad $\rho(|V_{us}|^2,|V_{ud}|^2)$ & \quad 0.08 \\
$|V_{cs}|^2$ & $0.983(30)$  \\
$|V_{cd}|^2$ & $0.0466(19)$ \\
\hline
$|V_{cb}|^2$ (ex) &  $0.00154(5)$  \\
$|V_{cb}|^2$  (in) & $0.00178(7)$ \\
$|V_{ub}|^2$ (ex) &  $1.26(9)\times10^{-5}$  \\
\quad $\rho(|V_{ub}|^2,|V_{cb}|^2)$ (ex) & \quad 0.14 \\
$|V_{ub}|^2$  (in) & $2.02(19)\times10^{-5}$ \\
$|V_{tb}|^2$  &  $1.01(7)$ \\
$|V_{ts}V_{tb}^*|^2$  & $0.00164(23)$ \\
$|V_{td}/V_{ts}|^2$  &  $0.047(5)$ \\
\hline
$|\epsilon|$ & $0.002228(11)$\\
$\sin2\beta$ & $0.691(17)$ \\
$\cos2\beta$ & $1.3(3)$ \\
$\alpha$ (degrees) & $88(5)$ or $268(5)$\\
$\gamma$ (degrees) & 71(8) or 251(8) \\
$\phi_s$ (degrees) & -3.3(2.6) \\
\hline\hline
\end{tabular}
\end{center}
\end{table}

It is of course of interest to investigate how consistent these measurements agree with the standard model
expectation of $3\times3$ unitarity, as well as to infer best estimates for the four standard model parameters. We perform a number of
studies to answer several questions of this nature. 

We remark that the error estimates in Table~\ref{tab:meas} sometimes include quadratically combined experimental and theoretical uncertainties. Our least squares fits thus treat these two sources of uncertainty symmetrically. In particular, the
interpretation of the $\chi^2$ statistic as being distributed according to a $\chi^2$ distribution depends on the quantities used as measurements being sampled from normal distributions with the error estimates taken to be the standard deviations. We have already discussed in Section~\ref{sec:fits} that such a treatment is controversial, since there is no expectation that theoretical errors arise in sampling from a normal distribution. Our derived interval estimates and $p$ values must be taken with caution. Compared with the CKMfitter or SCAN methods of Section~\ref{sec:fits}, we may expect our interval estimates to tend to be smaller (and may overcover or undercover) as well as our $p$ values. However, if we find that the standard model
is consistent with the data, according to our $p$ values, then we have confidence that this is a valid conclusion. On the other hand, if it looks
like there is an inconsistency, further investigation may be required, especially if the conclusion is marginal.

A simple first question is the consistency of the upper left $2\times2$ matrix with the Cabibbo model. We thus perform a least-squares fit to the
first four quantities in Table~\ref{tab:meas} to a $2\times2$ orthognal matrix with one parameter. As the measurements are the squares of the matrix elements, this is a strictly linear fit for $\sin^2\theta_C$, where $\theta_C$ is the Cabibbo angle. The result is
\begin{equation}
\sin^2\theta_C =0.05152(19)\qquad \sin\theta_C = 0.2248(4).
\end{equation}
According to convention, we pick the positive sign for $\sin\theta_C$.
The $p(\chi^2)$ value is 8\%, that is, the Cabibbo model remains
consistent with the magnitudes of the elements in the upper two-by-two submatrix of the CKM matrix.

Turning to the full $3\times3$ matrix, we investigate first what we learn from only measurements of the magnitudes of the elements. Our fits are all in the context of the standard model. The parameterization we adopt is $A,\lambda,\rho,\eta$. In terms of these parameters, we use equations~\ref{eq:Vangles}, and~\ref{eq:Wpars} to 
obtain the expressions for the expectation values of our measured quantities (there is no need to use the approximations in~Eq.\ref{eq:Vexpansion} for this). With the fit results we also make the transformation to $\bar\rho,\bar\eta$ according to
Eq.~\ref{eq:Vapex}, as well as the Jarlskog invariant, J (Eq.~\ref{eq:Jwolf}), and the CKM phase $\delta$ in the parameterization of Eq.~\ref{eq:Vangles}.

Since there is inconsistency between exclusive and inclusive measurements for $|V_{cb}|^2$ and  $|V_{ub}|^2$, we investigate the two types of measurement separately. It will be assumed that either the exclusive measurements are ``right'' in both cases, or the inclusive. This seems to be the most plausible scenario since the problem could arise from similar origins for both quantities. The fit results are shown in Table~\ref{tab:fcpfits}. 
The magnitude alone are sufficient to determine all four Wolfenstein parameters, well enough to conclude that $CP$ violation must be present. We do not learn anything about whether
the inclusive or exclusive measurements are preferred from this exercise. There is enough freedom in the standard model to accommodate  either set of measurements with acceptable and comparable $p$ values.

The linear correlation coefficients for the fitted parameters are shown in
Table~\ref{tab:fitcorr} for the fit using exclusive values for $|V_{cb}|^2$ and  $|V_{ub}|^2$. The pattern is the same for the fit using the inclusive measurements, although the values move around substantially in some cases. As can be expected, the $\bar\rho-\bar\eta$ correlation is negative, due to the presence of $\bar\rho^2+\bar\eta^2$ terms in the
expectation values.

\begin{table}
\caption{Fit results, either based on the measurement of the magnitudes of the CKM elements alone, or including also the measurements of the phases.
\label{tab:fcpfits}}
\begin{center}
\begin{tabular}[h]{cllll}
\hline\hline
Parameter & \multicolumn{2}{c}{Using magnitudes}&\multicolumn{2}{c}{Using magnitudes and phases} \\
& Exclusive & Inclusive & Exclusive & Inclusive\\
\hline
$A$           & 0.778(12)    &  0.834(16) & 0.790(10) &0.841(13)\\
$\lambda$  & $0.2248(4)$ & 0.2248(4) & 0.2248(4) & 0.2248(4) \\
$\rho$       & 0.14(5)        & 0.18(5)     & 0.115(22) & 0.143(25)\\
$\eta$       & $0.375(23)$ & 0.440(28) & 0.374(10) & 0.369(11)\\
\hline
$\bar\rho$ & 0.14(5)        & 0.17(5)     & 0.112(22) & 0.140(24)\\
$\bar\eta$ & $0.365(22)$ & 0.429(27) & 0.364(9) & 0.359(11)\\
$J$           & $2.93(18)10^{-5}$ & $3.95(25)10^{-5}$ & $3.01(7)10^{-5}$& $3.37(8)10^{-5}$\\
$\delta$    & $69(7)^\circ$ & $68(6)^\circ$ & $73(3)^\circ$ & $69(4)^\circ$ \\
\hline
$p$ value & 20\% & 26\% & 14\% & 1.8\% \\
\hline\hline
\end{tabular}
\end{center}
\end{table}

We may repeat this exercise by including as well all of the phase measurements in Table~\ref{tab:meas}. To do this, it is necessary to add auxillary parameters to incorporate the $|\epsilon|$ measurement and its expectation according to Eq.~\ref{eq:epsilonK} (including the $k_\epsilon$ factor). Referring to Section~\ref{sec:epsilon}, we choose to define four auxiliary parameters:
\begin{equation}
\begin{split}
C_T &\equiv C_\epsilon\hat B_K k_\epsilon = 2.64(7)\times10^{4}\\
C_{tt} & \equiv \eta_{tt}S_0(x_t) = 1.336(19)\\
C_{ct} &\equiv \eta_{ct}S_0(x_c,x_t) = 1.10(11)\times10^{-3}\\
C_{cc} &\equiv \eta_{cc}S_0(x_c) = 4.7(1.9)\times10^{-4}\\
\end{split}
\end{equation}
Four terms are added to the $\chi^2$ for these, with expectation values given by the point estimates.
Again, even though there are considerable theoretical components to the
error estimates for these quantities, we treat them symmetrically with the 
experimental errors and the above discussion applies. At the current level of precision, Eq.~\ref{eq:epsilonKrhoeta} could equally well
have been used as Eq.~\ref{eq:epsilonK}.

\begin{table}[h]
\caption{Linear correlation coefficients among the CKM parameters for selected fits. Results from the fits with exclusive values for $|V_{cb}|^2$ and  $|V_{ub}|^2$ are shown.
\label{tab:fitcorr}}
\begin{center}
\begin{tabular}[h]{clll}
\hline\hline
 & $\lambda$ & $\rho$ & $\eta$\\
\hline
\multicolumn{4}{c}{Using magnitudes}\\
 $A$ & -0.22 & -0.036 & -0.15\\
$\lambda$    & & 0.030  & -0.058  \\
$\rho$  &  &  & -0.70\\
\hline
\multicolumn{4}{c}{Using magnitudes and phases}\\
$A$ & -0.32 & 0.25 & -0.56\\
$\lambda$  &  & 0.050  & -0.072  \\
$\rho$ &  &   & -0.51\\
\hline\hline
\end{tabular}
\end{center}
\end{table}

With the phase information, we encounter a preference for the exclusive $|V_{ub}|^2$ and $|V_{cb}|^2$ measurements under the standard model. With the exclusive measurements, the data is consistent with
the standard model $3\times3$ unitarity (even given the questionable treatment of theoretical uncertainties). On the other hand, the inclusive measurements result in a low $p$ value for consistency, less than 2\%. However, as discussed above, the treatment of the theoretical uncertainties is an
issue in the interpretation of this. If the theoretical uncertainties are to be treated as ranges of allowed values, as in the CKMfitter and SCAN method approaches, somewhat larger $p$ values can be achieved. Even so, the
inclusive determination of  $|V_{ub}|^2$ and $|V_{cb}|^2$ 
is disfavored. Because 2\% (or a few percent, allowing for the theoretical errors) is not extremely unlikely, we cannot make
a strong statement. We have already remarked in Section~\ref{sec:Vcb} that
explaining the discrepancy in terms of new physics is difficult. Thus, unless new physics is affecting other measurements in a somewhat perverse way, we conclude that the presently somewhat favored scenario is that there is a problem with the inclusive determinations of  $|V_{ub}|^2$ and $|V_{cb}|^2$.

The $CP$-violation phase of Eq.~\ref{eq:Vangles} is approximately 70$^\circ$, i.e., not far
below the maximal $CP$-violating $90^\circ$. However, another way to look at the
amount of $CP$ violation is the Jarlskog invariant. In the parameterization of
Eq.~\ref{eq:Vangles}, striking out the second row and column and using Eq.~\ref{eq:Jinvariant} we evaluate
\begin{equation}
J = c_{12}s_{12}c_{13}^2s_{13}c_{23}s_{23}\sin\delta.
\end{equation}
This is maximal for $\sin\theta_{12}=\sin\theta_{23}=1/\sqrt{2}$, $\sin\theta_{13}=1/\sqrt{3}$, and $\sin\delta =1$, for a maximum of $\max J = 1/6\sqrt{3}\approx0.0962$.
Our measured value is $J\approx3\times 10^{-5}$, 3.5 orders of magnitude smaller.
The smallness of this measure of $CP$ violation arises from the uneven magnitudes of
the elements of $V$; the matrix is approximately diagonal.

We may obtain additional insight into the fit results by looking at the residuals. The normalized residuals are given by
\begin{equation}
r_i = \frac{x_i-f_i(\hat p)}{\sigma_i},
\end{equation}
where $x$ is the measurement vector, $f$ is the vector of expectation
values for the measurements based on the parameter estimates $\hat p$,
and $\sigma$ is the standard deviation vector for the measurements. Except for the small correlations, the residuals are the square roots of the
individual measurement contributions to the $\chi^2$ for the fit. The normalized residuals are summarized in Table~\ref{tab:normRes} for the
fits using the exclusive measurements of $|V_{ub}|^2$ and $|V_{cb}|^2$.
There are no obvious difficulties seen in these residuals. For the most part, adding the phase measurements doesn't move the residuals for the magnitudes by very much.

\begin{table}[h]
\caption{Normalized residuals of measurements related to the CKM matrix. Column ``A'' is from the fit using the magnitudes only; column ``B'' is from the fit using both magnitude and phase measurements. Both fits use the exlcusive averages for $|V_{ub}|^2$ and $|V_{cb}|^2$.
\label{tab:normRes}}
\begin{center}
\begin{tabular}[h]{lrr}
\hline\hline
Quantity & \multicolumn{2}{c}{Normalized residual}  \\
& A & B \\
\hline
$|V_{ud}|^2$ & -1.1 & -1.0  \\
$|V_{us}|^2$  & -0.4 & -0.4  \\
$|V_{cs}|^2$ & 1.1 & 1.1  \\
$|V_{cd}|^2$ & -2.0 & -2.0 \\
$|V_{cb}|^2$ & -0.2 & -1.2  \\
$|V_{ub}|^2$ & -0.0 & -0.3 \\
$|V_{tb}|^2$  & 0.2 & 0.2 \\
$|V_{ts}V_{tb}^*|^2$  & 0.8 & 0.6 \\
$|V_{td}/V_{ts}|^2$  &  -0.0 & -0.6 \\
$|\epsilon|$ &  & 0.1\\
$C_T$ & & -0.5\\
$C_{tt}$ & & -0.2\\
$C_{cc}$ & & 1.3 \\
$C_{ct}$ & & -0.7 \\
$\sin2\beta$ & & -0.7 \\
$\cos2\beta$ &  & $2.0$ \\
$\alpha$ (degrees) & & 0.6\\
$\gamma$ (degrees) & & -0.2 \\
$\phi_s$ (degrees) & & -0.4 \\
\hline\hline
\end{tabular}
\end{center}
\end{table}

The experimental and theoretical effort to understand the CKM matrix has been vast, reflecting its fundamental status in the standard model and as a potential window into physics beyond.
The precision of the measurements and theoretical input continues to advance.
Here, we have taken a snapshot of the experimental information measuring the CKM matrix. There are some issues, such as the
disagreement between inclusive and exclusive determinations of
$|V_{ub}|^2$ and $|V_{cb}|^2$. However, the general picture of $V$ as a $3\times3$ unitary matrix remains a remarkable achievement. 

\section*{Acknowledgments}

I am deeply indebted to my collaborators on the BaBar experiment.
I would like to thank Gerald Eigen, Brian Meadows, Arantza Oyanguren, Markus R\"ohrken, and Karim Trabelsi for helpful responses to my questions. I am also grateful for helpful comments from  Chao-Qiang Geng, Benjamin Grinstein, Michael Gronau, Yu-Kuo Hsiao, Xian-Wei Kang, and Zoltan Ligeti.
This work is supported in part by the US Department of Energy under grant number DE-SC0011925.

\vfil\break

\section*{References}

\bibliography{CKMexp}

\begin{thebibliography}{100}
\expandafter\ifx\csname url\endcsname\relax
  \def\url#1{\texttt{#1}}\fi
\expandafter\ifx\csname urlprefix\endcsname\relax\def\urlprefix{URL }\fi
\expandafter\ifx\csname href\endcsname\relax
  \def\href#1#2{#2} \def\path#1{#1}\fi

\bibitem{Cabibbo1963}
N.~Cabibbo, Unitary symmetry and leptonic decays, Phys. Rev. Lett. 10~(12)
  (1963) 531--533.
\newblock \href {http://dx.doi.org/10.1103/PhysRevLett.10.531}
  {\path{doi:10.1103/PhysRevLett.10.531}}.

\bibitem{Kobayashi1973}
M.~Kobayashi, T.~Maskawa, {$CP$}-violation in the renormalizable theory of weak
  interaction, Progress of Theoretical Physics 49~(2) (1973) 652--657.
\newblock \href {http://dx.doi.org/10.1143/PTP.49.652}
  {\path{doi:10.1143/PTP.49.652}}.

\bibitem{Olive2014}
K.~Olive, et~al. (Particle Data~Group), {Review of particle properties}, Chin.
  Phys. C38 (2014) 090001.
\newblock \href {http://dx.doi.org/10.1088/1674-1137/38/9/090001}
  {\path{doi:10.1088/1674-1137/38/9/090001}}.

\bibitem{Aoki2014}
S.~Aoki, Y.~Aoki, C.~Bernard, T.~Blum, G.~Colangelo, et~al., {Review of lattice
  results concerning low-energy particle physics}, Eur. Phys. J. C74 (2014)
  2890.
\newblock \href {http://dx.doi.org/10.1140/epjc/s10052-014-2890-7}
  {\path{doi:10.1140/epjc/s10052-014-2890-7}}.

\bibitem{Antonelli2010}
M.~Antonelli, et~al., {An evaluation of $|V_{us}|$ and precise tests of the
  standard model from world data on leptonic and semileptonic kaon decays},
  Eur.\ Phys.\ J. C69 (2010) 399--424.
\newblock \href {http://dx.doi.org/10.1140/epjc/s10052-010-1406-3}
  {\path{doi:10.1140/epjc/s10052-010-1406-3}}.

\bibitem{Amhis2012}
Y.~Amhis, et~al., {Averages of $b$-hadron, $c$-hadron, and $\tau$-lepton
  properties as of early 2012,} (2012).
\newblock \href {http://arxiv.org/abs/1207.1158} {\path{arXiv:1207.1158}}.

\bibitem{Amhis2014}
Y.~Amhis, et~al., {Averages of $b$-hadron, $c$-hadron, and $\tau$-lepton
  properties as of summer 2014} (2014).
\newblock \href {http://arxiv.org/abs/1412.7515} {\path{arXiv:1412.7515}}.

\bibitem{Gershon2015}
T.~Gershon,
  \url{http://www.slac.stanford.edu/xorg/hfag/triangle/summer2015/index.shtml},
  (HFAG collaboration) [Online; accessed 6-Jan-2016] (2015).

\bibitem{Bevan2014}
A.~J. Bevan, et~al., {The physics of the $B$ factories}, Eur. Phys. J. C74
  (2014) 3026.
\newblock \href {http://arxiv.org/abs/1406.6311} {\path{arXiv:1406.6311}},
  \href {http://dx.doi.org/10.1140/epjc/s10052-014-3026-9}
  {\path{doi:10.1140/epjc/s10052-014-3026-9}}.

\bibitem{Chau1984}
L.~L. Chau, W.~Y. Keung, Comments on the parameterization of the
  {Kobayshi-Maskawa} matrix, Phys. Rev. Let. 53~(19) (1984) 1802--1805.
\newblock \href {http://dx.doi.org/10.1103/PhysRevLett.53.1802}
  {\path{doi:10.1103/PhysRevLett.53.1802}}.

\bibitem{Wolfenstein1983}
L.~Wolfenstein, Parametrization of the {Kobayashi-Maskawa} matrix, Phys. Rev.
  Lett. 51~(21) (1983) 1945--1947.
\newblock \href {http://dx.doi.org/10.1103/PhysRevLett.51.1945}
  {\path{doi:10.1103/PhysRevLett.51.1945}}.

\bibitem{Buras1994}
A.~Buras, M.~Lautenbacher, G.~Ostermaier, Waiting for the top quark mass,
  {$K^+\to\pi^+\nu\bar\nu$, $B_s^0-\bar B_s^0$} mixing, and {$CP$} asymmetries
  in {$B$} decays, Phys. Rev. D 50~(5) (1994) 3433--3446.
\newblock \href {http://dx.doi.org/10.1103/PhysRevD.50.3433}
  {\path{doi:10.1103/PhysRevD.50.3433}}.

\bibitem{Schmidtler1992}
M.~Schmidtler, K.~R. Schubert, Experimental constraints on the phase in the
  {Cabibbo-Kobayashi-Maskawa} matrix, Z. Phys. C53 (1992) 347--354.
\newblock \href {http://dx.doi.org/10.1007/BF01597574}
  {\path{doi:10.1007/BF01597574}}.

\bibitem{Hoecker2001}
A.~Hoecker, H.~Lacker, S.~Laplace, F.~Le~Diberder, {A new approach to a global
  fit of the CKM matrix}, Eur. Phys. J. C21 (2001) 225--259.
\newblock \href {http://dx.doi.org/10.1007/s100520100729}
  {\path{doi:10.1007/s100520100729}}.

\bibitem{Jarlskog1985}
C.~Jarlskog, Commutator of the quark mass matrices in the standard electroweak
  model and a measure of maximal {$CP$} nonconservation, Phys. Rev. Lett.
  55~(10) (1985) 1039--1042, erratum: ibid. {\bf 58} (1987) 1698.
\newblock \href {http://dx.doi.org/10.1103/PhysRevLett.55.1039}
  {\path{doi:10.1103/PhysRevLett.55.1039}}.

\bibitem{Kim2012}
J.~E. Kim, M.-S. Seo, {A simple expression of the Jarlskog determinant}\href
  {http://arxiv.org/abs/1201.3005} {\path{arXiv:1201.3005}}.

\bibitem{Kim2015}
J.~E. Kim, D.~Y. Mo, S.~Nam, {Final state interaction phases obtained by data
  from CP asymmetries}, J. Korean Phys. Soc. 66~(6) (2015) 894--899.
\newblock \href {http://arxiv.org/abs/1402.2978} {\path{arXiv:1402.2978}},
  \href {http://dx.doi.org/10.3938/jkps.66.894}
  {\path{doi:10.3938/jkps.66.894}}.

\bibitem{Marciano1993}
W.~J. Marciano, A.~Sirlin, {Radiative corrections to $\pi_{\ell 2}$ decays},
  Phys. Rev. Lett. 71 (1993) 3629--3632.
\newblock \href {http://dx.doi.org/10.1103/PhysRevLett.71.3629}
  {\path{doi:10.1103/PhysRevLett.71.3629}}.

\bibitem{Rosner2015}
J.~L. Rosner, S.~Stone, R.~S. Van~de Water, {Leptonic decays of charged
  pseudoscalar mesons -- 2015}\href {http://arxiv.org/abs/1509.02220}
  {\path{arXiv:1509.02220}}.

\bibitem{Marciano1987}
W.~J. Marciano, A.~Sirlin, Constraint on additional neutral gauge bosons from
  electroweak radiative corrections, Phys. Rev. D 35~(5) (1987) 1672--1676.
\newblock \href {http://dx.doi.org/10.1103/PhysRevD.35.1672}
  {\path{doi:10.1103/PhysRevD.35.1672}}.

\bibitem{pocanic2004}
D.~Pocanic, E.~Frlez, V.~Baranov, W.~H. Bertl, C.~Bronnimann, et~al., {Precise
  measurement of the $\pi^+ \to \pi^0 e^+ \nu$ branching ratio}, Phys.Rev.Lett.
  93 (2004) 181803.
\newblock \href {http://dx.doi.org/10.1103/PhysRevLett.93.181803}
  {\path{doi:10.1103/PhysRevLett.93.181803}}.

\bibitem{Towner2010}
I.~Towner, J.~Hardy, {The evaluation of $V_{ud}$ and its impact on the
  unitarity of the Cabibbo-Kobayashi-Maskawa quark-mixing matrix}, Rept. Prog.
  Phys. 73 (2010) 046301.
\newblock \href {http://dx.doi.org/10.1088/0034-4885/73/4/046301}
  {\path{doi:10.1088/0034-4885/73/4/046301}}.

\bibitem{Hardy2015}
J.~Hardy, I.~Towner, {Superallowed $0^+\to 0^+$ nuclear β decays: 2014
  critical survey, with precise results for $V_{ud}$ and CKM unitarity}, Phys.
  Rev. C91~(2) (2015) 025501.
\newblock \href {http://dx.doi.org/10.1103/PhysRevC.91.025501}
  {\path{doi:10.1103/PhysRevC.91.025501}}.

\bibitem{Marciano2006}
W.~J. Marciano, A.~Sirlin, {Improved calculation of electroweak radiative
  corrections and the value of $V_{ud}$}, Phys. Rev. Lett. 96 (2006) 032002.
\newblock \href {http://dx.doi.org/10.1103/PhysRevLett.96.032002}
  {\path{doi:10.1103/PhysRevLett.96.032002}}.

\bibitem{Cirigliano2007}
V.~Cirigliano, I.~Rosell, {$\pi/K \to e \bar\nu_e$ branching ratios to $O(e^2
  p^4)$ in chiral perturbation theory}, JHEP 0710 (2007) 005.
\newblock \href {http://dx.doi.org/10.1088/1126-6708/2007/10/005}
  {\path{doi:10.1088/1126-6708/2007/10/005}}.

\bibitem{Marciano2004}
W.~J. Marciano, {Precise determination of $|V_{us}|$ from lattice calculations
  of pseudoscalar decay constants}, Phys. Rev. Lett. 93 (2004) 231803.
\newblock \href {http://dx.doi.org/10.1103/PhysRevLett.93.231803}
  {\path{doi:10.1103/PhysRevLett.93.231803}}.

\bibitem{Bazavov2013}
A.~Bazavov, et~al., {Leptonic decay-constant ratio $f_{K^+}/f_{\pi^+}$ from
  lattice QCD with physical light quarks}, Phys.Rev.Lett. 110 (2013) 172003.
\newblock \href {http://dx.doi.org/10.1103/PhysRevLett.110.172003}
  {\path{doi:10.1103/PhysRevLett.110.172003}}.

\bibitem{Dowdall2013}
R.~Dowdall, C.~Davies, G.~Lepage, C.~McNeile, {$V_{us}$ from $\pi$ and $K$
  decay constants in full lattice QCD with physical $u$, $d$, $s$ and $c$
  quarks}, Phys. Rev. D88 (2013) 074504.
\newblock \href {http://dx.doi.org/10.1103/PhysRevD.88.074504}
  {\path{doi:10.1103/PhysRevD.88.074504}}.

\bibitem{Bazavov2014a}
A.~Bazavov, et~al., {Charmed and light pseudoscalar meson decay constants from
  four-flavor lattice QCD with physical light quarks}, Phys. Rev. D90~(7)
  (2014) 074509.
\newblock \href {http://dx.doi.org/10.1103/PhysRevD.90.074509}
  {\path{doi:10.1103/PhysRevD.90.074509}}.

\bibitem{Leutwyler1984}
H.~Leutwyler, M.~Roos, {Determination of the elements $V_{us}$ and $V_{ud}$ of
  the Kobayashi-Maskawa matrix}, Z.Phys. C25 (1984) 91.
\newblock \href {http://dx.doi.org/10.1007/BF01571961}
  {\path{doi:10.1007/BF01571961}}.

\bibitem{Cirigliano2008}
V.~Cirigliano, M.~Giannotti, H.~Neufeld, {Electromagnetic effects in
  $K_{\ell3}$ decays}, JHEP 0811 (2008) 006.
\newblock \href {http://dx.doi.org/10.1088/1126-6708/2008/11/006}
  {\path{doi:10.1088/1126-6708/2008/11/006}}.

\bibitem{Moulson2014}
M.~Moulson, {Experimental determination of $V_{us}$ from kaon decays} (2014).
\newblock \href {http://arxiv.org/abs/1411.5252} {\path{arXiv:1411.5252}}.

\bibitem{Ambrosino2011}
F.~Ambrosino, et~al., {Precision measurement of $K_S$ meson lifetime with the
  KLOE detector}, Eur. Phys. J. C71 (2011) 1604.
\newblock \href {http://arxiv.org/abs/1011.2668} {\path{arXiv:1011.2668}},
  \href {http://dx.doi.org/10.1140/epjc/s10052-011-1604-7}
  {\path{doi:10.1140/epjc/s10052-011-1604-7}}.

\bibitem{Abouzaid2011}
E.~Abouzaid, et~al., {Precise Measurements of Direct CP Violation, CPT
  Symmetry, and Other Parameters in the Neutral Kaon System}, Phys. Rev. D83
  (2011) 092001.
\newblock \href {http://arxiv.org/abs/1011.0127} {\path{arXiv:1011.0127}},
  \href {http://dx.doi.org/10.1103/PhysRevD.83.092001}
  {\path{doi:10.1103/PhysRevD.83.092001}}.

\bibitem{Wanke2012}
R.~Wanke, {High precision measurement of the form factors of the semileptonic
  decays $K^\pm \to \pi^0 \ell^\pm \nu$ ($K_{\ell3}$)}, PoS HQL2012 (2012) 007.

\bibitem{Bazavov2014}
A.~Bazavov, C.~Bernard, C.~Bouchard, C.~DeTar, D.~Du, et~al., {Determination of
  $|V_{us}|$ from a lattice-QCD calculation of the $K\to\pi\ell\nu$
  semileptonic form factor with physical quark masses}, Phys. Rev. Lett.
  112~(11) (2014) 112001.
\newblock \href {http://dx.doi.org/10.1103/PhysRevLett.112.112001}
  {\path{doi:10.1103/PhysRevLett.112.112001}}.

\bibitem{Lusiani2014}
A.~Lusiani,
  \href{http://inspirehep.net/record/1328496/files/arXiv:1411.4526.pdf}{{Determination
  of $|V_{us}|$ from the tau lepton branching fractions}}, in: {8th
  International Workshop on the CKM Unitarity Triangle (CKM2014) Vienna,
  Austria, September 8-12, 2014}, 2014.
\newblock \href {http://arxiv.org/abs/1411.4526} {\path{arXiv:1411.4526}}.
\newline\urlprefix\url{http://inspirehep.net/record/1328496/files/arXiv:1411.4526.pdf}

\bibitem{Gamiz2007}
E.~Gamiz, M.~Jamin, A.~Pich, J.~Prades, F.~Schwab, {$|V_{us}|$ and $m_s$ from
  hadronic tau decays}, Nucl. Phys. Proc. Suppl. 169 (2007) 85--89.
\newblock \href {http://arxiv.org/abs/hep-ph/0612154}
  {\path{arXiv:hep-ph/0612154}}, \href
  {http://dx.doi.org/10.1016/j.nuclphysbps.2007.02.053}
  {\path{doi:10.1016/j.nuclphysbps.2007.02.053}}.

\bibitem{Maltman2015}
K.~Maltman, R.~J. Hudspith, R.~Lewis, C.~E. Wolfe, J.~Zanotti,
  \href{http://inspirehep.net/record/1406937/files/arXiv:1511.08514.pdf}{{A
  resolution of the puzzle of low $V_{u s}$ values from inclusive
  flavor-breaking sum rule analyses of hadronic tau decay}}, in: {10th
  International Workshop on $e^+e^-$ collisions from Phi to Psi (PHIPSI15)
  Hefei, Anhui, China, September 23-26, 2015}, 2015.
\newblock \href {http://arxiv.org/abs/1511.08514} {\path{arXiv:1511.08514}}.
\newline\urlprefix\url{http://inspirehep.net/record/1406937/files/arXiv:1511.08514.pdf}

\bibitem{Adametz2011}
A.~Adametz, \href{http://www.ub.uni-heidelberg.de/archiv/12325}{{Measurement of
  $\tau$ decays into a charged hadron accompanied by neutral $\pi$-mesons and
  determination of the CKM matrix element $|V_{us}|$}}, Ph.D. thesis,
  Heidelberg U. (2011).
\newline\urlprefix\url{http://www.ub.uni-heidelberg.de/archiv/12325}

\bibitem{delAmoSanchez2010b}
P.~del Amo~Sanchez, et~al., {Measurement of the absolute branching fractions
  for $D^-_s\rightarrow \ell^-\bar{\nu}_{\ell}$ and extraction of the decay
  constant $f_{D_s}$}, Phys. Rev. D82 (2010) 091103, [Erratum: Phys.
  Rev.D91,no.1,019901(2015)].
\newblock \href {http://arxiv.org/abs/1008.4080} {\path{arXiv:1008.4080}},
  \href {http://dx.doi.org/10.1103/PhysRevD.82.091103,
  10.1103/PhysRevD.91.019901} {\path{doi:10.1103/PhysRevD.82.091103,
  10.1103/PhysRevD.91.019901}}.

\bibitem{Zupanc2013}
A.~Zupanc, et~al., {Measurements of branching fractions of leptonic and
  hadronic $D_{s}^{+}$ meson decays and extraction of the $D_{s}^{+}$ meson
  decay constant}, JHEP 09 (2013) 139.
\newblock \href {http://arxiv.org/abs/1307.6240} {\path{arXiv:1307.6240}},
  \href {http://dx.doi.org/10.1007/JHEP09(2013)139}
  {\path{doi:10.1007/JHEP09(2013)139}}.

\bibitem{Alexander2009}
J.~Alexander, et~al., {Measurement of $B(D_s^+ \to \ell^+ \nu)$ and the decay
  constant $f_{D_s}^+$ from 600 $\hbox{pb}^{-1}$ of $e^+e^-$ annihilation data
  near 4170 MeV}, Phys.\ Rev. D79 (2009) 052001.
\newblock \href {http://dx.doi.org/10.1103/PhysRevD.79.052001}
  {\path{doi:10.1103/PhysRevD.79.052001}}.

\bibitem{Onyisi2009}
P.~U.~E. Onyisi, et~al., {Improved Measurement of Absolute Branching Fraction
  of D(s)+ ---> tau+ nu(tau)}, Phys. Rev. D79 (2009) 052002.
\newblock \href {http://arxiv.org/abs/0901.1147} {\path{arXiv:0901.1147}},
  \href {http://dx.doi.org/10.1103/PhysRevD.79.052002}
  {\path{doi:10.1103/PhysRevD.79.052002}}.

\bibitem{Naik2009}
P.~Naik, et~al., {Measurement of the pseudoscalar decay constant $f_{D_s}$
  using $D_s^+ \to \tau^+ \nu$, $\tau^+ \to \rho^+ \bar\nu$ decays}, Phys. Rev.
  D80 (2009) 112004.
\newblock \href {http://arxiv.org/abs/0910.3602} {\path{arXiv:0910.3602}},
  \href {http://dx.doi.org/10.1103/PhysRevD.80.112004}
  {\path{doi:10.1103/PhysRevD.80.112004}}.

\bibitem{Davies2010}
C.~Davies, C.~McNeile, E.~Follana, G.~Lepage, H.~Na, et~al., {Update: precision
  $D_s$ decay constant from full lattice QCD using very fine lattices}, Phys.
  Rev. D82 (2010) 114504.
\newblock \href {http://arxiv.org/abs/1008.4018} {\path{arXiv:1008.4018}},
  \href {http://dx.doi.org/10.1103/PhysRevD.82.114504}
  {\path{doi:10.1103/PhysRevD.82.114504}}.

\bibitem{Na2012}
H.~Na, C.~T. Davies, E.~Follana, G.~P. Lepage, J.~Shigemitsu, {$|V_{cd}|$ from
  $D$ meson leptonic decays}, Phys. Rev. D86 (2012) 054510.
\newblock \href {http://dx.doi.org/10.1103/PhysRevD.86.054510}
  {\path{doi:10.1103/PhysRevD.86.054510}}.

\bibitem{Burdman1997}
G.~Burdman, J.~Kambor, {Dispersive approach to semileptonic form-factors in
  heavy to light meson decays}, Phys. Rev. D55 (1997) 2817--2826.
\newblock \href {http://arxiv.org/abs/hep-ph/9602353}
  {\path{arXiv:hep-ph/9602353}}, \href
  {http://dx.doi.org/10.1103/PhysRevD.55.2817}
  {\path{doi:10.1103/PhysRevD.55.2817}}.

\bibitem{Becirevic2000}
D.~Becirevic, A.~B. Kaidalov, {Comment on the heavy $\to$ light form-factors},
  Phys.Lett. B478 (2000) 417--423.
\newblock \href {http://arxiv.org/abs/hep-ph/9904490}
  {\path{arXiv:hep-ph/9904490}}, \href
  {http://dx.doi.org/10.1016/S0370-2693(00)00290-2}
  {\path{doi:10.1016/S0370-2693(00)00290-2}}.

\bibitem{Hill2006}
R.~J. Hill, {Constraints on the form factors for $K \to \pi \ell \nu$ and
  implications for $|V_{us}|$}, Phys.Rev. D74 (2006) 096006.
\newblock \href {http://arxiv.org/abs/hep-ph/0607108}
  {\path{arXiv:hep-ph/0607108}}, \href
  {http://dx.doi.org/10.1103/PhysRevD.74.096006}
  {\path{doi:10.1103/PhysRevD.74.096006}}.

\bibitem{Scora1995}
D.~Scora, N.~Isgur, {Semileptonic meson decays in the quark model: An update},
  Phys.Rev. D52 (1995) 2783--2812.
\newblock \href {http://arxiv.org/abs/hep-ph/9503486}
  {\path{arXiv:hep-ph/9503486}}, \href
  {http://dx.doi.org/10.1103/PhysRevD.52.2783}
  {\path{doi:10.1103/PhysRevD.52.2783}}.

\bibitem{Ablikim2015a}
M.~Ablikim, et~al., {Study of dynamics of $D^0 \to K^- e^+ \nu_{e}$ and
  $D^0\to\pi^- e^+ \nu_{e}$ decays}, Phys. Rev. D92~(7) (2015) 072012.
\newblock \href {http://arxiv.org/abs/1508.07560} {\path{arXiv:1508.07560}},
  \href {http://dx.doi.org/10.1103/PhysRevD.92.072012}
  {\path{doi:10.1103/PhysRevD.92.072012}}.

\bibitem{Aubert2007}
B.~Aubert, et~al., {Measurement of the hadronic form-factor in $D^0 \to K^{-}
  e^{+} \nu_{e}$ decays}, Phys.Rev. D76 (2007) 052005.
\newblock \href {http://arxiv.org/abs/0704.0020} {\path{arXiv:0704.0020}},
  \href {http://dx.doi.org/10.1103/PhysRevD.76.052005}
  {\path{doi:10.1103/PhysRevD.76.052005}}.

\bibitem{Widhalm2006}
L.~Widhalm, et~al., {Measurement of $D^0 \to \pi \ell \nu (K\ell \nu)$ form
  factors and absolute branching fractions}, Phys.Rev.Lett. 97 (2006) 061804.
\newblock \href {http://arxiv.org/abs/hep-ex/0604049}
  {\path{arXiv:hep-ex/0604049}}, \href
  {http://dx.doi.org/10.1103/PhysRevLett.97.061804}
  {\path{doi:10.1103/PhysRevLett.97.061804}}.

\bibitem{Ablikim2015}
M.~Ablikim, et~al., {Study of decay dynamics and $CP$ asymmetry in $D^+ \to
  K^0_L e^+ \nu_e$ decay}, Phys. Rev. D92~(11) (2015) 112008.
\newblock \href {http://arxiv.org/abs/1510.00308} {\path{arXiv:1510.00308}},
  \href {http://dx.doi.org/10.1103/PhysRevD.92.112008}
  {\path{doi:10.1103/PhysRevD.92.112008}}.

\bibitem{Dobbs2008}
S.~Dobbs, et~al., {A study of the semileptonic charm decays $D^0 \to\pi^- e^+
  \nu_e$, $D^+ \to\pi^0 e^+ \nu_e$, $D^0 \to K^- e^+ \nu_e$, and $D^+ \to \bar
  K^0 e^+ \nu_e$}, Phys.Rev. D77 (2008) 112005.
\newblock \href {http://arxiv.org/abs/0712.1020} {\path{arXiv:0712.1020}},
  \href {http://dx.doi.org/10.1103/PhysRevD.77.112005}
  {\path{doi:10.1103/PhysRevD.77.112005}}.

\bibitem{Besson2009}
D.~Besson, et~al., {Improved measurements of $D$ meson semileptonic decays to
  $\pi$ and $K$ mesons}, Phys.Rev. D80 (2009) 032005.
\newblock \href {http://arxiv.org/abs/0906.2983} {\path{arXiv:0906.2983}},
  \href {http://dx.doi.org/10.1103/PhysRevD.80.032005}
  {\path{doi:10.1103/PhysRevD.80.032005}}.

\bibitem{Ge2009}
J.~Ge, et~al., {Study of $D^0 \to \pi^- e^+ \nu_e$, $D^+ \to \pi^0 e^+ \nu_e$ ,
  $D^0 \to K^- e^+ \nu_e$, and $D^+ \to \bar K^0 e^+ \nu_e$ in tagged decays of
  the $\psi(3770)$ resonance}, Phys.Rev. D79 (2009) 052010.
\newblock \href {http://arxiv.org/abs/0810.3878} {\path{arXiv:0810.3878}},
  \href {http://dx.doi.org/10.1103/PhysRevD.79.052010}
  {\path{doi:10.1103/PhysRevD.79.052010}}.

\bibitem{Atwood1990}
D.~Atwood, W.~J. Marciano, {Radiative corrections and semileptonic $B$ Decays},
  Phys. Rev. D41 (1990) 1736.
\newblock \href {http://dx.doi.org/10.1103/PhysRevD.41.1736}
  {\path{doi:10.1103/PhysRevD.41.1736}}.

\bibitem{Na2010}
H.~Na, C.~T. Davies, E.~Follana, G.~P. Lepage, J.~Shigemitsu, {The $D
  \rightarrow K \ell \nu$ semileptonic decay scalar form factor and $|V_{cs}|$
  from lattice QCD}, Phys. Rev. D82 (2010) 114506.
\newblock \href {http://dx.doi.org/10.1103/PhysRevD.82.114506}
  {\path{doi:10.1103/PhysRevD.82.114506}}.

\bibitem{Koponen2013}
J.~Koponen, C.~Davies, G.~Donald, E.~Follana, G.~Lepage, et~al., {The shape of
  the $D \to K$ semileptonic form factor from full lattice QCD and $V_{cs}$}
  (2013).
\newblock \href {http://arxiv.org/abs/1305.1462} {\path{arXiv:1305.1462}}.

\bibitem{Donald2014}
G.~Donald, C.~Davies, J.~Koponen, G.~Lepage, {$V_{cs}$ from $D_s \to \phi \ell
  \nu$ semileptonic decay and full lattice QCD}, Phys. Rev. D90~(7) (2014)
  074506.
\newblock \href {http://arxiv.org/abs/1311.6669} {\path{arXiv:1311.6669}},
  \href {http://dx.doi.org/10.1103/PhysRevD.90.074506}
  {\path{doi:10.1103/PhysRevD.90.074506}}.

\bibitem{Aubert2008}
B.~Aubert, et~al., {Study of the decay $D^+_{s} \to K^{+} K^{-} e^{+}
  \nu_{e}$}, Phys.Rev. D78 (2008) 051101.
\newblock \href {http://arxiv.org/abs/0807.1599} {\path{arXiv:0807.1599}},
  \href {http://dx.doi.org/10.1103/PhysRevD.78.051101}
  {\path{doi:10.1103/PhysRevD.78.051101}}.

\bibitem{Gilman2004}
F.~Gilman, K.~Kleinknecht, B.~Renk, {The Cabibbo-Kobayashi-Maskawa quark-mixing
  matrix}, Phys.\ Lett. B592 (2004) 130--135.
\newblock \href {http://dx.doi.org/10.1016/j.physletb.2004.06.001}
  {\path{doi:10.1016/j.physletb.2004.06.001}}.

\bibitem{Abramowicz1982}
H.~Abramowicz, et~al., {Experimental study of opposite sign dimuons produced in
  neutrino and anti-neutrinos interactions}, Z. Phys. C15 (1982) 19.
\newblock \href {http://dx.doi.org/10.1007/BF01573422}
  {\path{doi:10.1007/BF01573422}}.

\bibitem{Rabinowitz1993}
S.~A. Rabinowitz, et~al., {Measurement of the strange sea distribution using
  neutrino charm production}, Phys. Rev. Lett. 70 (1993) 134--137.
\newblock \href {http://dx.doi.org/10.1103/PhysRevLett.70.134}
  {\path{doi:10.1103/PhysRevLett.70.134}}.

\bibitem{Bazarko1995}
A.~O. Bazarko, et~al., {Determination of the strange quark content of the
  nucleon from a next-to-leading order QCD analysis of neutrino charm
  production}, Z. Phys. C65 (1995) 189--198.
\newblock \href {http://arxiv.org/abs/hep-ex/9406007}
  {\path{arXiv:hep-ex/9406007}}, \href {http://dx.doi.org/10.1007/BF01571875}
  {\path{doi:10.1007/BF01571875}}.

\bibitem{Vilain1999}
P.~Vilain, et~al., {Leading order QCD analysis of neutrino induced dimuon
  events}, Eur. Phys. J. C11 (1999) 19--34.
\newblock \href {http://dx.doi.org/10.1007/s100520050611}
  {\path{doi:10.1007/s100520050611}}.

\bibitem{Lellis2004}
G.~D. Lellis, P.~Migliozzi, P.~Santorelli, {Charm physics with neutrinos},
  Phys. Rept. 399 (2004) 227--320.
\newblock \href {http://dx.doi.org/10.1016/j.physrep.2005.02.001}
  {\path{doi:10.1016/j.physrep.2005.02.001}}.

\bibitem{KayisTopaksu2005}
A.~Kayis-Topaksu, et~al., {Measurement of topological muonic branching ratios
  of charmed hadrons produced in neutrino-induced charged-current
  interactions}, Phys.\ Lett. B626 (2005) 24--34.
\newblock \href {http://dx.doi.org/10.1016/j.physletb.2005.08.082}
  {\path{doi:10.1016/j.physletb.2005.08.082}}.

\bibitem{KayisTopaksu2008}
A.~Kayis-Topaksu, et~al., {Leading order analysis of neutrino induced dimuon
  events in the CHORUS experiment}, Nucl.\ Phys. B798 (2008) 1--16.
\newblock \href {http://dx.doi.org/10.1016/j.nuclphysb.2008.02.013}
  {\path{doi:10.1016/j.nuclphysb.2008.02.013}}.

\bibitem{Eisenstein2008}
B.~Eisenstein, et~al., {Precision measurement of $B(D^+ \to \mu^+ \nu)$ and the
  pseudoscalar decay constant $f_{D^+}$}, Phys. Rev. D78 (2008) 052003.
\newblock \href {http://dx.doi.org/10.1103/PhysRevD.78.052003}
  {\path{doi:10.1103/PhysRevD.78.052003}}.

\bibitem{Ablikim2014}
M.~Ablikim, et~al., {Precision measurements of $B(D^+ \rightarrow \mu^+
  \nu_{\mu})$, the pseudoscalar decay constant $f_{D^+}$, and the quark mixing
  matrix element $|V_{cd}|$}, Phys. Rev. D89~(5) (2014) 051104.
\newblock \href {http://dx.doi.org/10.1103/PhysRevD.89.051104}
  {\path{doi:10.1103/PhysRevD.89.051104}}.

\bibitem{Lees2015}
J.~Lees, et~al., {Measurement of the $D^0 \to \pi^- e^+ \nu_e$ differential
  decay branching fraction as a function of $q^2$ and study of form factor
  parameterizations}, Phys.Rev. D91~(5) (2015) 052022.
\newblock \href {http://arxiv.org/abs/1412.5502} {\path{arXiv:1412.5502}},
  \href {http://dx.doi.org/10.1103/PhysRevD.91.052022}
  {\path{doi:10.1103/PhysRevD.91.052022}}.

\bibitem{Na2011}
H.~Na, C.~T. Davies, E.~Follana, J.~Koponen, G.~P. Lepage, et~al., {$D
  \rightarrow \pi\ell \nu$ semileptonic decays, $|V_{cd}|$ and 2$^{nd}$ row
  unitarity from lattice QCD}, Phys.Rev. D84 (2011) 114505.
\newblock \href {http://arxiv.org/abs/1109.1501} {\path{arXiv:1109.1501}},
  \href {http://dx.doi.org/10.1103/PhysRevD.84.114505}
  {\path{doi:10.1103/PhysRevD.84.114505}}.

\bibitem{Caprini1998}
I.~Caprini, L.~Lellouch, M.~Neubert, {Dispersive bounds on the shape of $\bar B
  \to D^{(*)}$ lepton anti-neutrino form-factors}, Nucl. Phys. B530 (1998)
  153--181.
\newblock \href {http://arxiv.org/abs/hep-ph/9712417}
  {\path{arXiv:hep-ph/9712417}}, \href
  {http://dx.doi.org/10.1016/S0550-3213(98)00350-2}
  {\path{doi:10.1016/S0550-3213(98)00350-2}}.

\bibitem{Neubert1994}
M.~Neubert, {Theoretical update on the model independent determination of
  $|V_{cb}|$ using heavy quark symmetry}, Phys. Lett. B338 (1994) 84--91.
\newblock \href {http://arxiv.org/abs/hep-ph/9408290}
  {\path{arXiv:hep-ph/9408290}}, \href
  {http://dx.doi.org/10.1016/0370-2693(94)91348-X}
  {\path{doi:10.1016/0370-2693(94)91348-X}}.

\bibitem{Bailey2014}
J.~A. Bailey, et~al., {Update of $|V_{cb}|$ from the $\bar{B}\to
  D^*\ell\bar{\nu}$ form factor at zero recoil with three-flavor lattice QCD},
  Phys. Rev. D89~(11) (2014) 114504.
\newblock \href {http://arxiv.org/abs/1403.0635} {\path{arXiv:1403.0635}},
  \href {http://dx.doi.org/10.1103/PhysRevD.89.114504}
  {\path{doi:10.1103/PhysRevD.89.114504}}.

\bibitem{Sirlin1982}
A.~Sirlin, {Large $m_W$, $m_Z$ behavior of the $O(\alpha)$ corrections to
  semileptonic processes mediated by $W$}, Nucl.Phys. B196 (1982) 83.
\newblock \href {http://dx.doi.org/10.1016/0550-3213(82)90303-0}
  {\path{doi:10.1016/0550-3213(82)90303-0}}.

\bibitem{Luke1990}
M.~E. Luke, {Effects of subleading operators in the heavy quark effective
  theory}, Phys. Lett. B252 (1990) 447--455.
\newblock \href {http://dx.doi.org/10.1016/0370-2693(90)90568-Q}
  {\path{doi:10.1016/0370-2693(90)90568-Q}}.

\bibitem{Boyd1995}
C.~G. Boyd, B.~Grinstein, R.~F. Lebed, {Model independent extraction of
  $|V_{cb}|$ using dispersion relations}, Phys. Lett. B353 (1995) 306--312.
\newblock \href {http://dx.doi.org/10.1016/0370-2693(95)00480-9}
  {\path{doi:10.1016/0370-2693(95)00480-9}}.

\bibitem{Boyd1996}
C.~G. Boyd, B.~Grinstein, R.~F. Lebed, {Model independent determinations of
  $\bar B \to D \ell\bar\nu$, $D^* \ell \bar \nu$ form-factors}, Nucl. Phys.
  B461 (1996) 493--511.
\newblock \href {http://dx.doi.org/10.1016/0550-3213(95)00653-2}
  {\path{doi:10.1016/0550-3213(95)00653-2}}.

\bibitem{Boyd1997}
C.~G. Boyd, B.~Grinstein, R.~F. Lebed, {Precision corrections to dispersive
  bounds on form-factors}, Phys. Rev. D56 (1997) 6895--6911.
\newblock \href {http://dx.doi.org/10.1103/PhysRevD.56.6895}
  {\path{doi:10.1103/PhysRevD.56.6895}}.

\bibitem{Gambino2016}
P.~Gambino, {$|V_{cb}|$ and $|V_{ub}|$ from semileptonic $B$ decays},
  \url{https://indico.cern.ch/event/352928/contributions/1757312/attachments/1267165/1876272/Beauty16.pdf},
  {Beauty 2016: Interntational Conference on $B$-physics at Frontier Machines,
  Marseille, 2-6 May 2016; [Online; accessed 20-May-2016]} (2016).

\bibitem{Glattauer2016}
R.~Glattauer, et~al., {Measurement of the decay $B\to D\ell\nu_\ell$ in fully
  reconstructed events and determination of the Cabibbo-Kobayashi-Maskawa
  matrix element $|V_{cb}|$}, Phys. Rev. D93~(3) (2016) 032006.
\newblock \href {http://arxiv.org/abs/1510.03657} {\path{arXiv:1510.03657}},
  \href {http://dx.doi.org/10.1103/PhysRevD.93.032006}
  {\path{doi:10.1103/PhysRevD.93.032006}}.

\bibitem{Aubert2010}
B.~Aubert, et~al., {Measurement of $|V_{cb}|$ and the form-factor slope in
  $\bar B \to D \ell^- \bar \nu$ decays in events tagged by a fully
  reconstructed $B$ meson}, Phys. Rev. Lett. 104 (2010) 011802.
\newblock \href {http://arxiv.org/abs/0904.4063} {\path{arXiv:0904.4063}},
  \href {http://dx.doi.org/10.1103/PhysRevLett.104.011802}
  {\path{doi:10.1103/PhysRevLett.104.011802}}.

\bibitem{Bailey2015a}
J.~A. Bailey, et~al., {$B\to D\ell\nu$ form factors at nonzero recoil and
  $|V_{cb}|$ from 2+1-flavor lattice QCD}, Phys. Rev. D92~(3) (2015) 034506.
\newblock \href {http://arxiv.org/abs/1503.07237} {\path{arXiv:1503.07237}},
  \href {http://dx.doi.org/10.1103/PhysRevD.92.034506}
  {\path{doi:10.1103/PhysRevD.92.034506}}.

\bibitem{Buskulic1997}
D.~Buskulic, et~al., {Measurements of $|V_{cb}|$, form-factors and branching
  fractions in the decays $\bar B^0 \to D^{*+} \ell^- \bar\nu_\ell$ and $\bar
  B^0 \to D^+ \ell^- \bar\nu_\ell$}, Phys. Lett. B395 (1997) 373--387.
\newblock \href {http://dx.doi.org/10.1016/S0370-2693(97)00071-3}
  {\path{doi:10.1016/S0370-2693(97)00071-3}}.

\bibitem{Aubert2008c}
B.~Aubert, et~al., {Measurement of the Decay $B^{-} \to D^{*0} e^{-}
  \bar\nu_e$}, Phys. Rev. Lett. 100 (2008) 231803.
\newblock \href {http://arxiv.org/abs/0712.3493} {\path{arXiv:0712.3493}},
  \href {http://dx.doi.org/10.1103/PhysRevLett.100.231803}
  {\path{doi:10.1103/PhysRevLett.100.231803}}.

\bibitem{Aubert2008d}
B.~Aubert, et~al., {Determination of the form-factors for the decay $B^0 \to
  D^{*-} \ell^{+} \nu_{l}$ and of the CKM matrix element $|V_{cb}|$}, Phys.
  Rev. D77 (2008) 032002.
\newblock \href {http://arxiv.org/abs/0705.4008} {\path{arXiv:0705.4008}},
  \href {http://dx.doi.org/10.1103/PhysRevD.77.032002}
  {\path{doi:10.1103/PhysRevD.77.032002}}.

\bibitem{Aubert2009d}
B.~Aubert, et~al., {Measurements of the semileptonic decays $\bar B \to D \ell
  \bar\nu$ and $\bar B \to D^* \ell \bar\nu$ using a global fit to $D X \ell
  \bar\nu$ final states}, Phys. Rev. D79 (2009) 012002.
\newblock \href {http://arxiv.org/abs/0809.0828} {\path{arXiv:0809.0828}},
  \href {http://dx.doi.org/10.1103/PhysRevD.79.012002}
  {\path{doi:10.1103/PhysRevD.79.012002}}.

\bibitem{Dungel2010}
W.~Dungel, et~al., {Measurement of the form factors of the decay $B^0 \to
  D^{*-} \ell^+ \nu$ and determination of the CKM matrix element $|V_{cb}|$},
  Phys. Rev. D82 (2010) 112007.
\newblock \href {http://arxiv.org/abs/1010.5620} {\path{arXiv:1010.5620}},
  \href {http://dx.doi.org/10.1103/PhysRevD.82.112007}
  {\path{doi:10.1103/PhysRevD.82.112007}}.

\bibitem{Adam2003}
N.~E. Adam, et~al., {Determination of the $\bar B \to D^* \ell \bar\nu$ decay
  width and $|V_{cb}|$}, Phys. Rev. D67 (2003) 032001.
\newblock \href {http://arxiv.org/abs/hep-ex/0210040}
  {\path{arXiv:hep-ex/0210040}}, \href
  {http://dx.doi.org/10.1103/PhysRevD.67.032001}
  {\path{doi:10.1103/PhysRevD.67.032001}}.

\bibitem{Abreu2001}
P.~Abreu, et~al., {Measurement of $V_{cb}$ from the decay process $\bar B^0 \to
  D^{*+} \ell^- \bar\nu$}, Phys. Lett. B510 (2001) 55--74.
\newblock \href {http://arxiv.org/abs/hep-ex/0104026}
  {\path{arXiv:hep-ex/0104026}}, \href
  {http://dx.doi.org/10.1016/S0370-2693(01)00569-X}
  {\path{doi:10.1016/S0370-2693(01)00569-X}}.

\bibitem{Abdallah2004}
J.~Abdallah, et~al., {Measurement of $|V_{cb}|$ using the semileptonic decay
  $\bar B^0_d \to D^{*+} \ell^- \bar\nu_\ell$}, Eur. Phys. J. C33 (2004)
  213--232.
\newblock \href {http://arxiv.org/abs/hep-ex/0401023}
  {\path{arXiv:hep-ex/0401023}}, \href
  {http://dx.doi.org/10.1140/epjc/s2004-01598-6}
  {\path{doi:10.1140/epjc/s2004-01598-6}}.

\bibitem{Abbiendi2000}
G.~Abbiendi, et~al., {Measurement of $|V_{cb}|$ using $\bar B^0 \to D^{*+}
  \ell^- \bar\nu$ decays}, Phys. Lett. B482 (2000) 15--30.
\newblock \href {http://arxiv.org/abs/hep-ex/0003013}
  {\path{arXiv:hep-ex/0003013}}, \href
  {http://dx.doi.org/10.1016/S0370-2693(00)00457-3}
  {\path{doi:10.1016/S0370-2693(00)00457-3}}.

\bibitem{Alberti2015}
A.~Alberti, P.~Gambino, K.~J. Healey, S.~Nandi, {Precision determination of the
  Cabibbo-Kobayashi-Maskawa element $V_{cb}$}, Phys. Rev. Lett. 114~(6) (2015)
  061802.
\newblock \href {http://arxiv.org/abs/1411.6560} {\path{arXiv:1411.6560}},
  \href {http://dx.doi.org/10.1103/PhysRevLett.114.061802}
  {\path{doi:10.1103/PhysRevLett.114.061802}}.

\bibitem{Gambino2014}
P.~Gambino, C.~Schwanda, {Inclusive semileptonic fits, heavy quark masses, and
  $V_{cb}$}, Phys. Rev. D89~(1) (2014) 014022.
\newblock \href {http://arxiv.org/abs/1307.4551} {\path{arXiv:1307.4551}},
  \href {http://dx.doi.org/10.1103/PhysRevD.89.014022}
  {\path{doi:10.1103/PhysRevD.89.014022}}.

\bibitem{Aubert2010a}
B.~Aubert, et~al., {Measurement and interpretation of moments in inclusive
  semileptonic decays $\bar B \to X_c \ell^- \bar\nu$}, Phys. Rev. D81 (2010)
  032003.
\newblock \href {http://arxiv.org/abs/0908.0415} {\path{arXiv:0908.0415}},
  \href {http://dx.doi.org/10.1103/PhysRevD.81.032003}
  {\path{doi:10.1103/PhysRevD.81.032003}}.

\bibitem{Aubert2004}
B.~Aubert, et~al., {Measurement of the electron energy spectrum and its moments
  in inclusive $B \to X e \nu$ decays}, Phys. Rev. D69 (2004) 111104.
\newblock \href {http://arxiv.org/abs/hep-ex/0403030}
  {\path{arXiv:hep-ex/0403030}}, \href
  {http://dx.doi.org/10.1103/PhysRevD.69.111104}
  {\path{doi:10.1103/PhysRevD.69.111104}}.

\bibitem{Schwanda2007}
C.~Schwanda, et~al., {Moments of the hadronic invariant mass spectrum in $B \to
  X_c \ell \nu$ decays at {Belle}}, Phys. Rev. D75 (2007) 032005.
\newblock \href {http://arxiv.org/abs/hep-ex/0611044}
  {\path{arXiv:hep-ex/0611044}}, \href
  {http://dx.doi.org/10.1103/PhysRevD.75.032005}
  {\path{doi:10.1103/PhysRevD.75.032005}}.

\bibitem{Urquijo2007}
P.~Urquijo, et~al., {Moments of the electron energy spectrum and partial
  branching fraction of $B \to X_c e \nu$ decays at Belle}, Phys. Rev. D75
  (2007) 032001.
\newblock \href {http://arxiv.org/abs/hep-ex/0610012}
  {\path{arXiv:hep-ex/0610012}}, \href
  {http://dx.doi.org/10.1103/PhysRevD.75.032001}
  {\path{doi:10.1103/PhysRevD.75.032001}}.

\bibitem{Acosta2005}
D.~Acosta, et~al., {Measurement of the moments of the hadronic invariant mass
  distribution in semileptonic $B$ decays}, Phys. Rev. D71 (2005) 051103.
\newblock \href {http://arxiv.org/abs/hep-ex/0502003}
  {\path{arXiv:hep-ex/0502003}}, \href
  {http://dx.doi.org/10.1103/PhysRevD.71.051103}
  {\path{doi:10.1103/PhysRevD.71.051103}}.

\bibitem{Csorna2004}
S.~E. Csorna, et~al., {Moments of the $B$ meson inclusive semileptonic decay
  rate using neutrino reconstruction}, Phys. Rev. D70 (2004) 032002.
\newblock \href {http://arxiv.org/abs/hep-ex/0403052}
  {\path{arXiv:hep-ex/0403052}}, \href
  {http://dx.doi.org/10.1103/PhysRevD.70.032002}
  {\path{doi:10.1103/PhysRevD.70.032002}}.

\bibitem{Abdallah2005}
J.~Abdallah, et~al., {Determination of heavy quark non-perturbative parameters
  from spectral moments in semileptonic $B$ decays}, Eur. Phys. J. C45 (2006)
  35--59.
\newblock \href {http://arxiv.org/abs/hep-ex/0510024}
  {\path{arXiv:hep-ex/0510024}}, \href
  {http://dx.doi.org/10.1140/epjc/s2005-02406-7}
  {\path{doi:10.1140/epjc/s2005-02406-7}}.

\bibitem{Ricciardi2014}
G.~Ricciardi,
  \href{https://inspirehep.net/record/1334372/files/arXiv:1412.4288.pdf}{{Status
  of $|V_{cb}|$ and $|V_{ub}|$ CKM matrix elements}}, in: {11th Conference on
  Quark Confinement and the Hadron Spectrum (Confinement XI) St. Petersburg,
  Russia, September 8-12, 2014}, 2014.
\newblock \href {http://arxiv.org/abs/1412.4288} {\path{arXiv:1412.4288}}.
\newline\urlprefix\url{https://inspirehep.net/record/1334372/files/arXiv:1412.4288.pdf}

\bibitem{Crivellin2015}
A.~Crivellin, S.~Pokorski, {Can the differences in the determinations of
  $V_{ub}$ and $V_{cb}$ be explained by new physics?}, Phys. Rev. Lett. 114~(1)
  (2015) 011802.
\newblock \href {http://arxiv.org/abs/1407.1320} {\path{arXiv:1407.1320}},
  \href {http://dx.doi.org/10.1103/PhysRevLett.114.011802}
  {\path{doi:10.1103/PhysRevLett.114.011802}}.

\bibitem{Lees2012c}
J.~P. Lees, et~al., {Evidence for an excess of $\bar{B} \to D^{(*)}
  \tau^-\bar{\nu}_\tau$ decays}, Phys. Rev. Lett. 109 (2012) 101802.
\newblock \href {http://arxiv.org/abs/1205.5442} {\path{arXiv:1205.5442}},
  \href {http://dx.doi.org/10.1103/PhysRevLett.109.101802}
  {\path{doi:10.1103/PhysRevLett.109.101802}}.

\bibitem{Huschle2015}
M.~Huschle, et~al., {Measurement of the branching ratio of $\bar{B} \to
  D^{(\ast)} \tau^- \bar{\nu}_\tau$ relative to $\bar{B} \to D^{(\ast)} \ell^-
  \bar{\nu}_\ell$ decays with hadronic tagging at Belle}, Phys. Rev. D92~(7)
  (2015) 072014.
\newblock \href {http://arxiv.org/abs/1507.03233} {\path{arXiv:1507.03233}},
  \href {http://dx.doi.org/10.1103/PhysRevD.92.072014}
  {\path{doi:10.1103/PhysRevD.92.072014}}.

\bibitem{Abdesselam2016}
A.~Abdesselam, et~al., {Measurement of the branching ratio of $\bar{B}^0
  \rightarrow D^{*+} \tau^- \bar{\nu}_{\tau}$ relative to $\bar{B}^0
  \rightarrow D^{*+} \ell^- \bar{\nu}_{\ell}$ decays with a semileptonic
  tagging method} (2016).
\newblock \href {http://arxiv.org/abs/1603.06711} {\path{arXiv:1603.06711}}.

\bibitem{Aaij2015i}
R.~Aaij, et~al., {Measurement of the ratio of branching fractions
  $\mathcal{B}(\bar{B}^0 \to
  D^{*+}\tau^{-}\bar{\nu}_{\tau})/\mathcal{B}(\bar{B}^0 \to
  D^{*+}\mu^{-}\bar{\nu}_{\mu})$}, Phys. Rev. Lett. 115~(11) (2015) 111803,
  [Addendum: Phys. Rev. Lett.115,no.15,159901(2015)].
\newblock \href {http://arxiv.org/abs/1506.08614} {\path{arXiv:1506.08614}},
  \href {http://dx.doi.org/10.1103/PhysRevLett.115.159901,
  10.1103/PhysRevLett.115.111803} {\path{doi:10.1103/PhysRevLett.115.159901,
  10.1103/PhysRevLett.115.111803}}.

\bibitem{Aaij2015d}
R.~Aaij, et~al., {Determination of the quark coupling strength $|V_{ub}|$ using
  baryonic decays}, Nature Phys. 11.
\newblock \href {http://arxiv.org/abs/1504.01568} {\path{arXiv:1504.01568}},
  \href {http://dx.doi.org/10.1038/nphys3415} {\path{doi:10.1038/nphys3415}}.

\bibitem{Detmold2015}
W.~Detmold, C.~Lehner, S.~Meinel, {$\Lambda_b \to p \ell^- \bar{\nu}_\ell$ and
  $\Lambda_b \to \Lambda_c \ell^- \bar{\nu}_\ell$ form factors from lattice QCD
  with relativistic heavy quarks}, Phys. Rev. D92~(3) (2015) 034503.
\newblock \href {http://arxiv.org/abs/1503.01421} {\path{arXiv:1503.01421}},
  \href {http://dx.doi.org/10.1103/PhysRevD.92.034503}
  {\path{doi:10.1103/PhysRevD.92.034503}}.

\bibitem{Bailey2015}
J.~A. Bailey, et~al., {$|V_{ub}|$ from $B\to\pi\ell\nu$ decays and (2+1)-flavor
  lattice QCD}, Phys. Rev. D92~(1) (2015) 014024.
\newblock \href {http://arxiv.org/abs/1503.07839} {\path{arXiv:1503.07839}},
  \href {http://dx.doi.org/10.1103/PhysRevD.92.014024}
  {\path{doi:10.1103/PhysRevD.92.014024}}.

\bibitem{Flynn2015}
J.~M. Flynn, T.~Izubuchi, T.~Kawanai, C.~Lehner, A.~Soni, R.~S. Van~de Water,
  O.~Witzel, {$B \to \pi \ell \nu$ and $B_s \to K \ell \nu$ form factors and
  $|V_{ub}|$ from 2+1-flavor lattice QCD with domain-wall light quarks and
  relativistic heavy quarks}, Phys. Rev. D91~(7) (2015) 074510.
\newblock \href {http://arxiv.org/abs/1501.05373} {\path{arXiv:1501.05373}},
  \href {http://dx.doi.org/10.1103/PhysRevD.91.074510}
  {\path{doi:10.1103/PhysRevD.91.074510}}.

\bibitem{delAmoSanchez2011}
P.~del Amo~Sanchez, et~al., {Study of $B \to \pi \ell \nu$ and $B \to \rho \ell
  \nu$ decays and determination of $|V_{ub}|$}, Phys. Rev. D83 (2011) 032007.
\newblock \href {http://arxiv.org/abs/1005.3288} {\path{arXiv:1005.3288}},
  \href {http://dx.doi.org/10.1103/PhysRevD.83.032007}
  {\path{doi:10.1103/PhysRevD.83.032007}}.

\bibitem{Lees2012}
J.~P. Lees, et~al., {Branching fraction and form-factor shape measurements of
  exclusive charmless semileptonic $B$ decays, and determination of
  $|V_{ub}|$}, Phys. Rev. D86 (2012) 092004.
\newblock \href {http://arxiv.org/abs/1208.1253} {\path{arXiv:1208.1253}},
  \href {http://dx.doi.org/10.1103/PhysRevD.86.092004}
  {\path{doi:10.1103/PhysRevD.86.092004}}.

\bibitem{Ha2011}
H.~Ha, et~al., {Measurement of the decay $B^0\to\pi^-\ell^+\nu$ and
  determination of $|V_{ub}|$}, Phys. Rev. D83 (2011) 071101.
\newblock \href {http://arxiv.org/abs/1012.0090} {\path{arXiv:1012.0090}},
  \href {http://dx.doi.org/10.1103/PhysRevD.83.071101}
  {\path{doi:10.1103/PhysRevD.83.071101}}.

\bibitem{Sibidanov2013}
A.~Sibidanov, et~al., {Study of exclusive $B \to X_u \ell \nu$ decays and
  extraction of $|V_{ub}|$ using full reconstruction tagging at the Belle
  experiment}, Phys. Rev. D88~(3) (2013) 032005.
\newblock \href {http://arxiv.org/abs/1306.2781} {\path{arXiv:1306.2781}},
  \href {http://dx.doi.org/10.1103/PhysRevD.88.032005}
  {\path{doi:10.1103/PhysRevD.88.032005}}.

\bibitem{Lees2013}
J.~P. Lees, et~al., {Measurement of the $B^+ \to \omega \ell^+ \nu$ branching
  fraction with semileptonically tagged $B$ mesons}, Phys. Rev. D88~(7) (2013)
  072006.
\newblock \href {http://arxiv.org/abs/1308.2589} {\path{arXiv:1308.2589}},
  \href {http://dx.doi.org/10.1103/PhysRevD.88.072006}
  {\path{doi:10.1103/PhysRevD.88.072006}}.

\bibitem{Kang2014}
X.-W. Kang, B.~Kubis, C.~Hanhart, U.-G. Meißner, {$B_{\ell4}$ decays and the
  extraction of $|V_{ub}|$}, Phys. Rev. D89 (2014) 053015.
\newblock \href {http://arxiv.org/abs/1312.1193} {\path{arXiv:1312.1193}},
  \href {http://dx.doi.org/10.1103/PhysRevD.89.053015}
  {\path{doi:10.1103/PhysRevD.89.053015}}.

\bibitem{Hsiao2016}
Y.~K. Hsiao, C.~Q. Geng, {Determination of $|V_{ub}|$ from exclusive baryonic
  $B$ decays}, Phys. Lett. B755 (2016) 418--420.
\newblock \href {http://arxiv.org/abs/1510.04414} {\path{arXiv:1510.04414}},
  \href {http://dx.doi.org/10.1016/j.physletb.2016.02.051}
  {\path{doi:10.1016/j.physletb.2016.02.051}}.

\bibitem{Aubert2010b}
B.~Aubert, et~al., {A search for $B^+ \to \ell^+ \nu_{\ell}$ recoiling against
  $B^{-}\to D^{0} \ell^{-}\bar{\nu} X$}, Phys. Rev. D81 (2010) 051101.
\newblock \href {http://arxiv.org/abs/0912.2453} {\path{arXiv:0912.2453}},
  \href {http://dx.doi.org/10.1103/PhysRevD.81.051101}
  {\path{doi:10.1103/PhysRevD.81.051101}}.

\bibitem{Lees2013a}
J.~P. Lees, et~al., {Evidence of $B^+ \to \tau^+\nu$ decays with hadronic $B$
  tags}, Phys. Rev. D88~(3) (2013) 031102.
\newblock \href {http://arxiv.org/abs/1207.0698} {\path{arXiv:1207.0698}},
  \href {http://dx.doi.org/10.1103/PhysRevD.88.031102}
  {\path{doi:10.1103/PhysRevD.88.031102}}.

\bibitem{Hara2013}
K.~Hara, et~al., {Measurement of $B^- \to \tau^- \bar{\nu}_\tau$ with a
  hadronic tagging method using the full data sample of Belle}, Phys. Rev.
  Lett. 110~(13) (2013) 131801.
\newblock \href {http://arxiv.org/abs/1208.4678} {\path{arXiv:1208.4678}},
  \href {http://dx.doi.org/10.1103/PhysRevLett.110.131801}
  {\path{doi:10.1103/PhysRevLett.110.131801}}.

\bibitem{Kronenbitter2015}
B.~Kronenbitter, et~al., {Measurement of the branching fraction of $B^+ \to
  \tau^+ \nu_\tau$ decays with the semileptonic tagging method} (2015).
\newblock \href {http://arxiv.org/abs/1503.05613} {\path{arXiv:1503.05613}}.

\bibitem{Dowdall2013a}
R.~J. Dowdall, C.~T.~H. Davies, R.~R. Horgan, C.~J. Monahan, J.~Shigemitsu,
  {$B$-meson decay constants from improved lattice nonrelativistic QCD with
  physical $u$, $d$, $s$, and $c$ quarks}, Phys. Rev. Lett. 110~(22) (2013)
  222003.
\newblock \href {http://arxiv.org/abs/1302.2644} {\path{arXiv:1302.2644}},
  \href {http://dx.doi.org/10.1103/PhysRevLett.110.222003}
  {\path{doi:10.1103/PhysRevLett.110.222003}}.

\bibitem{Aglietti2004}
U.~Aglietti, G.~Ricciardi, {A model for next-to-leading order threshold
  resummed form-factors}, Phys. Rev. D70 (2004) 114008.
\newblock \href {http://arxiv.org/abs/hep-ph/0407225}
  {\path{arXiv:hep-ph/0407225}}, \href
  {http://dx.doi.org/10.1103/PhysRevD.70.114008}
  {\path{doi:10.1103/PhysRevD.70.114008}}.

\bibitem{Aglietti2007}
U.~Aglietti, G.~Ferrera, G.~Ricciardi, {Semi-Inclusive $B$ decays and a model
  for soft-gluon effects}, Nucl. Phys. B768 (2007) 85--115.
\newblock \href {http://arxiv.org/abs/hep-ph/0608047}
  {\path{arXiv:hep-ph/0608047}}, \href
  {http://dx.doi.org/10.1016/j.nuclphysb.2007.01.014}
  {\path{doi:10.1016/j.nuclphysb.2007.01.014}}.

\bibitem{Aglietti2009}
U.~Aglietti, F.~Di~Lodovico, G.~Ferrera, G.~Ricciardi, {Inclusive measure of
  $|V_{ub}|$ with the analytic coupling model}, Eur. Phys. J. C59 (2009)
  831--840.
\newblock \href {http://arxiv.org/abs/0711.0860} {\path{arXiv:0711.0860}},
  \href {http://dx.doi.org/10.1140/epjc/s10052-008-0817-x}
  {\path{doi:10.1140/epjc/s10052-008-0817-x}}.

\bibitem{Bauer2001}
C.~W. Bauer, Z.~Ligeti, M.~E. Luke, {Precision determination of $|V_{ub}|$ from
  inclusive decays}, Phys. Rev. D64 (2001) 113004.
\newblock \href {http://arxiv.org/abs/hep-ph/0107074}
  {\path{arXiv:hep-ph/0107074}}, \href
  {http://dx.doi.org/10.1103/PhysRevD.64.113004}
  {\path{doi:10.1103/PhysRevD.64.113004}}.

\bibitem{Bosch2004}
S.~W. Bosch, B.~O. Lange, M.~Neubert, G.~Paz, {Factorization and shape function
  effects in inclusive $B$ meson decays}, Nucl. Phys. B699 (2004) 335--386.
\newblock \href {http://arxiv.org/abs/hep-ph/0402094}
  {\path{arXiv:hep-ph/0402094}}, \href
  {http://dx.doi.org/10.1016/j.nuclphysb.2004.07.041}
  {\path{doi:10.1016/j.nuclphysb.2004.07.041}}.

\bibitem{Bosch2004a}
S.~W. Bosch, M.~Neubert, G.~Paz, {Subleading shape functions in inclusive $B$
  decays}, JHEP 11 (2004) 073.
\newblock \href {http://arxiv.org/abs/hep-ph/0409115}
  {\path{arXiv:hep-ph/0409115}}, \href
  {http://dx.doi.org/10.1088/1126-6708/2004/11/073}
  {\path{doi:10.1088/1126-6708/2004/11/073}}.

\bibitem{Lange2005}
B.~O. Lange, M.~Neubert, G.~Paz, {Theory of charmless inclusive $B$ decays and
  the extraction of $V_{ub}$}, Phys. Rev. D72 (2005) 073006.
\newblock \href {http://arxiv.org/abs/hep-ph/0504071}
  {\path{arXiv:hep-ph/0504071}}, \href
  {http://dx.doi.org/10.1103/PhysRevD.72.073006}
  {\path{doi:10.1103/PhysRevD.72.073006}}.

\bibitem{Andersen2006}
J.~R. Andersen, E.~Gardi, {Inclusive spectra in charmless semileptonic $B$
  decays by dressed gluon exponentiation}, JHEP 01 (2006) 097.
\newblock \href {http://arxiv.org/abs/hep-ph/0509360}
  {\path{arXiv:hep-ph/0509360}}, \href
  {http://dx.doi.org/10.1088/1126-6708/2006/01/097}
  {\path{doi:10.1088/1126-6708/2006/01/097}}.

\bibitem{Gambino2007}
P.~Gambino, P.~Giordano, G.~Ossola, N.~Uraltsev, {Inclusive semileptonic $B$
  decays and the determination of $|V_{ub}|$}, JHEP 10 (2007) 058.
\newblock \href {http://arxiv.org/abs/0707.2493} {\path{arXiv:0707.2493}},
  \href {http://dx.doi.org/10.1088/1126-6708/2007/10/058}
  {\path{doi:10.1088/1126-6708/2007/10/058}}.

\bibitem{Aubert2005}
B.~Aubert, et~al., {Determination of $|V_{ub}|$ from measurements of the
  electron and neutrino momenta in inclusive semileptonic $B$ decays}, Phys.
  Rev. Lett. 95 (2005) 111801, [Erratum: Phys. Rev. Lett.97,019903(2006)].
\newblock \href {http://arxiv.org/abs/hep-ex/0506036}
  {\path{arXiv:hep-ex/0506036}}, \href
  {http://dx.doi.org/10.1103/PhysRevLett.95.111801}
  {\path{doi:10.1103/PhysRevLett.95.111801}}.

\bibitem{Aubert2006}
B.~Aubert, et~al., {Measurement of the inclusive electron spectrum in charmless
  semileptonic $B$ decays near the kinematic endpoint and determination of
  $|V_{ub}|$}, Phys. Rev. D73 (2006) 012006.
\newblock \href {http://arxiv.org/abs/hep-ex/0509040}
  {\path{arXiv:hep-ex/0509040}}, \href
  {http://dx.doi.org/10.1103/PhysRevD.73.012006}
  {\path{doi:10.1103/PhysRevD.73.012006}}.

\bibitem{Kakuno2004}
H.~Kakuno, et~al., {Measurement of $|V_{ub}|$ using inclusive $B \to X_u \ell
  \nu$ decays with a novel $X_u$ reconstruction method}, Phys. Rev. Lett. 92
  (2004) 101801.
\newblock \href {http://arxiv.org/abs/hep-ex/0311048}
  {\path{arXiv:hep-ex/0311048}}, \href
  {http://dx.doi.org/10.1103/PhysRevLett.92.101801}
  {\path{doi:10.1103/PhysRevLett.92.101801}}.

\bibitem{Limosani2005}
A.~Limosani, et~al., {Measurement of inclusive charmless semileptonic $B$-meson
  decays at the endpoint of the electron momentum spectrum}, Phys. Lett. B621
  (2005) 28--40.
\newblock \href {http://arxiv.org/abs/hep-ex/0504046}
  {\path{arXiv:hep-ex/0504046}}, \href
  {http://dx.doi.org/10.1016/j.physletb.2005.06.011}
  {\path{doi:10.1016/j.physletb.2005.06.011}}.

\bibitem{Bizjak2005}
I.~Bizjak, et~al., {Determination of $|V_{ub}|$ from measurements of the
  inclusive charmless semileptonic partial rates of $B$ mesons using full
  reconstruction tags}, Phys. Rev. Lett. 95 (2005) 241801.
\newblock \href {http://arxiv.org/abs/hep-ex/0505088}
  {\path{arXiv:hep-ex/0505088}}, \href
  {http://dx.doi.org/10.1103/PhysRevLett.95.241801}
  {\path{doi:10.1103/PhysRevLett.95.241801}}.

\bibitem{Urquijo2010}
P.~Urquijo, et~al., {Measurement of $|V_{ub}|$ from inclusive charmless
  semileptonic $B$ decays}, Phys. Rev. Lett. 104 (2010) 021801.
\newblock \href {http://arxiv.org/abs/0907.0379} {\path{arXiv:0907.0379}},
  \href {http://dx.doi.org/10.1103/PhysRevLett.104.021801}
  {\path{doi:10.1103/PhysRevLett.104.021801}}.

\bibitem{Bornheim2002}
A.~Bornheim, et~al., {Improved measurement of $|V_{ub}|$ with inclusive
  semileptonic $B$ decays}, Phys. Rev. Lett. 88 (2002) 231803.
\newblock \href {http://arxiv.org/abs/hep-ex/0202019}
  {\path{arXiv:hep-ex/0202019}}, \href
  {http://dx.doi.org/10.1103/PhysRevLett.88.231803}
  {\path{doi:10.1103/PhysRevLett.88.231803}}.

\bibitem{Kidonakis2010}
N.~Kidonakis, {NNLL resummation for $s$-channel single top quark production},
  Phys. Rev. D81 (2010) 054028.
\newblock \href {http://arxiv.org/abs/1001.5034} {\path{arXiv:1001.5034}},
  \href {http://dx.doi.org/10.1103/PhysRevD.81.054028}
  {\path{doi:10.1103/PhysRevD.81.054028}}.

\bibitem{Kidonakis2010a}
N.~Kidonakis, {Two-loop soft anomalous dimensions for single top quark
  associated production with a $W^-$ or $H^-$}, Phys. Rev. D82 (2010) 054018.
\newblock \href {http://arxiv.org/abs/1005.4451} {\path{arXiv:1005.4451}},
  \href {http://dx.doi.org/10.1103/PhysRevD.82.054018}
  {\path{doi:10.1103/PhysRevD.82.054018}}.

\bibitem{Kidonakis2011}
N.~Kidonakis, {Next-to-next-to-leading-order collinear and soft gluon
  corrections for $t$-channel single top quark production}, Phys. Rev. D83
  (2011) 091503.
\newblock \href {http://arxiv.org/abs/1103.2792} {\path{arXiv:1103.2792}},
  \href {http://dx.doi.org/10.1103/PhysRevD.83.091503}
  {\path{doi:10.1103/PhysRevD.83.091503}}.

\bibitem{Martin2009}
A.~D. Martin, W.~J. Stirling, R.~S. Thorne, G.~Watt, {Parton distributions for
  the LHC}, Eur. Phys. J. C63 (2009) 189--285.
\newblock \href {http://arxiv.org/abs/0901.0002} {\path{arXiv:0901.0002}},
  \href {http://dx.doi.org/10.1140/epjc/s10052-009-1072-5}
  {\path{doi:10.1140/epjc/s10052-009-1072-5}}.

\bibitem{Aaltonen2014}
T.~A. Aaltonen, et~al., {Updated measurement of the single top quark production
  cross section and $V_{tb}$ in the missing transverse energy plus jets
  topology in $p\bar{p}$ collisions at $\sqrt{s} = 1.96$ TeV} (2014).
\newblock \href {http://arxiv.org/abs/1410.4909} {\path{arXiv:1410.4909}}.

\bibitem{Aaltonen2014a}
T.~A. Aaltonen, et~al., {Observation of $s$-channel production of single top
  quarks at the Tevatron}, Phys. Rev. Lett. 112 (2014) 231803.
\newblock \href {http://arxiv.org/abs/1402.5126} {\path{arXiv:1402.5126}},
  \href {http://dx.doi.org/10.1103/PhysRevLett.112.231803}
  {\path{doi:10.1103/PhysRevLett.112.231803}}.

\bibitem{Aaltonen2014b}
T.~A. Aaltonen, et~al., {Search for $s$-channel single-top-quark production in
  events with missing energy plus jets in $p\bar p$ collisions at $\sqrt{s} =
  1.96$ TeV}, Phys. Rev. Lett. 112~(23) (2014) 231805.
\newblock \href {http://arxiv.org/abs/1402.3756} {\path{arXiv:1402.3756}},
  \href {http://dx.doi.org/10.1103/PhysRevLett.112.231805}
  {\path{doi:10.1103/PhysRevLett.112.231805}}.

\bibitem{Aaltonen2014d}
T.~A. Aaltonen, et~al., {Evidence for $s$-channel single-top-quark production
  in events with one charged lepton and two jets at CDF}, Phys. Rev. Lett. 112
  (2014) 231804.
\newblock \href {http://arxiv.org/abs/1402.0484} {\path{arXiv:1402.0484}},
  \href {http://dx.doi.org/10.1103/PhysRevLett.112.231804}
  {\path{doi:10.1103/PhysRevLett.112.231804}}.

\bibitem{Abazov2013}
V.~M. Abazov, et~al., {Evidence for $s$-channel single top quark production in
  $p\bar{p}$ collisions at $\sqrt{s}$ = 1.96 TeV}, Phys.Lett. B726 (2013)
  656--664.
\newblock \href {http://arxiv.org/abs/1307.0731} {\path{arXiv:1307.0731}},
  \href {http://dx.doi.org/10.1016/j.physletb.2013.09.048}
  {\path{doi:10.1016/j.physletb.2013.09.048}}.

\bibitem{Aaltonen2015}
T.~A. Aaltonen, et~al., {Tevatron combination of single-top-quark cross
  sections and determination of the magnitude of the Cabibbo-Kobayashi-Maskawa
  matrix element $V_{tb}$}, Phys. Rev. Lett. 115~(15) (2015) 152003.
\newblock \href {http://arxiv.org/abs/1503.05027} {\path{arXiv:1503.05027}},
  \href {http://dx.doi.org/10.1103/PhysRevLett.115.152003}
  {\path{doi:10.1103/PhysRevLett.115.152003}}.

\bibitem{Aad2015}
G.~Aad, et~al., {Search for $s$-channel single top-quark production in
  proton-proton collisions at $\sqrt s=8$ TeV with the ATLAS detector}, Phys.
  Lett. B740 (2015) 118--136.
\newblock \href {http://arxiv.org/abs/1410.0647} {\path{arXiv:1410.0647}},
  \href {http://dx.doi.org/10.1016/j.physletb.2014.11.042}
  {\path{doi:10.1016/j.physletb.2014.11.042}}.

\bibitem{ATLASCMS2013}
{Combination of single top-quark cross-sections measurements in the $t$-channel
  at $\sqrt{s}=8$ TeV with the ATLAS and CMS experiments}, {The ATLAS and CMS
  Collaborations, CMS-PAS-TOP-12-002, ATLAS-COM-CONF-2013-061,
  ATLAS-CONF-2013-098} (2013).

\bibitem{Khachatryan2014a}
V.~Khachatryan, et~al., {Measurement of the $t$-channel single-top-quark
  production cross section and of the $|V_{tb}|$ CKM matrix element in pp
  collisions at $\sqrt{s}$= 8 TeV}, JHEP 06 (2014) 090.
\newblock \href {http://arxiv.org/abs/1403.7366} {\path{arXiv:1403.7366}},
  \href {http://dx.doi.org/10.1007/JHEP06(2014)090}
  {\path{doi:10.1007/JHEP06(2014)090}}.

\bibitem{Aad2014}
G.~Aad, et~al., {Comprehensive measurements of $t$-channel single top-quark
  production cross sections at $\sqrt{s} = 7$ TeV with the ATLAS detector},
  Phys. Rev. D90~(11) (2014) 112006.
\newblock \href {http://arxiv.org/abs/1406.7844} {\path{arXiv:1406.7844}},
  \href {http://dx.doi.org/10.1103/PhysRevD.90.112006}
  {\path{doi:10.1103/PhysRevD.90.112006}}.

\bibitem{Aad2012a}
G.~Aad, et~al., {Evidence for the associated production of a $W$ boson and a
  top quark in ATLAS at $\sqrt{s}=7$ TeV}, Phys. Lett. B716 (2012) 142--159.
\newblock \href {http://arxiv.org/abs/1205.5764} {\path{arXiv:1205.5764}},
  \href {http://dx.doi.org/10.1016/j.physletb.2012.08.011}
  {\path{doi:10.1016/j.physletb.2012.08.011}}.

\bibitem{ATLAS2014}
{Measurement of the inclusive and fiducial cross-section of single top-quark
  $t$-channel events in $pp$ collisions at $\sqrt{s}$ = 8 TeV}, {The ATLAS
  collaboration, ATLAS-CONF-2014-007, ATLAS-COM-CONF-2014-008} (2014).

\bibitem{ATLAS2013}
{Measurement of the cross-section for associated production of a top quark and
  a $W$ boson at $\sqrt{s}=8$ TeV with the ATLAS detector}, {The ATLAS
  collaboration, ATLAS-CONF-2013-100, ATLAS-COM-CONF-2013-116} (2013).

\bibitem{Chatrchyan2012}
S.~Chatrchyan, et~al., {Measurement of the single-top-quark $t$-channel cross
  section in $pp$ collisions at $\sqrt{s}=7$ TeV}, JHEP 12 (2012) 035.
\newblock \href {http://arxiv.org/abs/1209.4533} {\path{arXiv:1209.4533}},
  \href {http://dx.doi.org/10.1007/JHEP12(2012)035}
  {\path{doi:10.1007/JHEP12(2012)035}}.

\bibitem{Chatrchyan2013}
S.~Chatrchyan, et~al., {Evidence for associated production of a single top
  quark and $W$ boson in $pp$ collisions at $\sqrt{s}$ = 7 TeV}, Phys. Rev.
  Lett. 110 (2013) 022003.
\newblock \href {http://arxiv.org/abs/1209.3489} {\path{arXiv:1209.3489}},
  \href {http://dx.doi.org/10.1103/PhysRevLett.110.022003}
  {\path{doi:10.1103/PhysRevLett.110.022003}}.

\bibitem{Chatrchyan2014}
S.~Chatrchyan, et~al., {Observation of the associated production of a single
  top quark and a $W$ boson in $pp$ collisions at $\sqrt s = $8 TeV}, Phys.
  Rev. Lett. 112~(23) (2014) 231802.
\newblock \href {http://arxiv.org/abs/1401.2942} {\path{arXiv:1401.2942}},
  \href {http://dx.doi.org/10.1103/PhysRevLett.112.231802}
  {\path{doi:10.1103/PhysRevLett.112.231802}}.

\bibitem{Aaltonen2013}
T.~Aaltonen, et~al., {Measurement of $R = \mathcal{B}(t \rightarrow
  Wb)/\mathcal{B}(t \rightarrow Wq)$ in top--quark--pair decays using
  lepton+jets events and the full CDF Run II data set}, Phys.Rev. D87~(11)
  (2013) 111101.
\newblock \href {http://arxiv.org/abs/1303.6142} {\path{arXiv:1303.6142}},
  \href {http://dx.doi.org/10.1103/PhysRevD.87.111101}
  {\path{doi:10.1103/PhysRevD.87.111101}}.

\bibitem{Aaltonen2014c}
T.~A. Aaltonen, et~al., {Measurement of $B(t \to Wb)/B(t \to Wq)$ in
  top-quark-pair decays using dilepton events and the full CDF Run II data
  set}, Phys. Rev. Lett. 112~(22) (2014) 221801.
\newblock \href {http://arxiv.org/abs/1404.3392} {\path{arXiv:1404.3392}},
  \href {http://dx.doi.org/10.1103/PhysRevLett.112.221801}
  {\path{doi:10.1103/PhysRevLett.112.221801}}.

\bibitem{Abazov2011}
V.~M. Abazov, et~al., {Precision measurement of the ratio ${\rm B}(t \to
  Wb)/{\rm B}(t \to Wq)$ and extraction of $V_{tb}$}, Phys. Rev. Lett. 107
  (2011) 121802.
\newblock \href {http://arxiv.org/abs/1106.5436} {\path{arXiv:1106.5436}},
  \href {http://dx.doi.org/10.1103/PhysRevLett.107.121802}
  {\path{doi:10.1103/PhysRevLett.107.121802}}.

\bibitem{Khachatryan2014}
V.~Khachatryan, et~al., {Constraints on the Higgs boson width from off-shell
  production and decay to $Z$-boson pairs}, Phys. Lett. B736 (2014) 64.
\newblock \href {http://arxiv.org/abs/1405.3455} {\path{arXiv:1405.3455}},
  \href {http://dx.doi.org/10.1016/j.physletb.2014.06.077}
  {\path{doi:10.1016/j.physletb.2014.06.077}}.

\bibitem{Buras1998}
A.~J. Buras, R.~Fleischer, {Quark mixing, $CP$ violation and rare decays after
  the top quark discovery}, Adv. Ser. Direct. High Energy Phys. 15 (1998)
  65--238.
\newblock \href {http://arxiv.org/abs/hep-ph/9704376}
  {\path{arXiv:hep-ph/9704376}}, \href
  {http://dx.doi.org/10.1142/9789812812667_0002}
  {\path{doi:10.1142/9789812812667_0002}}.

\bibitem{Inami1981}
T.~Inami, C.~S. Lim, {Effects of Superheavy Quarks and Leptons in Low-Energy
  Weak Processes $K_L \to \mu \bar\mu$, $K^+ \to\pi^+ \nu \bar\nu$ and $K^0
  \leftrightarrow \bar K^0$}, Prog. Theor. Phys. 65 (1981) 297, [Erratum: Prog.
  Theor. Phys.65,1772(1981)].
\newblock \href {http://dx.doi.org/10.1143/PTP.65.297}
  {\path{doi:10.1143/PTP.65.297}}.

\bibitem{Buchalla1996}
G.~Buchalla, A.~J. Buras, M.~E. Lautenbacher, {Weak decays beyond leading
  logarithms}, Rev. Mod. Phys. 68 (1996) 1125--1144.
\newblock \href {http://arxiv.org/abs/hep-ph/9512380}
  {\path{arXiv:hep-ph/9512380}}, \href
  {http://dx.doi.org/10.1103/RevModPhys.68.1125}
  {\path{doi:10.1103/RevModPhys.68.1125}}.

\bibitem{Charles2005}
J.~Charles, et~al., {$CP$} violation and the {CKM} matrix: assessing the impact
  of the asymmetric {$B$} factories, Eur. Phys. J. C 41 (2005) 1--131,
  ({C}KMfitter group).
\newblock \href {http://dx.doi.org/10.1140/s2005-02169-1}
  {\path{doi:10.1140/s2005-02169-1}}.

\bibitem{Alekhin2014}
S.~Alekhin, J.~Blumlein, S.~Moch, {The ABM parton distributions tuned to LHC
  data}, Phys. Rev. D89~(5) (2014) 054028.
\newblock \href {http://arxiv.org/abs/1310.3059} {\path{arXiv:1310.3059}},
  \href {http://dx.doi.org/10.1103/PhysRevD.89.054028}
  {\path{doi:10.1103/PhysRevD.89.054028}}.

\bibitem{Abulencia2006}
A.~Abulencia, et~al., {Observation of $B^0_s - \bar{B}^0_s$ Oscillations},
  Phys. Rev. Lett. 97 (2006) 242003.
\newblock \href {http://arxiv.org/abs/hep-ex/0609040}
  {\path{arXiv:hep-ex/0609040}}, \href
  {http://dx.doi.org/10.1103/PhysRevLett.97.242003}
  {\path{doi:10.1103/PhysRevLett.97.242003}}.

\bibitem{Aaij2012e}
R.~Aaij, et~al., {Measurement of the $B^0_s - \bar{B}^0_s$ oscillation
  frequency $\Delta m_s$ in $B^0_s \to D_s^-(3) \pi$ decays}, Phys. Lett. B709
  (2012) 177--184.
\newblock \href {http://arxiv.org/abs/1112.4311} {\path{arXiv:1112.4311}},
  \href {http://dx.doi.org/10.1016/j.physletb.2012.02.031}
  {\path{doi:10.1016/j.physletb.2012.02.031}}.

\bibitem{Aaij2013e}
R.~Aaij, et~al., {Precision measurement of the $B^{0}_{s}$-$\bar{B}^{0}_{s}$
  oscillation frequency with the decay $B^{0}_{s}\rightarrow
  D^{-}_{s}\pi^{+}$}, New J. Phys. 15 (2013) 053021.
\newblock \href {http://arxiv.org/abs/1304.4741} {\path{arXiv:1304.4741}},
  \href {http://dx.doi.org/10.1088/1367-2630/15/5/053021}
  {\path{doi:10.1088/1367-2630/15/5/053021}}.

\bibitem{Aaij2015c}
R.~Aaij, et~al., {Precision measurement of $CP$ violation in $B_s^0 \to J/\psi
  K^+K^-$ decays}, Phys. Rev. Lett. 114~(4) (2015) 041801.
\newblock \href {http://arxiv.org/abs/1411.3104} {\path{arXiv:1411.3104}},
  \href {http://dx.doi.org/10.1103/PhysRevLett.114.041801}
  {\path{doi:10.1103/PhysRevLett.114.041801}}.

\bibitem{Aaij2013f}
R.~Aaij, et~al., {Observation of $B^0_s$-$\bar{B}^0_s$ mixing and measurement
  of mixing frequencies using semileptonic $B$ decays}, Eur. Phys. J. C73~(12)
  (2013) 2655.
\newblock \href {http://arxiv.org/abs/1308.1302} {\path{arXiv:1308.1302}},
  \href {http://dx.doi.org/10.1140/epjc/s10052-013-2655-8}
  {\path{doi:10.1140/epjc/s10052-013-2655-8}}.

\bibitem{Gamiz2009}
E.~Gamiz, C.~T.~H. Davies, G.~P. Lepage, J.~Shigemitsu, M.~Wingate, {Neutral
  $B$ meson mixing in unquenched lattice QCD}, Phys. Rev. D80 (2009) 014503.
\newblock \href {http://arxiv.org/abs/0902.1815} {\path{arXiv:0902.1815}},
  \href {http://dx.doi.org/10.1103/PhysRevD.80.014503}
  {\path{doi:10.1103/PhysRevD.80.014503}}.

\bibitem{Bazavov2016}
A.~Bazavov, et~al., {$B^0_{s}$-mixing matrix elements from lattice QCD for the
  standard model and beyond} (2016).
\newblock \href {http://arxiv.org/abs/1602.03560} {\path{arXiv:1602.03560}}.

\bibitem{Du2016}
D.~Du, A.~X. El-Khadra, S.~Gottlieb, A.~S. Kronfeld, J.~Laiho, E.~Lunghi, R.~S.
  Van~de Water, R.~Zhou, {Phenomenology of semileptonic $B$-meson decays with
  form factors from lattice QCD}, Phys. Rev. D93 (2016) 034005, [Phys.
  Rev.D93,034005(2016)].
\newblock \href {http://arxiv.org/abs/1510.02349} {\path{arXiv:1510.02349}},
  \href {http://dx.doi.org/10.1103/PhysRevD.93.034005}
  {\path{doi:10.1103/PhysRevD.93.034005}}.

\bibitem{Aaij2014g}
R.~Aaij, et~al., {Differential branching fractions and isospin asymmetries of
  $B \to K^{(*)} \mu^+ \mu^-$ decays}, JHEP 06 (2014) 133.
\newblock \href {http://arxiv.org/abs/1403.8044} {\path{arXiv:1403.8044}},
  \href {http://dx.doi.org/10.1007/JHEP06(2014)133}
  {\path{doi:10.1007/JHEP06(2014)133}}.

\bibitem{Aaij2015g}
R.~Aaij, et~al., {First measurement of the differential branching fraction and
  $CP$ asymmetry of the $B^\pm\to\pi^\pm\mu^+\mu^-$ decay}, JHEP 10 (2015) 034.
\newblock \href {http://arxiv.org/abs/1509.00414} {\path{arXiv:1509.00414}},
  \href {http://dx.doi.org/10.1007/JHEP10(2015)034}
  {\path{doi:10.1007/JHEP10(2015)034}}.

\bibitem{Fleischer2004}
R.~Fleischer, \href{http://doc.cern.ch/yellowrep/CERN-PH-TH-2004-085}{{Flavor
  physics and $CP$ violation}}, in: {High-energy physics. Proceedings, European
  School, Tsakhkadzor, Armenia, August 24-September 6, 2003}, 2004, pp.
  81--150.
\newblock \href {http://arxiv.org/abs/hep-ph/0405091}
  {\path{arXiv:hep-ph/0405091}}.
\newline\urlprefix\url{http://doc.cern.ch/yellowrep/CERN-PH-TH-2004-085}

\bibitem{Nir2005}
Y.~Nir, {$CP$ violation in meson decays}, in: {Particle physics beyond the
  standard model. Proceedings, Summer School on Theoretical Physics, 84th
  Session, Les Houches, France, August 1-26, 2005}, 2005, pp. 79--145,
  [,79(2005)].
\newblock \href {http://arxiv.org/abs/hep-ph/0510413}
  {\path{arXiv:hep-ph/0510413}}.

\bibitem{Nierste2009}
U.~Nierste,
  \href{https://inspirehep.net/record/817820/files/arXiv:0904.1869.pdf}{{Three
  lectures on meson mixing and CKM phenomenology}}, in: {Heavy quark physics.
  Proceedings, Helmholtz International School, HQP08, Dubna, Russia, August
  11-21, 2008}, 2009, pp. 1--38.
\newblock \href {http://arxiv.org/abs/0904.1869} {\path{arXiv:0904.1869}}.
\newline\urlprefix\url{https://inspirehep.net/record/817820/files/arXiv:0904.1869.pdf}

\bibitem{Winstein1993}
B.~Winstein, L.~Wolfenstein, {The search for direct $CP$ violation}, Rev. Mod.
  Phys. 65 (1993) 1113--1148.
\newblock \href {http://dx.doi.org/10.1103/RevModPhys.65.1113}
  {\path{doi:10.1103/RevModPhys.65.1113}}.

\bibitem{Gaiser1981}
B.~D. Gaiser, T.~Tsao, M.~B. Wise, {Parameters of the six quark model}, Annals
  Phys. 132 (1981) 66.
\newblock \href {http://dx.doi.org/10.1016/0003-4916(81)90269-4}
  {\path{doi:10.1016/0003-4916(81)90269-4}}.

\bibitem{Vladikas2015}
A.~Vladikas, {FLAG: Lattice QCD tests of the standard model and foretaste for
  beyond}, PoS FPCP2015 (2015) 016.
\newblock \href {http://arxiv.org/abs/1509.01155} {\path{arXiv:1509.01155}}.

\bibitem{Gerard2011}
J.-M. Gerard, {An upper bound on the kaon $B$-parameter and Re($\epsilon_K$)},
  JHEP 02 (2011) 075.
\newblock \href {http://arxiv.org/abs/1012.2026} {\path{arXiv:1012.2026}},
  \href {http://dx.doi.org/10.1007/JHEP02(2011)075}
  {\path{doi:10.1007/JHEP02(2011)075}}.

\bibitem{Buras2014}
A.~J. Buras, J.-M. Gérard, W.~A. Bardeen, {Large $N$ approach to kaon decays
  and mixing 28 years later: $\Delta I = 1/2$ rule, $\hat B_K$ and $\Delta
  M_K$}, Eur. Phys. J. C74 (2014) 2871.
\newblock \href {http://arxiv.org/abs/1401.1385} {\path{arXiv:1401.1385}},
  \href {http://dx.doi.org/10.1140/epjc/s10052-014-2871-x}
  {\path{doi:10.1140/epjc/s10052-014-2871-x}}.

\bibitem{Blanke2016}
M.~Blanke, A.~J. Buras, {Universal unitarity triangle 2016 and the tension
  between $\Delta M_{s,d}$ and $\varepsilon_K$ in CMFV models} (2016).
\newblock \href {http://arxiv.org/abs/1602.04020} {\path{arXiv:1602.04020}}.

\bibitem{Brod2012}
J.~Brod, M.~Gorbahn, {Next-to-next-to-leading-order charm-quark contribution to
  the $CP$ violation parameter $\epsilon_K$ and $\Delta M_K$}, Phys. Rev. Lett.
  108 (2012) 121801.
\newblock \href {http://arxiv.org/abs/1108.2036} {\path{arXiv:1108.2036}},
  \href {http://dx.doi.org/10.1103/PhysRevLett.108.121801}
  {\path{doi:10.1103/PhysRevLett.108.121801}}.

\bibitem{Brod2010}
J.~Brod, M.~Gorbahn, {$\epsilon_K$ at Next-to-next-to-leading order: The
  charm-top-quark contribution}, Phys. Rev. D82 (2010) 094026.
\newblock \href {http://arxiv.org/abs/1007.0684} {\path{arXiv:1007.0684}},
  \href {http://dx.doi.org/10.1103/PhysRevD.82.094026}
  {\path{doi:10.1103/PhysRevD.82.094026}}.

\bibitem{Buras2008}
A.~J. Buras, D.~Guadagnoli, {Correlations among new $CP$ violating effects in
  $\Delta F = 2$ observables}, Phys. Rev. D78 (2008) 033005.
\newblock \href {http://arxiv.org/abs/0805.3887} {\path{arXiv:0805.3887}},
  \href {http://dx.doi.org/10.1103/PhysRevD.78.033005}
  {\path{doi:10.1103/PhysRevD.78.033005}}.

\bibitem{Buras2010}
A.~J. Buras, D.~Guadagnoli, G.~Isidori, {On $\epsilon_K$ beyond lowest order in
  the operator product expansion}, Phys. Lett. B688 (2010) 309--313.
\newblock \href {http://arxiv.org/abs/1002.3612} {\path{arXiv:1002.3612}},
  \href {http://dx.doi.org/10.1016/j.physletb.2010.04.017}
  {\path{doi:10.1016/j.physletb.2010.04.017}}.

\bibitem{Cirigliano2012}
V.~Cirigliano, G.~Ecker, H.~Neufeld, A.~Pich, J.~Portoles, {Kaon decays in the
  standard model}, Rev. Mod. Phys. 84 (2012) 399.
\newblock \href {http://arxiv.org/abs/1107.6001} {\path{arXiv:1107.6001}},
  \href {http://dx.doi.org/10.1103/RevModPhys.84.399}
  {\path{doi:10.1103/RevModPhys.84.399}}.

\bibitem{Ligeti2016}
Z.~Ligeti, F.~Sala, {A new look at the theory uncertainty of $\epsilon_K$}\href
  {http://arxiv.org/abs/1602.08494} {\path{arXiv:1602.08494}}.

\bibitem{Bigi1981}
I.~I.~Y. Bigi, A.~I. Sanda, {Notes on the observability of $CP$ violations in
  $B$ decays}, Nucl. Phys. B193 (1981) 85.
\newblock \href {http://dx.doi.org/10.1016/0550-3213(81)90519-8}
  {\path{doi:10.1016/0550-3213(81)90519-8}}.

\bibitem{Aubert2006c}
B.~Aubert, et~al., {Search for $T$, $CP$ and $CPT$ violation in $B^0 \bar B^0$
  mixing with inclusive dilepton events}, Phys. Rev. Lett. 96 (2006) 251802.
\newblock \href {http://arxiv.org/abs/hep-ex/0603053}
  {\path{arXiv:hep-ex/0603053}}, \href
  {http://dx.doi.org/10.1103/PhysRevLett.96.251802}
  {\path{doi:10.1103/PhysRevLett.96.251802}}.

\bibitem{Aubert2004b}
B.~Aubert, et~al., {Limits on the decay rate difference of neutral-$B$ mesons
  and on $CP$, $T$, and $CPT$ violation in $B^0 \bar B^0$ oscillations}, Phys.
  Rev. D70 (2004) 012007.
\newblock \href {http://arxiv.org/abs/hep-ex/0403002}
  {\path{arXiv:hep-ex/0403002}}, \href
  {http://dx.doi.org/10.1103/PhysRevD.70.012007}
  {\path{doi:10.1103/PhysRevD.70.012007}}.

\bibitem{Higuchi2012}
T.~Higuchi, et~al., {Search for time-dependent $CPT$ violation in hadronic and
  semileptonic $B$ decays}, Phys. Rev. D85 (2012) 071105(R).
\newblock \href {http://dx.doi.org/10.1103/PhysRevD.85.071105}
  {\path{doi:10.1103/PhysRevD.85.071105}}.

\bibitem{Aaij2012d}
R.~Aaij, et~al., {Determination of the sign of the decay width difference in
  the $B_s$ system}, Phys. Rev. Lett. 108 (2012) 241801.
\newblock \href {http://arxiv.org/abs/1202.4717} {\path{arXiv:1202.4717}},
  \href {http://dx.doi.org/10.1103/PhysRevLett.108.241801}
  {\path{doi:10.1103/PhysRevLett.108.241801}}.

\bibitem{Hagelin1981}
J.~S. Hagelin, {Mass mixing and $CP$ violation in the $B^0-\bar{B}^0$ system},
  Nucl. Phys. B193 (1981) 123--149.
\newblock \href {http://dx.doi.org/10.1016/0550-3213(81)90521-6}
  {\path{doi:10.1016/0550-3213(81)90521-6}}.

\bibitem{Lenz2007}
A.~Lenz, U.~Nierste, {Theoretical update of $B_s - \bar{B}_s$ mixing}, JHEP 06
  (2007) 072.
\newblock \href {http://arxiv.org/abs/hep-ph/0612167}
  {\path{arXiv:hep-ph/0612167}}, \href
  {http://dx.doi.org/10.1088/1126-6708/2007/06/072}
  {\path{doi:10.1088/1126-6708/2007/06/072}}.

\bibitem{Lees2016}
J.~P. Lees, et~al., {Search for mixing-induced $CP$ violation using partial
  reconstruction of $\bar B^0 \to D^{*+} X\ell^- \bar \nu_{\ell}$ and kaon
  tagging}, Phys. Rev. D93~(3) (2016) 032001.
\newblock \href {http://arxiv.org/abs/1506.00234} {\path{arXiv:1506.00234}},
  \href {http://dx.doi.org/10.1103/PhysRevD.93.032001}
  {\path{doi:10.1103/PhysRevD.93.032001}}.

\bibitem{Aaij2015h}
R.~Aaij, et~al., {Measurement of the semileptonic $CP$ asymmetry in
  $B^0-\overline{B}{}^0$ mixing}, Phys. Rev. Lett. 114 (2015) 041601.
\newblock \href {http://arxiv.org/abs/1409.8586} {\path{arXiv:1409.8586}},
  \href {http://dx.doi.org/10.1103/PhysRevLett.114.041601}
  {\path{doi:10.1103/PhysRevLett.114.041601}}.

\bibitem{Aaij2014h}
R.~Aaij, et~al., {Measurement of the flavour-specific $CP$-violating asymmetry
  $a_{sl}^s$ in $B_s^0$ decays}, Phys. Lett. B728 (2014) 607--615.
\newblock \href {http://arxiv.org/abs/1308.1048} {\path{arXiv:1308.1048}},
  \href {http://dx.doi.org/10.1016/j.physletb.2013.12.030}
  {\path{doi:10.1016/j.physletb.2013.12.030}}.

\bibitem{Abazov2014}
V.~M. Abazov, et~al., {Study of $CP$-violating charge asymmetries of single
  muons and like-sign dimuons in $p\bar p$ collisions}, Phys. Rev. D89~(1)
  (2014) 012002.
\newblock \href {http://arxiv.org/abs/1310.0447} {\path{arXiv:1310.0447}},
  \href {http://dx.doi.org/10.1103/PhysRevD.89.012002}
  {\path{doi:10.1103/PhysRevD.89.012002}}.

\bibitem{Buras1984}
A.~J. Buras, W.~Slominski, H.~Steger, {$B^0 \bar B^0$ mixing, $CP$ violation
  and the $B$ meson decay}, Nucl. Phys. B245 (1984) 369.
\newblock \href {http://dx.doi.org/10.1016/0550-3213(84)90437-1}
  {\path{doi:10.1016/0550-3213(84)90437-1}}.

\bibitem{Gronau1993}
M.~Gronau, A.~Nippe, J.~L. Rosner, {Method for flavor tagging in neutral $B$
  meson decays}, Phys. Rev. D47 (1993) 1988--1993.
\newblock \href {http://arxiv.org/abs/hep-ph/9211311}
  {\path{arXiv:hep-ph/9211311}}, \href
  {http://dx.doi.org/10.1103/PhysRevD.47.1988}
  {\path{doi:10.1103/PhysRevD.47.1988}}.

\bibitem{Abe1998}
F.~Abe, et~al., {Measurement of the $B^0 \overline{B}^0$ oscillation frequency
  using $\pi B$ meson charge-flavor correlations in $p\bar{p}$ collisions at
  $\sqrt{s} = 1.8$ TeV}, Phys. Rev. Lett. 80 (1998) 2057--2062.
\newblock \href {http://arxiv.org/abs/hep-ex/9712004}
  {\path{arXiv:hep-ex/9712004}}, \href
  {http://dx.doi.org/10.1103/PhysRevLett.80.2057}
  {\path{doi:10.1103/PhysRevLett.80.2057}}.

\bibitem{Affolder2000}
T.~Affolder, et~al., {A measurement of $\sin(2\beta)$ from $B \to J/\psi K^0_S$
  with the {CDF} detector}, Phys. Rev. D61 (2000) 072005.
\newblock \href {http://arxiv.org/abs/hep-ex/9909003}
  {\path{arXiv:hep-ex/9909003}}, \href
  {http://dx.doi.org/10.1103/PhysRevD.61.072005}
  {\path{doi:10.1103/PhysRevD.61.072005}}.

\bibitem{Aaij2015a}
R.~Aaij, et~al., {Measurement of $CP$ violation in $B^0 \rightarrow J/\psi
  K^0_S$ decays}, Phys. Rev. Lett. 115~(3) (2015) 031601.
\newblock \href {http://arxiv.org/abs/1503.07089} {\path{arXiv:1503.07089}},
  \href {http://dx.doi.org/10.1103/PhysRevLett.115.031601}
  {\path{doi:10.1103/PhysRevLett.115.031601}}.

\bibitem{Sato2012}
Y.~Sato, et~al., {Measurement of the $CP$-violation parameter sin2$\phi_1$ with
  a new tagging method at the $\Upsilon(5S)$ resonance}, Phys. Rev. Lett. 108
  (2012) 171801.
\newblock \href {http://arxiv.org/abs/1201.3502} {\path{arXiv:1201.3502}},
  \href {http://dx.doi.org/10.1103/PhysRevLett.108.171801}
  {\path{doi:10.1103/PhysRevLett.108.171801}}.

\bibitem{Aubert2004a}
B.~Aubert, et~al., {Measurement of $\sin2\beta$ using hadronic $J/\psi$
  decays}, Phys. Rev. D69 (2004) 052001.
\newblock \href {http://arxiv.org/abs/hep-ex/0309039}
  {\path{arXiv:hep-ex/0309039}}, \href
  {http://dx.doi.org/10.1103/PhysRevD.69.052001}
  {\path{doi:10.1103/PhysRevD.69.052001}}.

\bibitem{Aubert2009}
B.~Aubert, et~al., {Measurement of time-dependent $CP$ asymmetry in $B^0 \to c
  \bar c K^{(*)0}$ decays}, Phys. Rev. D79 (2009) 072009.
\newblock \href {http://arxiv.org/abs/0902.1708} {\path{arXiv:0902.1708}},
  \href {http://dx.doi.org/10.1103/PhysRevD.79.072009}
  {\path{doi:10.1103/PhysRevD.79.072009}}.

\bibitem{Aubert2009a}
B.~Aubert, et~al., {Time-dependent amplitude analysis of $B^0 \to K^0_S \pi^+
  \pi^-$}, Phys. Rev. D80 (2009) 112001.
\newblock \href {http://arxiv.org/abs/0905.3615} {\path{arXiv:0905.3615}},
  \href {http://dx.doi.org/10.1103/PhysRevD.80.112001}
  {\path{doi:10.1103/PhysRevD.80.112001}}.

\bibitem{Adachi2012}
I.~Adachi, et~al., {Precise measurement of the $CP$ violation parameter
  $\sin2\phi_1$ in $B^0\to (c\bar c)K^0$ decays}, Phys. Rev. Lett. 108 (2012)
  171802.
\newblock \href {http://arxiv.org/abs/1201.4643} {\path{arXiv:1201.4643}},
  \href {http://dx.doi.org/10.1103/PhysRevLett.108.171802}
  {\path{doi:10.1103/PhysRevLett.108.171802}}.

\bibitem{Barate2000}
R.~Barate, et~al., {Study of the $CP$ asymmetry of $B^0 \to J/\psi K^0_{S}$
  decays in ALEPH}, Phys. Lett. B492 (2000) 259--274.
\newblock \href {http://arxiv.org/abs/hep-ex/0009058}
  {\path{arXiv:hep-ex/0009058}}, \href
  {http://dx.doi.org/10.1016/S0370-2693(00)01091-1}
  {\path{doi:10.1016/S0370-2693(00)01091-1}}.

\bibitem{Ackerstaff1998}
K.~Ackerstaff, et~al., {Investigation of CP violation in $B^0 \to J/\psi
  K^0_{S}$ decays at LEP}, Eur. Phys. J. C5 (1998) 379--388.
\newblock \href {http://arxiv.org/abs/hep-ex/9801022}
  {\path{arXiv:hep-ex/9801022}}, \href
  {http://dx.doi.org/10.1007/s100520050284} {\path{doi:10.1007/s100520050284}}.

\bibitem{Aaij2013}
R.~Aaij, et~al., {Measurement of the time-dependent $CP$ asymmetry in $B^0 \to
  J/\psi K^0_{S}$ decays}, Phys. Lett. B721 (2013) 24--31.
\newblock \href {http://arxiv.org/abs/1211.6093} {\path{arXiv:1211.6093}},
  \href {http://dx.doi.org/10.1016/j.physletb.2013.02.054}
  {\path{doi:10.1016/j.physletb.2013.02.054}}.

\bibitem{Gronau1989}
M.~Gronau, {CP Violation in Neutral B Decays to CP Eigenstates}, Phys. Rev.
  Lett. 63 (1989) 1451.
\newblock \href {http://dx.doi.org/10.1103/PhysRevLett.63.1451}
  {\path{doi:10.1103/PhysRevLett.63.1451}}.

\bibitem{Fleischer1999}
R.~Fleischer, {Extracting $\gamma$ from $B(s/d) \to J/\psi K_{S}$ and $B(d/s)
  \to D^+(d/s) D^-(d/s)$}, Eur. Phys. J. C10 (1999) 299--306.
\newblock \href {http://arxiv.org/abs/hep-ph/9903455}
  {\path{arXiv:hep-ph/9903455}}, \href
  {http://dx.doi.org/10.1007/s100529900099} {\path{doi:10.1007/s100529900099}}.

\bibitem{Ciuchini2005}
M.~Ciuchini, M.~Pierini, L.~Silvestrini, {The effect of penguins in the $B_d
  \to J / \psi K^0\ CP$ asymmetry}, Phys. Rev. Lett. 95 (2005) 221804.
\newblock \href {http://arxiv.org/abs/hep-ph/0507290}
  {\path{arXiv:hep-ph/0507290}}, \href
  {http://dx.doi.org/10.1103/PhysRevLett.95.221804}
  {\path{doi:10.1103/PhysRevLett.95.221804}}.

\bibitem{Faller2009}
S.~Faller, M.~Jung, R.~Fleischer, T.~Mannel, {The golden modes $B^0 \to J/\psi
  K_{S,L}$ in the era of precision flavour physics}, Phys. Rev. D79 (2009)
  014030.
\newblock \href {http://arxiv.org/abs/0809.0842} {\path{arXiv:0809.0842}},
  \href {http://dx.doi.org/10.1103/PhysRevD.79.014030}
  {\path{doi:10.1103/PhysRevD.79.014030}}.

\bibitem{Gronau2009}
M.~Gronau, J.~L. Rosner, {Doubly CKM-suppressed corrections to $CP$ asymmetries
  in $B^0 \to J/\psi K^0$}, Phys. Lett. B672 (2009) 349--353.
\newblock \href {http://arxiv.org/abs/0812.4796} {\path{arXiv:0812.4796}},
  \href {http://dx.doi.org/10.1016/j.physletb.2009.01.049}
  {\path{doi:10.1016/j.physletb.2009.01.049}}.

\bibitem{Ligeti2015}
Z.~Ligeti, D.~J. Robinson, {Towards more precise determinations of the quark
  mixing phase $\beta$}, Phys. Rev. Lett. 115~(25) (2015) 251801.
\newblock \href {http://arxiv.org/abs/1507.06671} {\path{arXiv:1507.06671}},
  \href {http://dx.doi.org/10.1103/PhysRevLett.115.251801}
  {\path{doi:10.1103/PhysRevLett.115.251801}}.

\bibitem{Beneke2005}
M.~Beneke, {Corrections to $\sin2 \beta$ from $CP$ asymmetries in $B^0 \to
  (\pi^0, \rho^0, \eta, \eta^\prime, \omega, \phi) K_S$ decays}, Phys. Lett.
  B620 (2005) 143--150.
\newblock \href {http://arxiv.org/abs/hep-ph/0505075}
  {\path{arXiv:hep-ph/0505075}}, \href
  {http://dx.doi.org/10.1016/j.physletb.2005.06.045}
  {\path{doi:10.1016/j.physletb.2005.06.045}}.

\bibitem{Frings2015}
P.~Frings, U.~Nierste, M.~Wiebusch, {Penguin contributions to $CP$ phases in
  $B_{d,s}$ decays to charmonium}, Phys. Rev. Lett. 115~(6) (2015) 061802.
\newblock \href {http://arxiv.org/abs/1503.00859} {\path{arXiv:1503.00859}},
  \href {http://dx.doi.org/10.1103/PhysRevLett.115.061802}
  {\path{doi:10.1103/PhysRevLett.115.061802}}.

\bibitem{Dadisman2016}
R.~Dadisman, S.~Gardner, X.~Yan, {Trapping penguins with entangled $B$ mesons},
  Phys. Lett. B754 (2016) 1--5.
\newblock \href {http://arxiv.org/abs/1409.6801} {\path{arXiv:1409.6801}},
  \href {http://dx.doi.org/10.1016/j.physletb.2016.01.002}
  {\path{doi:10.1016/j.physletb.2016.01.002}}.

\bibitem{Lees2012b}
J.~P. Lees, et~al., {Observation of time reversal violation in the $B^0$ meson
  system}, Phys. Rev. Lett. 109 (2012) 211801.
\newblock \href {http://arxiv.org/abs/1207.5832} {\path{arXiv:1207.5832}},
  \href {http://dx.doi.org/10.1103/PhysRevLett.109.211801}
  {\path{doi:10.1103/PhysRevLett.109.211801}}.

\bibitem{Abdesselam2015}
A.~Abdesselam, et~al., {First observation of $CP$ violation in $\overline{B}^0
  \to D^{(*)}_{\rm CP} h^0$ decays by a combined time-dependent analysis of
  BaBar and Belle data}, Phys. Rev. Lett. 115~(12) (2015) 121604.
\newblock \href {http://arxiv.org/abs/1505.04147} {\path{arXiv:1505.04147}},
  \href {http://dx.doi.org/10.1103/PhysRevLett.115.121604}
  {\path{doi:10.1103/PhysRevLett.115.121604}}.

\bibitem{Chua2006}
C.-K. Chua, {Standard model expectations on $\sin2\beta(\phi_1)$ from $b \to s$
  penguins}, eConf C060409 (2006) 008.
\newblock \href {http://arxiv.org/abs/hep-ph/0605301}
  {\path{arXiv:hep-ph/0605301}}.

\bibitem{Dunietz1991}
I.~Dunietz, H.~R. Quinn, A.~Snyder, W.~Toki, H.~J. Lipkin, {How to extract $CP$
  violating asymmetries from angular correlations}, Phys. Rev. D43 (1991)
  2193--2208.
\newblock \href {http://dx.doi.org/10.1103/PhysRevD.43.2193}
  {\path{doi:10.1103/PhysRevD.43.2193}}.

\bibitem{Bondar2005}
A.~Bondar, T.~Gershon, P.~Krokovny, {A method to measure $\phi_1$ using $\bar
  B^0 \to D h^0$ with multibody $D$ decay}, Phys. Lett. B624 (2005) 1--10.
\newblock \href {http://arxiv.org/abs/hep-ph/0503174}
  {\path{arXiv:hep-ph/0503174}}, \href
  {http://dx.doi.org/10.1016/j.physletb.2005.07.053}
  {\path{doi:10.1016/j.physletb.2005.07.053}}.

\bibitem{Browder2000}
T.~E. Browder, A.~Datta, P.~J. O'Donnell, S.~Pakvasa, {Measuring $\beta$ in $B
  \to D^{(*)+} D^{(*)-} K_S$ decays}, Phys. Rev. D61 (2000) 054009.
\newblock \href {http://arxiv.org/abs/hep-ph/9905425}
  {\path{arXiv:hep-ph/9905425}}, \href
  {http://dx.doi.org/10.1103/PhysRevD.61.054009}
  {\path{doi:10.1103/PhysRevD.61.054009}}.

\bibitem{Aubert2006a}
B.~Aubert, et~al., {Measurement of the branching fraction and time-dependent
  $CP$ asymmetry in the decay $B^0 \to D^{*+} D^{*-} K^0_{s}$}, Phys. Rev. D74
  (2006) 091101.
\newblock \href {http://arxiv.org/abs/hep-ex/0608016}
  {\path{arXiv:hep-ex/0608016}}, \href
  {http://dx.doi.org/10.1103/PhysRevD.74.091101}
  {\path{doi:10.1103/PhysRevD.74.091101}}.

\bibitem{Dalseno2007}
J.~Dalseno, et~al., {Measurement of branching fraction and time-dependent $CP$
  asymmetry parameters in $B^0 \to D^{*+} D^{*-} K^0_S$ decays}, Phys. Rev. D76
  (2007) 072004.
\newblock \href {http://arxiv.org/abs/0706.2045} {\path{arXiv:0706.2045}},
  \href {http://dx.doi.org/10.1103/PhysRevD.76.072004}
  {\path{doi:10.1103/PhysRevD.76.072004}}.

\bibitem{Aubert2005a}
B.~Aubert, et~al., {Ambiguity-free measurement of $\cos(2\beta)$:
  time-integrated and time-dependent angular analyses of $B \to J/\psi K \pi$},
  Phys. Rev. D71 (2005) 032005.
\newblock \href {http://arxiv.org/abs/hep-ex/0411016}
  {\path{arXiv:hep-ex/0411016}}, \href
  {http://dx.doi.org/10.1103/PhysRevD.71.032005}
  {\path{doi:10.1103/PhysRevD.71.032005}}.

\bibitem{Itoh2005}
R.~Itoh, et~al., {Studies of $CP$ violation in $B \to J/\psi K^*$ decays},
  Phys. Rev. Lett. 95 (2005) 091601.
\newblock \href {http://arxiv.org/abs/hep-ex/0504030}
  {\path{arXiv:hep-ex/0504030}}, \href
  {http://dx.doi.org/10.1103/PhysRevLett.95.091601}
  {\path{doi:10.1103/PhysRevLett.95.091601}}.

\bibitem{Aubert2007a}
B.~Aubert, et~al., {Measurement of $\cos 2 \beta$ in $B^0 \to D^{(*)} h^0$
  decays with a time-dependent Dalitz plot analysis of $D \to K^0_{S} \pi^{+}
  \pi^{-}$}, Phys. Rev. Lett. 99 (2007) 231802.
\newblock \href {http://arxiv.org/abs/0708.1544} {\path{arXiv:0708.1544}},
  \href {http://dx.doi.org/10.1103/PhysRevLett.99.231802}
  {\path{doi:10.1103/PhysRevLett.99.231802}}.

\bibitem{Krokovny2006}
P.~Krokovny, et~al., {Measurement of the quark mixing parameter $\cos 2\phi_1$
  using time-dependent Dalitz analysis of $\bar B^0 \to D [K_S^0 \pi^+ \pi^-]
  h^0$}, Phys. Rev. Lett. 97 (2006) 081801.
\newblock \href {http://arxiv.org/abs/hep-ex/0605023}
  {\path{arXiv:hep-ex/0605023}}, \href
  {http://dx.doi.org/10.1103/PhysRevLett.97.081801}
  {\path{doi:10.1103/PhysRevLett.97.081801}}.

\bibitem{Gronau1990}
M.~Gronau, D.~London, {Isospin analysis of $CP$ asymmetries in $B$ decays},
  Phys. Rev. Lett. 65 (1990) 3381--3384.
\newblock \href {http://dx.doi.org/10.1103/PhysRevLett.65.3381}
  {\path{doi:10.1103/PhysRevLett.65.3381}}.

\bibitem{Lees2013b}
J.~P. Lees, et~al., {Measurement of $CP$ asymmetries and branching fractions in
  charmless two-body $B$-meson decays to pions and kaons}, Phys. Rev. D87~(5)
  (2013) 052009.
\newblock \href {http://arxiv.org/abs/1206.3525} {\path{arXiv:1206.3525}},
  \href {http://dx.doi.org/10.1103/PhysRevD.87.052009}
  {\path{doi:10.1103/PhysRevD.87.052009}}.

\bibitem{Dalseno2013}
J.~Dalseno, et~al., {Measurement of the $CP$ violation parameters in $B^0 \to
  \pi^+ \pi^-$ decays}, Phys. Rev. D88~(9) (2013) 092003.
\newblock \href {http://arxiv.org/abs/1302.0551} {\path{arXiv:1302.0551}},
  \href {http://dx.doi.org/10.1103/PhysRevD.88.092003}
  {\path{doi:10.1103/PhysRevD.88.092003}}.

\bibitem{Aaij2013a}
R.~Aaij, et~al., {First measurement of time-dependent $CP$ violation in $B^0_s
  \to K^+K^-$ decays}, JHEP 10 (2013) 183.
\newblock \href {http://arxiv.org/abs/1308.1428} {\path{arXiv:1308.1428}},
  \href {http://dx.doi.org/10.1007/JHEP10(2013)183}
  {\path{doi:10.1007/JHEP10(2013)183}}.

\bibitem{Lipkin1991}
H.~J. Lipkin, Y.~Nir, H.~R. Quinn, A.~Snyder, {Penguin trapping with isospin
  analysis and $CP$ asymmetries in $B$ decays}, Phys. Rev. D44 (1991)
  1454--1460.
\newblock \href {http://dx.doi.org/10.1103/PhysRevD.44.1454}
  {\path{doi:10.1103/PhysRevD.44.1454}}.

\bibitem{Snyder1993}
A.~E. Snyder, H.~R. Quinn, {Measuring $CP$ asymmetry in $B \to \rho \pi$ decays
  without ambiguities}, Phys. Rev. D48 (1993) 2139--2144.
\newblock \href {http://dx.doi.org/10.1103/PhysRevD.48.2139}
  {\path{doi:10.1103/PhysRevD.48.2139}}.

\bibitem{Quinn2000}
H.~R. Quinn, J.~P. Silva, {The use of early data on $B \to \rho \pi$ decays},
  Phys. Rev. D62 (2000) 054002.
\newblock \href {http://arxiv.org/abs/hep-ph/0001290}
  {\path{arXiv:hep-ph/0001290}}, \href
  {http://dx.doi.org/10.1103/PhysRevD.62.054002}
  {\path{doi:10.1103/PhysRevD.62.054002}}.

\bibitem{Lees2013c}
J.~P. Lees, et~al., {Measurement of $CP$-violating asymmetries in $B^0 \to
  (\rho \pi)^0$ decays using a time-dependent Dalitz plot analysis}, Phys. Rev.
  D88~(1) (2013) 012003.
\newblock \href {http://arxiv.org/abs/1304.3503} {\path{arXiv:1304.3503}},
  \href {http://dx.doi.org/10.1103/PhysRevD.88.012003}
  {\path{doi:10.1103/PhysRevD.88.012003}}.

\bibitem{Kusaka2007}
A.~Kusaka, et~al., {Measurement of $CP$ asymmetry in a time-dependent Dalitz
  analysis of $B^0 \to (\rho \pi)^0$ and a constraint on the CKM angle
  $\phi_2$}, Phys. Rev. Lett. 98 (2007) 221602.
\newblock \href {http://arxiv.org/abs/hep-ex/0701015}
  {\path{arXiv:hep-ex/0701015}}, \href
  {http://dx.doi.org/10.1103/PhysRevLett.98.221602}
  {\path{doi:10.1103/PhysRevLett.98.221602}}.

\bibitem{Kusaka2008}
A.~Kusaka, et~al., {Measurement of $CP$ asymmetries and branching fractions in
  a time-dependent Dalitz analysis of $B^0 \to (\rho \pi)^0$ and a constraint
  on the quark mixing angle $\phi_2$}, Phys. Rev. D77 (2008) 072001.
\newblock \href {http://arxiv.org/abs/0710.4974} {\path{arXiv:0710.4974}},
  \href {http://dx.doi.org/10.1103/PhysRevD.77.072001}
  {\path{doi:10.1103/PhysRevD.77.072001}}.

\bibitem{Aaij2015e}
R.~Aaij, et~al., {Observation of the ${B^0 \to \rho^0 \rho^0}$ decay from an
  amplitude analysis of ${B^0 \to (\pi^+\pi^-)(\pi^+\pi^-)}$ decays}, Phys.
  Lett. B747 (2015) 468--478.
\newblock \href {http://arxiv.org/abs/1503.07770} {\path{arXiv:1503.07770}},
  \href {http://dx.doi.org/10.1016/j.physletb.2015.06.027}
  {\path{doi:10.1016/j.physletb.2015.06.027}}.

\bibitem{Aubert2005b}
B.~Aubert, et~al., {Improved measurement of the {CKM} angle $\alpha$ using $B^0
  \to \rho^+ \rho^-$ decays}, Phys. Rev. Lett. 95 (2005) 041805.
\newblock \href {http://arxiv.org/abs/hep-ex/0503049}
  {\path{arXiv:hep-ex/0503049}}, \href
  {http://dx.doi.org/10.1103/PhysRevLett.95.041805}
  {\path{doi:10.1103/PhysRevLett.95.041805}}.

\bibitem{Aubert2008a}
B.~Aubert, et~al., {Measurement of the branching fraction, polarization, and
  $CP$ asymmetries in $B^0 \to \rho^0 \rho^0$ decay, and implications for the
  CKM angle $\alpha$}, Phys. Rev. D78 (2008) 071104.
\newblock \href {http://arxiv.org/abs/0807.4977} {\path{arXiv:0807.4977}},
  \href {http://dx.doi.org/10.1103/PhysRevD.78.071104}
  {\path{doi:10.1103/PhysRevD.78.071104}}.

\bibitem{Aubert2009b}
B.~Aubert, et~al., {Improved measurement of $B^+ \to\rho^+\rho^0$ and
  determination of the quark-mixing phase angle $\alpha$}, Phys. Rev. Lett. 102
  (2009) 141802.
\newblock \href {http://arxiv.org/abs/0901.3522} {\path{arXiv:0901.3522}},
  \href {http://dx.doi.org/10.1103/PhysRevLett.102.141802}
  {\path{doi:10.1103/PhysRevLett.102.141802}}.

\bibitem{Zhang2003}
J.~Zhang, et~al., {Observation of $B^+ \to \rho^+ \rho^0$}, Phys. Rev. Lett. 91
  (2003) 221801.
\newblock \href {http://arxiv.org/abs/hep-ex/0306007}
  {\path{arXiv:hep-ex/0306007}}, \href
  {http://dx.doi.org/10.1103/PhysRevLett.91.221801}
  {\path{doi:10.1103/PhysRevLett.91.221801}}.

\bibitem{Somov2006}
A.~Somov, et~al., {Measurement of the branching fraction, polarization, and
  $CP$ asymmetry for $B^0 \to \rho^+ \rho^-$ decays, and determination of the
  CKM phase $\phi_2$}, Phys. Rev. Lett. 96 (2006) 171801.
\newblock \href {http://arxiv.org/abs/hep-ex/0601024}
  {\path{arXiv:hep-ex/0601024}}, \href
  {http://dx.doi.org/10.1103/PhysRevLett.96.171801}
  {\path{doi:10.1103/PhysRevLett.96.171801}}.

\bibitem{Somov2007}
A.~Somov, et~al., {Improved measurement of $CP$-violating parameters in $B^0
  \to \rho^+ \rho^-$ decays}, Phys. Rev. D76 (2007) 011104.
\newblock \href {http://arxiv.org/abs/hep-ex/0702009}
  {\path{arXiv:hep-ex/0702009}}, \href
  {http://dx.doi.org/10.1103/PhysRevD.76.011104}
  {\path{doi:10.1103/PhysRevD.76.011104}}.

\bibitem{Vanhoefer2014}
P.~Vanhoefer, et~al., {Study of $B^0 \to \rho^0 \rho^0$ decays, implications
  for the CKM angle $\phi_2$ and search for other $B^0$ decay modes with a
  four-pion final state}, Phys. Rev. D89 (2014) 072008, [Addendum: Phys.
  Rev.D89, no.11, 119903(2014)].
\newblock \href {http://arxiv.org/abs/1212.4015} {\path{arXiv:1212.4015}},
  \href {http://dx.doi.org/10.1103/PhysRevD.89.072008,
  10.1103/PhysRevD.89.119903} {\path{doi:10.1103/PhysRevD.89.072008,
  10.1103/PhysRevD.89.119903}}.

\bibitem{Vanhoefer2016}
P.~Vanhoefer, et~al., {Study of $B^{0}\rightarrow\rho^{+}\rho^{-}$ decays and
  implications for the CKM angle $\phi_2$}, Phys. Rev. D93~(3) (2016) 032010.
\newblock \href {http://arxiv.org/abs/1510.01245} {\path{arXiv:1510.01245}},
  \href {http://dx.doi.org/10.1103/PhysRevD.93.032010}
  {\path{doi:10.1103/PhysRevD.93.032010}}.

\bibitem{Gronau2006}
M.~Gronau, J.~Zupan, {Weak phase $\alpha$ from $B^0 \to a_1^\pm(1260)
  \pi^\mp$}, Phys. Rev. D73 (2006) 057502.
\newblock \href {http://arxiv.org/abs/hep-ph/0512148}
  {\path{arXiv:hep-ph/0512148}}, \href
  {http://dx.doi.org/10.1103/PhysRevD.73.057502}
  {\path{doi:10.1103/PhysRevD.73.057502}}.

\bibitem{Aubert2007b}
B.~Aubert, et~al., {Measurements of $CP$-violating asymmetries in $B^0 \to
  a^\pm_1 (1260) \pi^\mp$ decays}, Phys. Rev. Lett. 98 (2007) 181803.
\newblock \href {http://arxiv.org/abs/hep-ex/0612050}
  {\path{arXiv:hep-ex/0612050}}, \href
  {http://dx.doi.org/10.1103/PhysRevLett.98.181803}
  {\path{doi:10.1103/PhysRevLett.98.181803}}.

\bibitem{Dalseno2012}
J.~Dalseno, et~al., {Measurement of branching fraction and first evidence of
  $CP$ violation in $B^0 \to a_1^{\pm}(1260) \pi^\mp$ decays}, Phys. Rev. D86
  (2012) 092012.
\newblock \href {http://arxiv.org/abs/1205.5957} {\path{arXiv:1205.5957}},
  \href {http://dx.doi.org/10.1103/PhysRevD.86.092012}
  {\path{doi:10.1103/PhysRevD.86.092012}}.

\bibitem{Carter1981}
A.~B. Carter, A.~I. Sanda, {$CP$ violation in $B$ meson decays}, Phys. Rev. D23
  (1981) 1567.
\newblock \href {http://dx.doi.org/10.1103/PhysRevD.23.1567}
  {\path{doi:10.1103/PhysRevD.23.1567}}.

\bibitem{Gronau1991a}
M.~Gronau, D.~London, {How to determine all the angles of the unitarity
  triangle from $B_d^0 \to D K_S$ and $B_s^0 \to D^0$}, Phys. Lett. B253 (1991)
  483--488.
\newblock \href {http://dx.doi.org/10.1016/0370-2693(91)91756-L}
  {\path{doi:10.1016/0370-2693(91)91756-L}}.

\bibitem{Gronau1991b}
M.~Gronau, D.~Wyler, {On determining a weak phase from $CP$ asymmetries in
  charged $B$ decays}, Phys. Lett. B265 (1991) 172--176.
\newblock \href {http://dx.doi.org/10.1016/0370-2693(91)90034-N}
  {\path{doi:10.1016/0370-2693(91)90034-N}}.

\bibitem{Amorim1999}
A.~Amorim, M.~G. Santos, J.~P. Silva, {New $CP$ violating parameters in cascade
  decays}, Phys. Rev. D59 (1999) 056001.
\newblock \href {http://arxiv.org/abs/hep-ph/9807364}
  {\path{arXiv:hep-ph/9807364}}, \href
  {http://dx.doi.org/10.1103/PhysRevD.59.056001}
  {\path{doi:10.1103/PhysRevD.59.056001}}.

\bibitem{Grossman2005}
Y.~Grossman, A.~Soffer, J.~Zupan, {The effect of $D - \bar D$ mixing on the
  measurement of $\gamma$ in $B \to DK$ decays}, Phys. Rev. D72 (2005) 031501.
\newblock \href {http://arxiv.org/abs/hep-ph/0505270}
  {\path{arXiv:hep-ph/0505270}}, \href
  {http://dx.doi.org/10.1103/PhysRevD.72.031501}
  {\path{doi:10.1103/PhysRevD.72.031501}}.

\bibitem{Giri2003}
A.~Giri, Y.~Grossman, A.~Soffer, J.~Zupan, {Determining gamma using $B^\pm \to
  DK^\pm$ with multibody $D$ decays}, Phys. Rev. D68 (2003) 054018.
\newblock \href {http://arxiv.org/abs/hep-ph/0303187}
  {\path{arXiv:hep-ph/0303187}}, \href
  {http://dx.doi.org/10.1103/PhysRevD.68.054018}
  {\path{doi:10.1103/PhysRevD.68.054018}}.

\bibitem{Atwood1997}
D.~Atwood, I.~Dunietz, A.~Soni, {Enhanced $CP$ violation with $B \to K D^0
  (\bar D^0)$ modes and extraction of the CKM angle $\gamma$}, Phys. Rev. Lett.
  78 (1997) 3257--3260.
\newblock \href {http://arxiv.org/abs/hep-ph/9612433}
  {\path{arXiv:hep-ph/9612433}}, \href
  {http://dx.doi.org/10.1103/PhysRevLett.78.3257}
  {\path{doi:10.1103/PhysRevLett.78.3257}}.

\bibitem{Atwood2001}
D.~Atwood, I.~Dunietz, A.~Soni, {Improved methods for observing $CP$ violation
  in $B^\pm \to K D$ and measuring the CKM phase $\gamma$}, Phys. Rev. D63
  (2001) 036005.
\newblock \href {http://arxiv.org/abs/hep-ph/0008090}
  {\path{arXiv:hep-ph/0008090}}, \href
  {http://dx.doi.org/10.1103/PhysRevD.63.036005}
  {\path{doi:10.1103/PhysRevD.63.036005}}.

\bibitem{Poluektov2004}
A.~Poluektov, et~al., {Measurement of $\phi_3$ with Dalitz plot analysis of
  $B^\pm \to D^{(*)} K^\pm$ decay}, Phys. Rev. D70 (2004) 072003.
\newblock \href {http://arxiv.org/abs/hep-ex/0406067}
  {\path{arXiv:hep-ex/0406067}}, \href
  {http://dx.doi.org/10.1103/PhysRevD.70.072003}
  {\path{doi:10.1103/PhysRevD.70.072003}}.

\bibitem{Gershon2009}
T.~Gershon, {On the measurement of the unitarity triangle angle $\gamma$ from
  $B^0 \to$ DK*0 decays}, Phys. Rev. D79 (2009) 051301.
\newblock \href {http://arxiv.org/abs/0810.2706} {\path{arXiv:0810.2706}},
  \href {http://dx.doi.org/10.1103/PhysRevD.79.051301}
  {\path{doi:10.1103/PhysRevD.79.051301}}.

\bibitem{Gershon2009a}
T.~Gershon, M.~Williams, {Prospects for the measurement of the unitarity
  triangle angle $\gamma$ from $B^0 \to DK^+ \pi^-$ Decays}, Phys. Rev. D80
  (2009) 092002.
\newblock \href {http://arxiv.org/abs/0909.1495} {\path{arXiv:0909.1495}},
  \href {http://dx.doi.org/10.1103/PhysRevD.80.092002}
  {\path{doi:10.1103/PhysRevD.80.092002}}.

\bibitem{Aaij2016}
R.~Aaij, et~al., {Constraints on the unitarity triangle angle $\gamma$ from
  Dalitz plot analysis of $B^0 \to D K^+ \pi^-$ decays}\href
  {http://arxiv.org/abs/1602.03455} {\path{arXiv:1602.03455}}.

\bibitem{Briere2009}
R.~A. Briere, et~al., {First model-independent determination of the relative
  strong phase between $D^0$ and $\bar D^0 \to K^0_S \pi^+ \pi^-$ and its
  impact on the CKM angle $\gamma/\phi_3$ measurement}, Phys. Rev. D80 (2009)
  032002.
\newblock \href {http://arxiv.org/abs/0903.1681} {\path{arXiv:0903.1681}},
  \href {http://dx.doi.org/10.1103/PhysRevD.80.032002}
  {\path{doi:10.1103/PhysRevD.80.032002}}.

\bibitem{Lowery2009}
N.~Lowrey, et~al., {Determination of the $D^0 \to K^- \pi^+ \pi^0$ and $D^0 \to
  K^-\pi^+\pi^+\pi^-$ coherence factors and average strong-phase differences
  using quantum-correlated measurements}, Phys. Rev. D80 (2009) 031105.
\newblock \href {http://arxiv.org/abs/0903.4853} {\path{arXiv:0903.4853}},
  \href {http://dx.doi.org/10.1103/PhysRevD.80.031105}
  {\path{doi:10.1103/PhysRevD.80.031105}}.

\bibitem{Libby2010}
J.~Libby, et~al., {Model-independent determination of the strong-phase
  difference between $D^0$ and $\bar{D}^0 \to K^0_{S,L} h^+ h^-$ ($h=\pi,K$)
  and its impact on the measurement of the CKM angle $\gamma/\phi_3$}, Phys.
  Rev. D82 (2010) 112006.
\newblock \href {http://arxiv.org/abs/1010.2817} {\path{arXiv:1010.2817}},
  \href {http://dx.doi.org/10.1103/PhysRevD.82.112006}
  {\path{doi:10.1103/PhysRevD.82.112006}}.

\bibitem{Charles2015}
J.~Charles, et~al., {Current status of the standard model CKM fit and
  constraints on $\Delta F=2$ new physics}, Phys. Rev. D91~(7) (2015) 073007.
\newblock \href {http://arxiv.org/abs/1501.05013} {\path{arXiv:1501.05013}},
  \href {http://dx.doi.org/10.1103/PhysRevD.91.073007}
  {\path{doi:10.1103/PhysRevD.91.073007}}.

\bibitem{Lees2013d}
J.~P. Lees, et~al., {Observation of direct $CP$ violation in the measurement of
  the Cabibbo-Kobayashi-Maskawa angle gamma with $B^\pm\to D^{(*)}K^{(*)\pm}$
  decays}, Phys. Rev. D87~(5) (2013) 052015.
\newblock \href {http://arxiv.org/abs/1301.1029} {\path{arXiv:1301.1029}},
  \href {http://dx.doi.org/10.1103/PhysRevD.87.052015}
  {\path{doi:10.1103/PhysRevD.87.052015}}.

\bibitem{Poluektov2010}
A.~Poluektov, et~al., {Evidence for direct $CP$ violation in the decay $B\to
  D^{(*)}K$, $D\to K_S \pi^+\pi^-$ and measurement of the CKM phase $\phi_3$},
  Phys. Rev. D81 (2010) 112002.
\newblock \href {http://arxiv.org/abs/1003.3360} {\path{arXiv:1003.3360}},
  \href {http://dx.doi.org/10.1103/PhysRevD.81.112002}
  {\path{doi:10.1103/PhysRevD.81.112002}}.

\bibitem{Horii2011}
Y.~Horii, et~al., {Evidence for the suppressed decay $B^- \to DK^-$, $D \to
  K^+\pi^-$}, Phys. Rev. Lett. 106 (2011) 231803.
\newblock \href {http://arxiv.org/abs/1103.5951} {\path{arXiv:1103.5951}},
  \href {http://dx.doi.org/10.1103/PhysRevLett.106.231803}
  {\path{doi:10.1103/PhysRevLett.106.231803}}.

\bibitem{Trabelsi2013}
K.~Trabelsi,
  \href{https://inspirehep.net/record/1210029/files/arXiv:1301.2033.pdf}{{Study
  of direct $CP$ in charmed $B$ decays and measurement of the CKM angle
  $\gamma$ at Belle}}, in: {7th Workshop on the CKM Unitarity Triangle (CKM
  2012) Cincinnati, Ohio, USA, September 28-October 2, 2012}, 2013.
\newblock \href {http://arxiv.org/abs/1301.2033} {\path{arXiv:1301.2033}}.
\newline\urlprefix\url{https://inspirehep.net/record/1210029/files/arXiv:1301.2033.pdf}

\bibitem{Aaij2012c}
R.~Aaij, et~al., {Observation of $CP$ violation in B$^\pm$→DK$^\pm$ decays},
  Phys. Lett. B712 (2012) 203--212, [Erratum: Phys. Lett.B713,351(2012)].
\newblock \href {http://arxiv.org/abs/1203.3662} {\path{arXiv:1203.3662}},
  \href {http://dx.doi.org/10.1016/j.physletb.2012.04.060,
  10.1016/j.physletb.2012.05.060} {\path{doi:10.1016/j.physletb.2012.04.060,
  10.1016/j.physletb.2012.05.060}}.

\bibitem{Aaij2013b}
R.~Aaij, et~al., {Observation of the suppressed ADS modes $B^\pm \to [\pi^\pm
  K^\mp \pi^+\pi^-]_D K^\pm$ and $B^\pm \to [\pi^\pm K^\mp \pi^+\pi^-]_D
  \pi^\pm$}, Phys. Lett. B723 (2013) 44--53.
\newblock \href {http://arxiv.org/abs/1303.4646} {\path{arXiv:1303.4646}},
  \href {http://dx.doi.org/10.1016/j.physletb.2013.05.009}
  {\path{doi:10.1016/j.physletb.2013.05.009}}.

\bibitem{Aaij2014a}
R.~Aaij, et~al., {Measurement of $CP$ asymmetry in $B^0_s \rightarrow D^{\mp}_s
  K^{\pm}$ decays}, JHEP 11 (2014) 060.
\newblock \href {http://arxiv.org/abs/1407.6127} {\path{arXiv:1407.6127}},
  \href {http://dx.doi.org/10.1007/JHEP11(2014)060}
  {\path{doi:10.1007/JHEP11(2014)060}}.

\bibitem{Aaij2014b}
R.~Aaij, et~al., {Measurement of the CKM angle $\gamma$ using $B^\pm \to D
  K^\pm$ with $D \to K^0_{\rm S} \pi^+\pi^-, K^0_{\rm S} K^+ K^-$ decays}, JHEP
  10 (2014) 097.
\newblock \href {http://arxiv.org/abs/1408.2748} {\path{arXiv:1408.2748}},
  \href {http://dx.doi.org/10.1007/JHEP10(2014)097}
  {\path{doi:10.1007/JHEP10(2014)097}}.

\bibitem{Aaij2014c}
R.~Aaij, et~al., {A study of $CP$ violation in $B^{\pm} \to DK^{\pm}$ and
  $B^{\pm} \to D\pi^{\pm}$ decays with $D \to K_S^0K^{\pm}\pi^{\mp}$ final
  states}, Phys. Lett. B733 (2014) 36--45.
\newblock \href {http://arxiv.org/abs/1402.2982} {\path{arXiv:1402.2982}},
  \href {http://dx.doi.org/10.1016/j.physletb.2014.03.051}
  {\path{doi:10.1016/j.physletb.2014.03.051}}.

\bibitem{Aaij2014d}
R.~Aaij, et~al., {Measurement of $CP$ violation parameters in $B^0 \to DK^{*0}$
  decays}, Phys. Rev. D90~(11) (2014) 112002.
\newblock \href {http://arxiv.org/abs/1407.8136} {\path{arXiv:1407.8136}},
  \href {http://dx.doi.org/10.1103/PhysRevD.90.112002}
  {\path{doi:10.1103/PhysRevD.90.112002}}.

\bibitem{LHCb2014}
{LHCb collaboration},
  \href{http://inspirehep.net/record/1388239/files/LHCb-CONF-2014-004.pdf}{{Improved
  constraints on $\gamma$: CKM2014 update}}, in: {8th International Workshop on
  the CKM Unitarity Triangle (CKM 2014) Vienna, Austria, September 8-12, 2014},
  CERN, CERN, 2014. Geneva, 2014.
\newline\urlprefix\url{http://inspirehep.net/record/1388239/files/LHCb-CONF-2014-004.pdf}

\bibitem{Dunietz1998}
I.~Dunietz, {Clean CKM information from $B_d(t) \to D^{*\mp} \pi^\pm$}, Phys.
  Lett. B427 (1998) 179--182.
\newblock \href {http://arxiv.org/abs/hep-ph/9712401}
  {\path{arXiv:hep-ph/9712401}}, \href
  {http://dx.doi.org/10.1016/S0370-2693(98)00304-9}
  {\path{doi:10.1016/S0370-2693(98)00304-9}}.

\bibitem{Aubert2005c}
B.~Aubert, et~al., {Measurement of time-dependent $CP$-violating asymmetries
  and constraints on $\sin(2\beta+\gamma)$ with partial reconstruction of $B
  \to D^{*\mp} \pi^\pm$ decays}, Phys. Rev. D71 (2005) 112003.
\newblock \href {http://arxiv.org/abs/hep-ex/0504035}
  {\path{arXiv:hep-ex/0504035}}, \href
  {http://dx.doi.org/10.1103/PhysRevD.71.112003}
  {\path{doi:10.1103/PhysRevD.71.112003}}.

\bibitem{Aubert2006b}
B.~Aubert, et~al., {Measurement of time-dependent $CP$ asymmetries in $B^0 \to
  D^{(*)\pm}\pi^\mp$ and $B^0 \to D^\pm \rho^\mp$ decays}, Phys. Rev. D73
  (2006) 111101.
\newblock \href {http://arxiv.org/abs/hep-ex/0602049}
  {\path{arXiv:hep-ex/0602049}}, \href
  {http://dx.doi.org/10.1103/PhysRevD.73.111101}
  {\path{doi:10.1103/PhysRevD.73.111101}}.

\bibitem{Bahinipati2011}
S.~Bahinipati, et~al., {Measurements of time-dependent $CP$ asymmetries in $B
  \to D^{*\mp} \pi^{\pm}$ decays using a partial reconstruction technique},
  Phys. Rev. D84 (2011) 021101.
\newblock \href {http://arxiv.org/abs/1102.0888} {\path{arXiv:1102.0888}},
  \href {http://dx.doi.org/10.1103/PhysRevD.84.021101}
  {\path{doi:10.1103/PhysRevD.84.021101}}.

\bibitem{Ronga2006}
F.~J. Ronga, et~al., {Measurements of $CP$ violation in $B^0 \to D^{*-} \pi^+$
  and $B^0 \to D^- \pi^+$ decays}, Phys. Rev. D73 (2006) 092003.
\newblock \href {http://arxiv.org/abs/hep-ex/0604013}
  {\path{arXiv:hep-ex/0604013}}, \href
  {http://dx.doi.org/10.1103/PhysRevD.73.092003}
  {\path{doi:10.1103/PhysRevD.73.092003}}.

\bibitem{Aubert2008b}
B.~Aubert, et~al., {Time-dependent Dalitz plot analysis of $B^0 \to D^\mp K^0
  \pi^\pm$ decays}, Phys. Rev. D77 (2008) 071102.
\newblock \href {http://arxiv.org/abs/0712.3469} {\path{arXiv:0712.3469}},
  \href {http://dx.doi.org/10.1103/PhysRevD.77.071102}
  {\path{doi:10.1103/PhysRevD.77.071102}}.

\bibitem{Amhis2015}
Y.~S. Amhis, T.~Aushev, M.~Jung,
  \href{https://inspirehep.net/record/1400799/files/arXiv:1510.07321.pdf}{{Mixing
  and mixing-related $CP$ violation in the $B$ system}}, in: {8th International
  Workshop on the CKM Unitarity Triangle (CKM 2014) Vienna, Austria, September
  8-12, 2014}, 2015.
\newblock \href {http://arxiv.org/abs/1510.07321} {\path{arXiv:1510.07321}}.
\newline\urlprefix\url{https://inspirehep.net/record/1400799/files/arXiv:1510.07321.pdf}

\bibitem{Aaij2014e}
R.~Aaij, et~al., {Measurement of the $CP$-violating phase $\phi_s$ in
  $\bar{B}^{0}_{s}\to D_{s}^{+}D_{s}^{-}$ decays}, Phys. Rev. Lett. 113~(21)
  (2014) 211801.
\newblock \href {http://arxiv.org/abs/1409.4619} {\path{arXiv:1409.4619}},
  \href {http://dx.doi.org/10.1103/PhysRevLett.113.211801}
  {\path{doi:10.1103/PhysRevLett.113.211801}}.

\bibitem{Aaij2014f}
R.~Aaij, et~al., {Measurement of the CP-violating phase $\phi_s$ in
  $\overline{B}^0_s\rightarrow J/\psi \pi^+\pi^-$ decays}, Phys. Lett. B736
  (2014) 186--195.
\newblock \href {http://arxiv.org/abs/1405.4140} {\path{arXiv:1405.4140}},
  \href {http://dx.doi.org/10.1016/j.physletb.2014.06.079}
  {\path{doi:10.1016/j.physletb.2014.06.079}}.

\bibitem{Aad2016}
G.~Aad, et~al., {Measurement of the $CP$-violating phase $\phi_s$ and the
  $B^0_s$ meson decay width difference with $B^0_s \to J/\psi\phi$ decays in
  ATLAS}\href {http://arxiv.org/abs/1601.03297} {\path{arXiv:1601.03297}}.

\bibitem{Khachatryan2015}
V.~Khachatryan, et~al., {Measurement of the $CP$-violating weak phase $ \phi_s
  $ and the decay width difference $ \Delta \Gamma_{ \mathrm{s} }$ using the $
  \mathrm{B^0_s} \to \mathrm{J} / \psi \phi(1020) $ decay channel in $pp$
  collisions at $\sqrt{s} =$ 8 TeV} (2015).
\newblock \href {http://arxiv.org/abs/1507.07527} {\path{arXiv:1507.07527}}.

\bibitem{Aad2014a}
G.~Aad, et~al., {Flavor tagged time-dependent angular analysis of the $B_s
  \rightarrow J/\psi \phi$ decay and extraction of $\Delta\Gamma$s and the weak
  phase $\phi_s$ in ATLAS}, Phys. Rev. D90~(5) (2014) 052007.
\newblock \href {http://arxiv.org/abs/1407.1796} {\path{arXiv:1407.1796}},
  \href {http://dx.doi.org/10.1103/PhysRevD.90.052007}
  {\path{doi:10.1103/PhysRevD.90.052007}}.

\bibitem{Aaltonen2012}
T.~Aaltonen, et~al., {Measurement of the bottom-strange meson mixing phase in
  the full CDF data set}, Phys. Rev. Lett. 109 (2012) 171802.
\newblock \href {http://arxiv.org/abs/1208.2967} {\path{arXiv:1208.2967}},
  \href {http://dx.doi.org/10.1103/PhysRevLett.109.171802}
  {\path{doi:10.1103/PhysRevLett.109.171802}}.

\bibitem{Abazov2012}
V.~M. Abazov, et~al., {Measurement of the $CP$-violating phase $\phi_s^{J/\psi
  \phi}$ using the flavor-tagged decay $B_s^0 \rightarrow J/\psi \phi$ in 8
  fb$^{-1}$ of $p \bar p$ collisions}, Phys. Rev. D85 (2012) 032006.
\newblock \href {http://arxiv.org/abs/1109.3166} {\path{arXiv:1109.3166}},
  \href {http://dx.doi.org/10.1103/PhysRevD.85.032006}
  {\path{doi:10.1103/PhysRevD.85.032006}}.

\bibitem{Eigen2014}
G.~Eigen, G.~Dubois-Felsmann, D.~G. Hitlin, F.~C. Porter, {Global CKM fits with
  the scan method}, Phys. Rev. D89~(3) (2014) 033004.
\newblock \href {http://arxiv.org/abs/1301.5867} {\path{arXiv:1301.5867}},
  \href {http://dx.doi.org/10.1103/PhysRevD.89.033004}
  {\path{doi:10.1103/PhysRevD.89.033004}}.

\bibitem{Eigen2015}
G.~Eigen, G.~Dubois-Felsmann, D.~G. Hitlin, F.~C. Porter,
  \href{http://inspirehep.net/record/1351218/files/arXiv:1503.02289.pdf}{{Global
  fits of the CKM matrix with the SCAN method}}, in: {50 Years of CP Violation
  London, UK, July 10-11, 2014}, 2015.
\newblock \href {http://arxiv.org/abs/1503.02289} {\path{arXiv:1503.02289}}.
\newline\urlprefix\url{http://inspirehep.net/record/1351218/files/arXiv:1503.02289.pdf}

\bibitem{Harrison1998}
D.~Boutigny, et~al.,
  \href{http://www-public.slac.stanford.edu/sciDoc/docMeta.aspx?slacPubNumber=SLAC-R-504}{{The
  BaBar physics book: Physics at an asymmetric $B$ factory}}, in: {Workshop on
  Physics at an Asymmetric $B$ Factory (BaBar Collaboration Meeting) Pasadena,
  California, September 22-24, 1997}, 1998.
\newline\urlprefix\url{http://www-public.slac.stanford.edu/sciDoc/docMeta.aspx?slacPubNumber=SLAC-R-504}

\bibitem{Ciuchini2001}
M.~Ciuchini, G.~D'Agostini, E.~Franco, V.~Lubicz, G.~Martinelli, F.~Parodi,
  P.~Roudeau, A.~Stocchi, {2000 CKM triangle analysis: A critical review with
  updated experimental inputs and theoretical parameters}, JHEP 07 (2001) 013.
\newblock \href {http://arxiv.org/abs/hep-ph/0012308}
  {\path{arXiv:hep-ph/0012308}}, \href
  {http://dx.doi.org/10.1088/1126-6708/2001/07/013}
  {\path{doi:10.1088/1126-6708/2001/07/013}}.

\bibitem{Bona2005}
M.~Bona, et~al., {The 2004 UTfit collaboration report on the status of the
  unitarity triangle in the standard model}, JHEP 07 (2005) 028.
\newblock \href {http://arxiv.org/abs/hep-ph/0501199}
  {\path{arXiv:hep-ph/0501199}}, \href
  {http://dx.doi.org/10.1088/1126-6708/2005/07/028}
  {\path{doi:10.1088/1126-6708/2005/07/028}}.

\bibitem{Bona2006}
M.~Bona, et~al., {The unitarity triangle fit in the standard model and hadronic
  parameters from lattice QCD: A reappraisal after the measurements of $\Delta
  m_s$ and BR$(B \to \tau \nu_\tau)$}, JHEP 10 (2006) 081.
\newblock \href {http://arxiv.org/abs/hep-ph/0606167}
  {\path{arXiv:hep-ph/0606167}}, \href
  {http://dx.doi.org/10.1088/1126-6708/2006/10/081}
  {\path{doi:10.1088/1126-6708/2006/10/081}}.

\bibitem{Ali2003}
A.~Ali, {CKM phenomenology and $B$ meson physics: Present status and current
  issues}, in: {31st International Meeting on Fundamental Physics (IMFP 2003)
  Cangas de Onis, Asturias, Spain, February 24-28, 2003}, 2003.
\newblock \href {http://arxiv.org/abs/hep-ph/0312303}
  {\path{arXiv:hep-ph/0312303}}.

\bibitem{Narsky2014}
I.~Narsky, F.~C. Porter,
  \href{http://www.wiley-vch.de/publish/dt/books/ISBN3-527-41086-4}{{Statistical
  analysis techniques in particle physics}}, Wiley-VCH, Weinheim, Germany,
  2014.
\newline\urlprefix\url{http://www.wiley-vch.de/publish/dt/books/ISBN3-527-41086-4}

\bibitem{Shao2003}
J.~Shao, {Mathematical Statistics}, 2nd Edition, Springer-Verlag, New York,
  USA, 2003.

\end{thebibliography}

\end{document}